%% file: torsion_diffusion_march12_10.tex
\documentclass[letterpaper, reqno, 12pt]{amsart}
\usepackage{graphicx, amssymb, setspace}
\usepackage{color}
\usepackage{amsmath}
\usepackage{psfrag}
\usepackage{fullpage}
\usepackage{subfig}
\usepackage{setspace}

\newcommand{\pfx}[2]{\dfrac{\partial{#1}}{\partial{#2}}}

\title[Degradation and healing in a generalized neo-Hookean solid]{Degradation and Healing in a generalized neo-Hookean solid due to infusion of a fluid}
\author{Satish Karra}
\address{Satish Karra,
Texas A\&M University\\
Department of Mechanical Engineering\\
3123 TAMU\\
College Station TX 77843-3123\\
United States of America
}
\email{satkarra@tamu.edu}

\author{K. R. Rajagopal}
\address{K. R. Rajagopal
Texas A\&M University\\
Department of Mechanical Engineering\\
3123 TAMU\\
College Station TX 77843-3123\\
United States of America
}
\email{krajagopal@tamu.edu}
\date{\today}

\begin{document}

\bibliographystyle{plain}

\begin{abstract}
The mechanical response and load bearing capacity of high performance polymer composites changes due to diffusion of a fluid, temperature, oxidation or the extent of the deformation. Hence, there is a need to study the response of bodies under such degradation mechanisms. In this paper, we study the effect of degradation and healing due to the  diffusion of a fluid on the response of a solid which prior to the diffusion can be described by the generalized neo-Hookean model. We show that a generalized neo-Hookean solid - which behaves like an elastic body (i.e., it does not produce entropy) within a purely mechanical context - creeps and stress relaxes when infused with a fluid and behaves like a body whose material properties are time dependent. We specifically investigate the torsion of a generalized neo-Hookean circular cylindrical annulus infused with a fluid. The equations of equilibrium for a generalized neo-Hookean solid are solved together with the convection-diffusion equation for the fluid concentration. Different boundary conditions for the fluid concentration are also considered. We also solve the problem for the case when the diffusivity of the fluid depends on the deformation of the generalized neo-Hookean solid. 


\end{abstract}

\maketitle

\section{Introduction}
As elastic bodies are incapable of producing any entropy, this would be the proper definition of an elastic body within a thermodynamic context. Elastic bodies also cannot stress relax, i.e., when the strain is kept constant in time the stresses within the body cannot change with time. However, the stresses in a body that is initially elastic, which subsequently undergoes chemical reactions due to interactions with the environment or which is subject to the effects of electro-magnetic radiation such as ultra-violet rays, could change with time and the body's response  ceases to be that of an elastic body. The chemical reactions or the interactions with the environment invariably produce entropy. However, it could happen that when the chemical reactions cease and the body is isolated from the environment the body ceases to produce entropy, that is, it becomes a  different elastic body. The body, due to its exposure to the environment can undergo deterioration or enhancement with respect to its load carrying capacity or other properties. While moisture diffusion in a polymer can cause degradation of the body in that its load carrying capacity goes down, a body such as biological matter can be strengthened due to a drug that is being injected. Though the stress in the body might decrease when the strain is held constant in bodies that are undergoing degradation, this phenomenon is quite different from the stress relaxation observed in viscoelastic solids (Christensen \cite{christensen.rm:theory}, Wineman and Rajagopal \cite{wineman.as.rajagopal.kr:mechanical}).

 There is  need for a good understanding of the degradation of materials as this is relevant to a plethora of applications. For instance, the mechanical properties of high performance polymer composites (like polyimides) used in hypersonic vehicles  vary due to the effect of high temperature, diffusion of moisture and the subsequent oxidation.  Chen and Tyler \cite{chen.r.tyler.dr:origin} have recently shown that polymers can also degrade due to deformation.  In this short paper we are interested in studying the response of a body whose properties are changing due to the presence of a fluid, the extent of the change depending on the concentration of the fluid.

There has been considerable interest in the damage and degradation of polymeric solids and polymer composites. However, most of these studies appeal to ad hoc constitutive equations. Also, many studies appeal to the notion of  a hygrothermal expansion coefficient (see Bouadi and Sun \cite{bouadi.h.sun.ct:hygrothermal}, Cai and Weitsman \cite{cai.lw.weitsman.y:nonfickian}) and such an approach would not be appropriate when large deformations due to swelling are involved. A general thermodynamic framework has been developed to describe damage in composites by Weitsman \cite{weitsman.y:coupled}. This work introduces the notion of a damage tensor which is essentially a tensor internal variable. While the study is very detailed and has a thermodynamic basis, the theory however involves several material moduli that depend on as many as 32 invariants in the case of transversely isotropic materials, making it impractical to put the theory to use as it is impossible to develop an experimental program to determine the numerous material functions that characterize the body. Moreover, such theories lead to intractable mathematical equations. While there have been several other studies concerning the diffusion of moisture in composites (e.g., Snead and Palazotto \cite{snead.jm.palazotto.an:moisture}, Lee and Yen \cite{lee.sy.yen.wj:hygrothermal}, Bouadi and Sun \cite{bouadi.h.sun.ct:hygrothermal1990}, and Kardomateas and Chung \cite{kardomateas.ga.chung.cb:boundary1994}), they are primarily parametric studies in which the material moduli are allowed to depend on the moisture content according to some pre-assigned manner and not as a consequence of a reasonable convection-diffusion equation; that is the coupled problem for the deformation of the composite and the diffusion of moisture is not solved simultaneously. In general, the diffusivity depends on the temperature, moisture content, strain, and stress. In fact, the diffusivity can depend on the history of these quantities. Weitsman \cite{weitsman.y:stress1987} and Roy et al. \cite{roy.s.vengadassalam.k.wang.y.park.s.liechi.km:characterization2006} have also addressed damage due to diffusion but not in the manner advocated here.

Using ideas in multi-network theories for polymeric materials (see Tobolsky and Andrews \cite{tobolsky.av.andrews.rd:systems1945}, Tobolsky et al. \cite{tobolsky.av.prettyman.ib.dillon.jh:stress1944}), Wineman and Rajagopal \cite{wineman.as.rajagopal.kr:constitutive1990} and Rajagopal and Wineman \cite{rajagopal.kr.wineman.as:constitutive1992} developed a theory applicable to the large deformation of polymeric solids that exhibit scisson and healing and permanent set, within a purely mechanical context. This work was extended in a series of papers by Huntley et al. \cite{huntley.he.wineman.as.rajagopal.kr:load2001}, \cite{huntley.he.wineman.as.rajagopal.kr:stress1997}, \cite{huntley.he.wineman.as.rajagopal.kr:chemorheological1996} to deal with a variety of deformations involving elastomeric solids that undergo permanent set. The scission and healing that takes place can be viewed within the context of deformation induced damage and due to cross-linkings that take place. These works within a purely mechanical context were later generalized to include thermal effects (see Wineman and Shaw \cite{wineman.as.shaw.j:scission2002}, Shaw et al. \cite{shaw.ja.jones.as.wineman.as:chemorheological2005}).

 A general thermodynamic framework which takes into account chemical reactions (with chemical kinetics), diffusion and thermal effects needs to be put into place in order to develop constitutive relations to study the degradation of materials (like polyimides) due to chemical reactions. As a first step towards such a goal, in this paper, we shall first solve the problem of torsion of a degrading body, which when there is no degradation taking place responds like a generalized neo-Hookean body. We also look at healing (or strengthening) of a generalized neo-Hookean body when a fluid is infused. We assume the body to be  a cylindrical annulus of finite length. We shall study  the torsion of a cylindrical annulus through which a fluid is infusing. We will assume that the infusion of the fluid is radial and thus the degradation or healing takes place radially. We introduce a parameter that is a measure of the degradation or healing which in virtue of the diffusion being radial also varies radially. The material parameters are assumed to be functions of the variables that quantify the degradation or healing. We need to then solve the convection-diffusion equation that governs the diffusion of the fluid in tandem with the  equilibrium equations for the torsion problem; we look at how the moment varies with time when the angle of twist of the cylinder is kept constant in time, and how the angle of twist of the cylinder varies when the moment applied to the cylinder is kept constant. We find that the moment that is needed to maintain the angular displacement decreases with time, that is the body stress relaxes, when the material is degrading, but as we observed earlier such a decrease of stress is very different from the stress relaxation observed in viscoelastic bodies. The stresses can decrease in the body due to a variety of reasons such as degradation of properties of the body due to chemical reactions, aging, etc., and it is important to recognize the reason for the ``stress relaxation". Rajagopal and Wineman \cite{rajagopal.kr.wineman.as:note2004} have studied stress relaxation due to aging and they find that in marked contrast to the stress relaxation observed in viscoelastic bodies, the decrease in stress is dependent on the geometry of the body. This aspect related to stress relaxation, namely its dependence on the geometry is what sets stress relaxation in viscoelastic materials apart from stress relaxation that manifests itself due to most degradation theories. In order to highlight this difference, Rajagopal and Wineman \cite{rajagopal.kr.wineman.as:note2004} considered the torsion of a viscoelastic cylinder that ages. They found that the stress relaxation can be split into two parts, one that is a consequence of the body being viscoelastic, which is independent of the size of the specimen, and another part due to the aging of the cylindrical specimen, this being dependent on the radius of the cylinder undergoing torsion. We also find that the angular displacement undergone by the body increases with time for the applied moment, when the material is degrading. When one looks at healing due to diffusion of a fluid, we find that that the moment needed to maintain a given angular displacement increases with time, whereas the body's angular displacement decreases with time when one applies a constant moment.

We will now turn to a discussion of the response characteristics of the undamaged elastic solid, namely the power-law neo-Hookean solid. The generalized (or power-law) neo-Hookean elastic model (see Knowles \cite{knowles.jk:finite1977})  allows for softening and stiffening under simple shear when the power-law parameter ($n$) is lesser or greater than unity, respectively. Softening and stiffening which are seen in real materials cannot be explained using the classical neo-Hookean model while some of it possibly can be explained in terms of the generalized neo-Hookean solid. In any event, one should recognize that such models are merely caricatures of reality and in this study we are mainly interested in obtaining some understanding of the degradation due to the infusion of fluid. On setting the power-law parameter ($n$) to 1, one obtains the classical neo-Hookean model. Knowles \cite{knowles.jk:finite1977} studied the anti-plane shear of a power-law neo-Hookean body that has a crack, and noted that for $n\geq\frac{1}{2}$, the equation of equilibrium for anti-plane shear is always elliptic and that the ellipticity is lost when $n<\frac{1}{2}$.  Hou and Zhang \cite{hou.hs.zhang.y:effect1990} have studied the stability of a power-law neo-Hookean cylinder under axial stretching and constant radial traction. The power-law model has also been used in several subsequent studies (see Saccomandi \cite{saccomandi.g:some2005}).

Rajagopal and Tao \cite{rajagopal.kr.tao.l:inhomogeneous1992} studied inhomogeneous deformations in a wedge of a generalized neo-Hookean material numerically, and later, Mcleod and Rajagopal  \cite{mcleod.jb.rajagopal.kr:inhomogeneous1999}  re-investigated the same problem with a view towards establishing existence of solutions to the governing equations.   Tao and Rajagopal \cite{tao.l.rajagopal.kr.wineman.as:circular1992} have analyzed the problem of inhomogeneous circular shear combined with torsion for a genearalized neo-Hookean material and have obtained exact solutions for certain values of the material constants. This study was followed by an analysis by Zhang and Rajagopal \cite{zhang.jp.rajagopal.kr:some1992} concerning steady and unsteady inhomogeneous shear of a slab and a cylindrical annulus. The reason for citing these studies is that in all of \cite{rajagopal.kr.tao.l:inhomogeneous1992,mcleod.jb.rajagopal.kr:inhomogeneous1999,tao.l.rajagopal.kr.wineman.as:circular1992,zhang.jp.rajagopal.kr:some1992}, ``boundary layer'' type solutions (that is, close to the boundary the deformations are inhomogeneous while the deformations are mainly homogenous in the inner core region of the body) have been observed for certain values of the material parameters and also the strain is concentrated in these boundary layers in that the strain gradients are very large in these layers. Thus, one would expect damage, degradation and failure to occur in these boundary layers. A systematic method to obtain approximate equations  within the boundary layer of  a generalized neo-Hookean solid similar to the boundary layer theory in fluid mechanics has been discussed by Rajagopal \cite{rajagopal.kr:deformations1996}. To show the efficacy of this method, the approximate boundary layer equations were solved for the circumferential shear problem and were compared to the full solution.  

While there have been several studies concerning generalized neo-Hookean materials within the context of a purely mechanical setting, as noted above, there is no study concerning the degradation or enhancement of the properties of generalized neo-Hookean materials, which would be important as materials such as elastomers for which the generalized neo-Hookean model is used, degrade or heal due to the diffusion of moisture or other chemicals.  

 Recently, there have been a few investigations of the deformation of an elastic solid that is undergoing degradation due to the influence of a diffusant. Muliana et al. \cite{muliana.a.rajagopal.kr.subramanian.sc:degradation2009} analyzed the response of a composite cylinder that is undergoing degradation due to the diffusion of a fluid, and Darbha and Rajagopal \cite{darbha.s.rajagopal.kr:unsteady2009} studied unsteady motions of a slab through which a fluid is diffusing. They investigated the unsteady shear of an infinite slab of finite thickness and a cylindrical annulus of infinite length under degradation due to diffusion of a chemical species. In both the studies the body was assumed to be a linearized elastic body and the material moduli were assumed to vary linearly with the concentration of the diffusing species. 

  Rajagopal \cite{rajagopal.kr:boundary1994} assumed the shear modulus of a generalized neo-Hookean material to depend on temperature and solved the problem of inhomogeneous shear of an infinite slab. Boundary layer type solutions were obtained based on the nature of the boundary conditions for the temperature. Although, this paper was not written within the context of understanding degradation due to temperature, one can extend this work by assuming that all the material moduli depend on temperature in a certain fashion and study the effect of temperature on the response of the body. It is expected that boundary layers would develop due to degradation and the failure of the body would be determined by the stresses in this boundary layer. 
 
The current paper is organized as follows. In section \ref{sec1}, the problem of torsion of a finite cylindrical annulus comprised of a generalized neo-Hookean material through which a fluid is diffusing, is set up. In section \ref{sec2}, we specifically assume that the degradation or healing is due to diffusion and also assume that material moduli vary linearly with the concentration of the diffusant, and obtain the solutions to the convection-diffusion equation for the diffusing species as well as the equilibrium equations. We also obtain an expression for the moment that is applied in terms of the the angular displacement and the material parameters. In section \ref{sec3}, the results are discussed in detail, followed by final conclusions in section \ref{sec4}.
  
\section{Torsion of a Cylindrical Annulus undergoing Degradation} \label{sec1}

Let $\kappa_R$ and $\kappa_t$ denote the reference and the current configuration of a body, respectively. The motion $\chi_{\kappa_R}$ is a one-one mapping that assigns to each point $\mathbf{X}\in \kappa_R$, a point $\mathbf{x}\in \kappa_t$, at a time $t$, i.e.,
\begin{equation}
 \mathbf{x} = \chi_{\kappa_R}(\mathbf{X},t).
\end{equation}
 The gradient of motion (or the deformation gradient) $\mathbf{F}$ is defined by
\begin{equation}
 \mathbf{F}:=\frac{\partial \chi_{\kappa_R}}{\partial \mathbf{X}},
\end{equation}
with the velocity $\mathbf{v}$ defined as
\begin{align}
\mathbf{v}:= \frac{\partial \chi_{\kappa_R}}{\partial t}.
\end{align}
Let $(R,\Theta, Z)$ and $(r, \theta, z)$ be cylindrical co-ordinates in the reference and current configuration, respectively. Consider a cylindrical annulus of height $H$ with inner radius $R_i$ and outer radius $R_o$  under torsion, whose motion is given by
\begin{align}
 r =  R, \quad \theta = \Theta + f(z, t), \quad z = Z. \label{mot}
\end{align}
The deformation gradient $\mathbf{F}$ and the left Cauchy-Green stretch tensor $\mathbf{B}$ associated with (\ref{mot}) are given by
\begin{align}
 \mathbf{F} = \left[ \begin{array}{ccc}
                                               1 &0 &0\\
						0& 1 & rf_z\\
						0 & 0 &1
                                              \end{array}\right], \quad 
 \mathbf{B} := \mathbf{F}\mathbf{F}^T = \left[ \begin{array}{ccc}
                                               1 &0 &0\\
						0& 1+\left(rf_z \right)^2 & rf_z\\
						0 & rf_z &1
                                              \end{array}\right].
\end{align}
Thus, the first invariant of $\mathbf{B}$, $I_1 = tr(\mathbf{B}) = 3+\left(rf_z \right)^2 $, where $f_z := \dfrac{\partial f(z, t)}{\partial z}$,  and $tr(.)$ is the trace of a second order tensor. 

The Cauchy stress tensor $\mathbf{T}$ in an incompressible generalized neo-Hookean material, is given by
\begin{align}
 \mathbf{T} = -p\mathbf{I} + \mu \left[1+\frac{b}{n}\left(I_1 -3 \right) \right]^{n-1}\mathbf{B}, \label{T1}
\end{align}
where $-p\mathbf{I}$ is the spherical part due to the constraint of incompressibility, and $\mu \left[1+\frac{b}{n}\left(I_1 -3 \right) \right]^{n-1}$ is the generalized shear modulus, $\mu$ being the shear modulus at zero stretch. In general, the degradation or healing of the generalized neo-Hookean material can be caused by diffusion, temperature, electromagnetic radiation, etc. We shall denote the variable that is a measure of the  degradation or healing by $\alpha$,  i.e., when degradation due to diffusion is considered, $\alpha$ would be the concentration of the diffusing species; $\alpha$ would be temperature if we have degradation or healing due to temperature, etc., with an appropriate governing equation for the variable. We shall refer to $\alpha$ as the degradation/healing parameter.  We shall assume that the material moduli are functions of the degradation/healing parameter i.e., $n = n(\alpha)$, $\mu = \mu(\alpha)$, $b =b(\alpha)$, which leads to changes in the response characteristics of the body. In view of the geometry of the body, and the assumed form for the deformation field, we shall further assume that the degradation/healing parameter varies only with the radius and time, i.e., $\alpha = \alpha(r, t)$. Then, the material moduli would be functions of the radius and time and (\ref{T1}) reduces to the form
\begin{align}
 \mathbf{T} = -p\mathbf{I} + \Phi(r, z, t)\mathbf{B}, \label{T}
\end{align}
where $\Phi(r, z, t) := \mu(r, t) \left[1+\frac{b(r, t)}{n(r, t)}\left(r f_z\right)^2 \right]^{n(r, t)-1}$.

 We shall consider the diffusion to be reasonably slow and that the inertial effects in the solid can be neglected. It is important to bear in mind that the temporal effects are not being ignored. One could choose to view time as a parameter. Neglecting the body force, the balance of linear momentum reduces to the equations of equilibrium,
\begin{align}
 \text{div} \ \mathbf{T} = 0, \label{linmom}
\end{align}
where div(.) denotes the divergence operator in the current configuration.
 We assume that the concentration of the diffusing species (c) is governed by the following convection-diffusion equation
\begin{align}
\pfx{c}{t} + \text{div}  \left(c  \mathbf{v} \right) = \text{div} \left(D \, \text{grad}(c) \right) , \label{con-diff}
\end{align}
where $D$ is the diffusivity and in general it could depend on the deformation as well as the concentration, grad$(.)$ is the gradient based on the current configuration. In this study, we assume that the diffusion is very slow so that the velocity of the fluid that is diffusing can be neglected. The second term on the left side of (\ref{con-diff}) involves the derivative of the velocity and of course it is possible that even if the velocity is small its derivatives need not be small. We shall however assume that the derivatives of the velocity are small and thus we neglect the second term in (\ref{con-diff}). We shall assume that $D$ depends on the deformation, thus the equation governing the diffusion of the fluid is coupled with the balance of linear momentum. Since the concentration is only a function of the radius $r$ and time $t$, equation (\ref{con-diff}) reduces to
\begin{align}
 \dfrac{\partial c(r, t)}{\partial t} = \dfrac{1}{r}\dfrac{\partial }{\partial r}\left(D r \dfrac{\partial c(r, t)}{\partial r} \right). \label{diff}
\end{align}

Next, we shall document the balance of energy for the generalized neo-Hookean solid and fluid as follows (see Truesdell et al. \cite{truesdell.c.noll.w.antmann.ss:non2004}):
\begin{align}
\rho^i \frac{d\varepsilon^i}{dt} = \mathbf{T}^i.\mathbf{L}^i - \text{div}\mathbf{q}^i + \rho^i r^i + E^i , \quad i = \text{solid}, \text{fluid}, \label{pie}
\end{align}
where $\varepsilon^i$, $\mathbf{q}^i$, $r^i$ are the specific internal energy, heat flux, radiant heating associated with the $i$-th component and $E^i$ is the energy supplied to the $i$-th constituent from the other constituents.
 We shall ignore the energy equation associated with the fluid and the contributions due to the interactions between the fluid and the solid ($E^s$) in the energy balance for the solid. We shall however later on incorporate the energy associated with fluid indirectly by assuming that the internal energy of the solid depends on the concentration of the diffusing fluid.  This is clearly a crude approximation. In such a case, the energy equation is merely,  
\begin{align}
 \rho \dot{\varepsilon} = \mathbf{T}.\mathbf{L} - \text{div} \mathbf{q} + \rho r, \label{EE}
\end{align}
where $\varepsilon$ is the internal energy of the solid, $\mathbf{L}$ is the velocity gradient of the solid, $\mathbf{q}$ is the heat flux, and $r$ is the specific radiant heating.  Further assumption that the internal energy of the solid depends on the temperature ($T$), concentration of the diffusing fluid and the deformation gradient of the solid, i.e.,
\begin{align}
 \varepsilon = \varepsilon \left(T, c, \mathbf{F}\right),
\end{align}
leads to 
\begin{align}
 \rho \pfx{\varepsilon}{T} \dot{T} + \rho \pfx{\varepsilon}{c} \dot{c}+ \rho \pfx{\varepsilon}{\mathbf{F}} \mathbf{F}^{T}.\mathbf{L}  = \mathbf{T}.\mathbf{L} - \text{div} \mathbf{q} + \rho r, \label{EE1}
\end{align}
where $(.)^T$ is the transpose of a second-order tensor.
We shall define the specific Helmholtz potential as $\Psi = \varepsilon - T \eta$, where $\eta$ is the specific entropy. Then (\ref{EE1}) reduces to
\begin{align}
 \rho \pfx{\varepsilon}{T} \dot{T} + \rho \pfx{\Psi}{\mathbf{F}} \mathbf{F}^{T}.\mathbf{L} - \rho T \dfrac{\partial^2 \Psi}{\partial T \partial \mathbf{F}} \mathbf{F}^T. \mathbf{L} + \rho \pfx{\Psi}{c}\dot{c} - \rho T \dfrac{\partial^2 \Psi}{\partial T \partial c} \dot{c} = \mathbf{T}.\mathbf{L} - \text{div} \mathbf{q} + \rho r. \label{EE2}
\end{align}
where we have also used that $\eta = -\pfx{\Psi}{T}$.
Now, if we set $\mathbf{T} = -\pi\mathbf{I} + \rho \pfx{\Psi}{\mathbf{F}} \mathbf{F}^{T}$, where $\pi$ is the Lagrange multiplier due to the constraint of incompressibility (given by $tr(\mathbf{L}) = 0$), then in the absence of radiation along with the assumption of Fourier's relation for heat conduction,
\begin{align}
 \mathbf{q} = -k \; \text{grad} (T),
\end{align}
(\ref{EE2})  reduces to
\begin{align}
  \rho C_v \dot{T} - \rho T \dfrac{\partial^2 \Psi}{\partial T \partial \mathbf{F}} \mathbf{F}^T. \mathbf{L}+ \rho \pfx{\Psi}{c}\dot{c} - \rho T \dfrac{\partial^2 \Psi}{\partial T \partial c} \dot{c}  =   \text{div}\left( k \; \text{grad}(T) \right) ,  \label{temp}
\end{align}
where $C_v =  \pfx{\varepsilon}{T} = -T \dfrac{\partial^2 \Psi}{\partial T^2}$ is the specific heat capacity and $k$ is the heat conductivity, both of which could depend on the deformation, temperature as well as concentration of the diffusing fluid (in general, it could depend on the degradation/healing parameter $\alpha$). We also note that for a generalized neo-Hookean body, the specific Helmholtz potential is given by
\begin{align}
 \Psi = \dfrac{\mu}{2\rho b} \left\lbrace \left[1+ \dfrac{b}{n} \left(I_1-3 \right)  \right]^n -1  \right\rbrace. \label{psineo}
\end{align}

In general, along with infusion of a fluid, the body could be subject to high temperature which could cause additional degradation, and so the material parameters could depend on temperature as well i.e., $n = n(c, T)$, $\mu = \mu(c, T)$, $b= b(c, T)$. In such a case, one ought to solve (\ref{linmom}), (\ref{diff}), (\ref{temp}), simultaneously along with (\ref{psineo}). In what follows, we shall ignore the thermal effects and study the problem of degradation or healing only due to the diffusion of a fluid.

\section{Degradation and Healing due to Diffusion}\label{sec2}

It follows from  (\ref{T}) and (\ref{linmom}) that
\begin{align}
 &\dfrac{\partial  }{\partial r}\left( -p+\Phi(r, z, t) \right) -r \Phi(r, z, t) \left(f_z \right) ^2 = 0 ,  \label{lin1}\\
&\dfrac{1}{r}\dfrac{\partial }{\partial \theta} \left( -p+\Phi(r, z, t)\right) + \dfrac{\partial}{\partial z}\left(\Phi(r, z, t) r f_z \right) = 0, \label{lin2} \\
&\dfrac{\partial}{\partial z}\left(-p+\Phi(r, z, t)\right) = 0 .\label{lin3}
\end{align}
 In solving (\ref{lin1}), (\ref{lin2}) and (\ref{lin3}), time $t$ is treated as a parameter i.e., we are solving the balance of linear momentum (\ref{linmom}) at every instant of time assuming that the problem is quasi-static.

Next, from (\ref{lin3}),  it follows that
\begin{align}
 -p+\Phi(r, z, t) = g(\theta, r) \label{g}.
\end{align}
Taking the derivative of (\ref{lin1}) with respect to $\theta$, and using (\ref{g}), we obtain that
\begin{align}
 &\pfx{}{r}\left(\pfx{g(r, \theta)}{\theta} \right) = 0, \notag \\
\Rightarrow & \pfx{g(r, \theta)}{\theta}  = h(\theta), \label{h}
\end{align}
and therefore  (\ref{g}), (\ref{h}) reduce to
\begin{align}
 \dfrac{\partial }{\partial \theta} \left( -p +\Phi(r, z, t) \right) = h(\theta), \label{lin4}
 \end{align}
where $h$ is a function of $\theta$.
From (\ref{lin2}) and (\ref{lin4}), we obtain that
\begin{align}
 -\dfrac{\partial }{\partial \theta} \left( -p +\Phi(r, z, t) \right) = r\dfrac{\partial}{\partial z} \left(\Phi(r, z, t) r f_z \right) =E, \label{p}
\end{align}
where $E$ is a constant. On the basis of periodicity, we can assume that $\pfx{p}{\theta}=0$, and hence from (\ref{p}) it follows that $E$ must be zero. Using the definition of $\Phi$ in (\ref{p}), we obtain that
\begin{align}
 \left\lbrace \left[1+\frac{b}{n}\left(r f_z\right)^2 \right]^{n-1} + \dfrac{2b (n-1)}{n} \left(r f_z \right)^2 \left[1+\frac{b}{n}\left(r f_z\right)^2 \right]^{n-2}\right\rbrace f_{zz}   = \dfrac{E}{\mu r^2}=0, \label{f}
\end{align}
where $f_{zz} := \dfrac{\partial^2 f(z, t)}{\partial z^2}$. Thus, (\ref{f}) reduces to either 
\begin{align}
 f_{zz} = 0, \label{sol1}
\end{align}
or
\begin{align} \left\lbrace \left[1+\frac{b}{n}\left(r f_z\right)^2 \right]^{n-1} + \dfrac{2b (n-1)}{n} \left(r f_z \right)^2 \left[1+\frac{b}{n}\left(r f_z\right)^2 \right]^{n-2}\right\rbrace = 0. \label{sol2}
\end{align}  Equation (\ref{sol1}) implies that 
\begin{align}
 f(z, t) = C_1(t) z+C_2(t), \label{final_sol}
\end{align} and  (\ref{sol2}) reduces to
\begin{align}
 1 + \dfrac{b (2n-1)}{n} \left(r f_z \right)^2  = 0,
\end{align}
which has no solution.
%
 Hence, (\ref{final_sol}) is the solution to (\ref{linmom}). Interestingly, the solution (\ref{final_sol}) does not depend on the material parameters $n$, $b $ or  $\mu$.

 If we assume that the angle of twist at $z=0$ is zero, i.e., if we assume that the bottom of the cylinder is fixed, then we obtain that
\begin{align}
 f(z, t) = \psi(t)z,
\end{align}
where $\psi(t)$ is the angle of twist per unit length of the cylinder  which is the time dependent generalization of the classical torsion solution. The twisting moment is given by
\begin{align}
 M(t) &= 2\pi\int_{R_i}^{R_0} T_{z\theta} r^2 dr \notag\\
&= 2\pi\int_{R_i}^{R_0} \mu(c(r, t)) \left[1+\frac{b(c(r, t))}{n(c(r, t))}\left(r \psi(t)\right)^2 \right]^{n(c(r, t))-1} \psi(t) r^3 dr. \label{mom} 
\end{align}

 In our work, we shall enforce the following two sets of initial and boundary conditions for the concentration of the diffusing species:

Case 1:

\begin{align}
& c(r, 0) = 0,  ~R_i \leq r \leq R_0,  \label{init2} \\
&\dfrac{\partial c}{\partial r}(R_i, t) = 0, ~\forall t >0,  \label{bc21} \\
&c(R_o, t) = 1, ~\forall t>0. \label{bc22}
\end{align}
That is, initially, the body is assumed to be in its virgin state and there is no diffusant in the body. Also, the boundary condition (\ref{bc21}) implies that the gradient of the concentration is zero at the inner boundary of the annular cylinder.  This can be achieved by having some sort of a membrane on the inside of the annulus which is impermeable to the diffusing species. The boundary condition (\ref{bc22}) means that the outer cylinder is always exposed to the diffusant. 

Case 2:
\begin{align}
& c(r, 0) = 0,  ~R_i \leq r \leq R_0,  \label{init} \\
&c(R_i, t) = c_0, ~\forall t >0,  \label{bc1} \\
&c(R_o, t) = 1, ~\forall t>0. \label{bc2}
\end{align}
Here, as opposed to case 1, the inner boundary of the annular cylinder is held at a  constant diffusant concentration (as reflected by (\ref{bc1})).  By constructing a mechanism which continuously removes the diffusing species from the inside of the  cylindrical annulus, a constant concentration boundary condition can be maintained. For example, if the diffusing species is moisture, then by blowing air inside the annulus continuously, one can control the concentration of the moisture. For convenience, we shall set $c_0$ to zero for this case.

In addition, we shall  assume that the material moduli change in a linear fashion due to the diffusion, that is the material parameters change with $c$ as follows:
\begin{align}
\mu &= \mu_0 \pm \mu_1c,\quad 0<\mu_1<\mu_0, \\
b &= b_0 \pm b_1c, \quad 0<b_1<b_0, \\
n &= n_0 \pm n_1c, \quad 0<n_1<n_0,
\end{align}
with the minus sign chosen when degradation is considered and plus sign for healing. The restrictions on the ranges of $\mu_1$, $b_1$, $n_1$ are being enforced to ensure that  $\mu$, $b$, $n$ are positive, in the case of degradation.

Before proceeding with the solutions, we render the governing equations dimensionless. The non-dimensionalization is carried out in the following manner. We define 
\begin{align}
 \bar{t} = \dfrac{t}{t_0}, \bar{r} = \dfrac{r}{R_o}, \bar{z} = \dfrac{z}{H}, \bar{\psi} = \dfrac{\psi H}{\theta_0}, \bar{\mu} = \dfrac{\mu}{\mu_0}, 
\end{align}
where $t_0$, $\theta_0$ are characterstic time and angle.
Then (\ref{mom}) and (\ref{diff}) reduce to
\begin{align}
 \underbrace{\dfrac{MH}{2\pi R_o^4 \theta_0 \mu_0}}_{\bar{M}} = \int_{0}^{1} \bar{\mu} \left[1+\dfrac{b}{n} \underbrace{\dfrac{R_o^2\theta_0^2}{H^2}}_{\bar{q}}(\bar{r}\bar{\psi})^2 \right]^{n-1} \bar{\psi} \bar{r}^3 d\bar{r}, \label{nonmom}
\end{align}
 
\begin{align}
 \dfrac{\partial c(\bar{r}, \bar{t})}{\partial \bar{t}} &=  \dfrac{1}{\bar{r}}\dfrac{\partial }{\partial \bar{r}}\left(\bar{r}\underbrace{\left(\dfrac{Dt_0}{R_o^2} \right)}_{\bar{D}} \dfrac{\partial c(\bar{r}, \bar{t})}{\partial \bar{r}} \right), \label{nondiff}
\end{align}
where $\bar{\mu} = 1 \pm \bar{\mu}_1c$, with $\bar{\mu}_1 = \dfrac{\mu_1}{\mu_0}$.

The initial and boundary conditions for $c$ after non-dimensionalization reduce to

Case 1:
\begin{align}
& c(\bar{r}, 0) = 0,  ~\bar{r}_i \leq \bar{r} \leq 1,  \label{case1_1}  \\ 
&\dfrac{\partial c}{\partial \bar{r}}(\bar{r}_i, \bar{t}) = 0, ~\forall \bar{t} >0,   \\
&c(1, \bar{t}) = 1, ~\forall \bar{t}>0.  \label{case1_3}
\end{align}

Case 2:
\begin{align}
& c(\bar{r}, 0) = 0,  ~\bar{r}_i \leq \bar{r} \leq 1, \label{case2_1}\\ 
&c(\bar{r}_i, \bar{t}) = 0, ~\forall \bar{t} >0, \\
&c(1, \bar{t}) = 1, ~\forall \bar{t}>0, \label{case2_3}
\end{align}
where $\bar{r}_i = \dfrac{R_i}{R_o}$.

\section{Discussion of Results}\label{sec3}
\begin{figure}
	\subfloat[]{
	\label{fig:concentration-profile}
 	\centering
	 \input{matlab-codes/figs/concentration-profile.tex}
	\includegraphics[scale = 0.6]{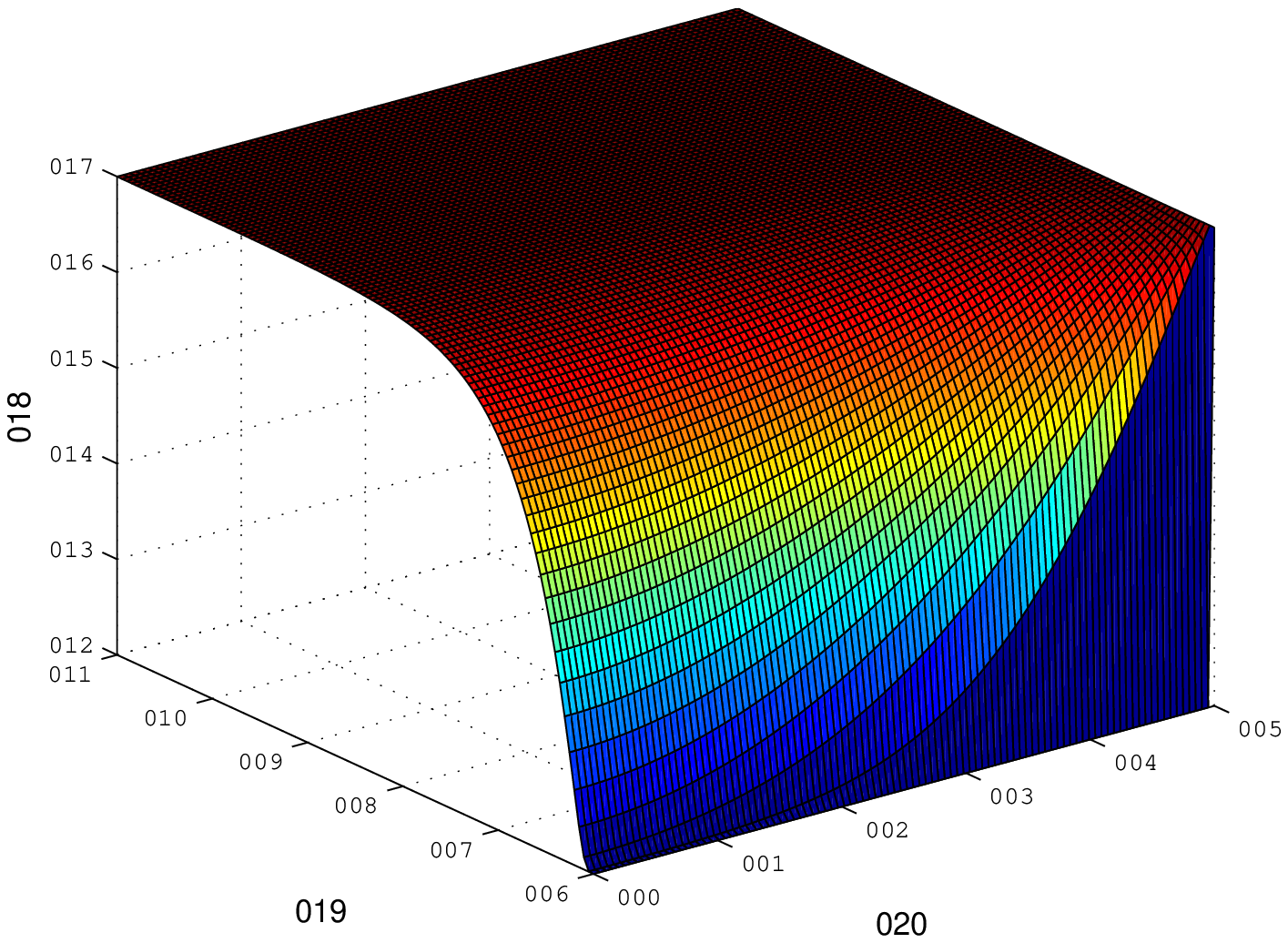}
  	}\\
	\subfloat[]{
	\label{fig:Moment-vs-t-mu1-changing}
 	\centering
	 \input{matlab-codes/figs/Moment-vs-t-mu1-changing.tex}
	\includegraphics[scale = 0.5]{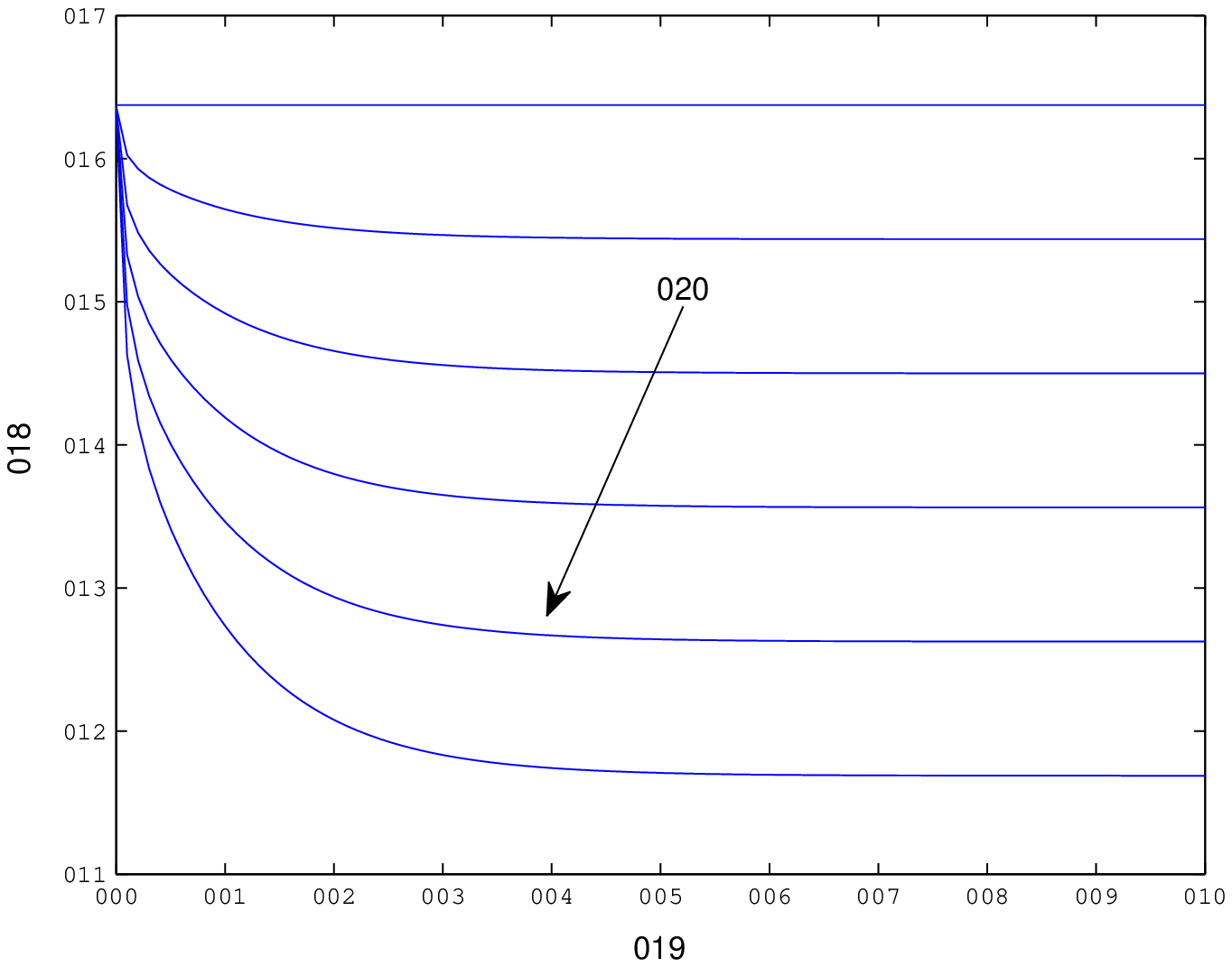}
  	}
	\subfloat[]{
	\label{fig:Moment-vs-t-mu1-changing-n1-nonzero}
 	\centering
	 \input{matlab-codes/figs/Moment-vs-t-mu1-changing-n1-nonzero.tex}
	\includegraphics[scale = 0.5]{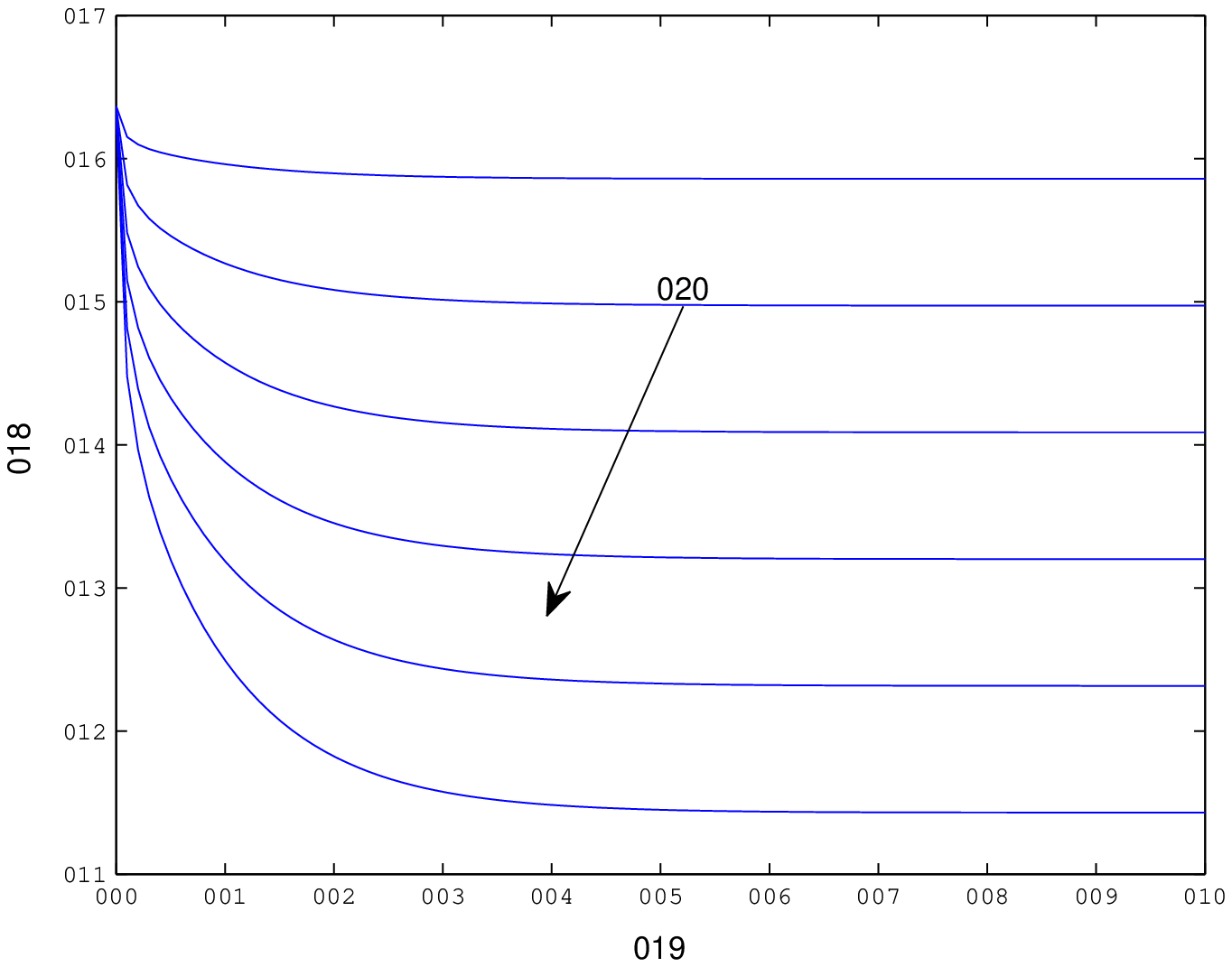}
  	}
\caption{(a) Solution to the convection diffusion equation. (b) Non-dimensional moment ($\bar{M}$) as a function of non-dimensional time ($\bar{t}$) for various values of $\bar{\mu}_1$ starting from 0 to 0.5 in increments of 0.1 for the degradation case. Values chosen were $R_i/R_o = 0.5$, $\bar{q} = 1$, $\bar{\psi} = 1$, $\bar{D} = 0.01$, $b_0 = n_0 = 1$, $b_1 = 0$, $n_1 = 0$. This corresponds to the neo-Hookean model since $n=1$. (c) Non-dimensional moment ($\bar{M}$) as a function of non-dimensional time ($\bar{t}$) with $\bar{\mu}_1$ varying from 0 to 0.5 in increments of 0.1 for the degradation case. Values chosen were $R_i/R_o = 0.5$, $\bar{q} = 1$, $\bar{\psi} = 1$, $\bar{D} = 0.01$, $b_0 = n_0 = 1$, $b_1 = 0.1$,  $n_1 = 0.1$.}
\label{fig:1}
\end{figure}

We now proceed to solve the non-dimensionalized problem, numerically. Figure \eqref{fig:concentration-profile} shows the solution to the convection-diffusion equation (\ref{nondiff}) for $\bar{D} = 0.01$, which has been solved using \texttt{pdepe} solver in MATLAB. We notice that after a certain time, the concentration profile reaches a steady state which implies that no further healing or degradation of the body takes place. For a given value of $\bar{\psi}$, the values of concentration thus obtained at various radii and times were used to numerically integrate \eqref{nonmom} using composite trapezoidal rule to find $\bar{M}$. On the other hand for a given value of $\bar{M}$, $\bar{\psi}$ was calculated using a combination of the bisection method and the composite trapezoidal rule of intergration on \eqref{nonmom}. The numerical values chosen for the non-dimensional quantities are shown in the figures. 

\begin{figure}
	\subfloat[]{
	\label{fig:Moment-vs-t-b-changing}
 	\centering
	 \input{matlab-codes/figs/Moment-vs-t-b-changing.tex}
	\includegraphics[scale = 0.5]{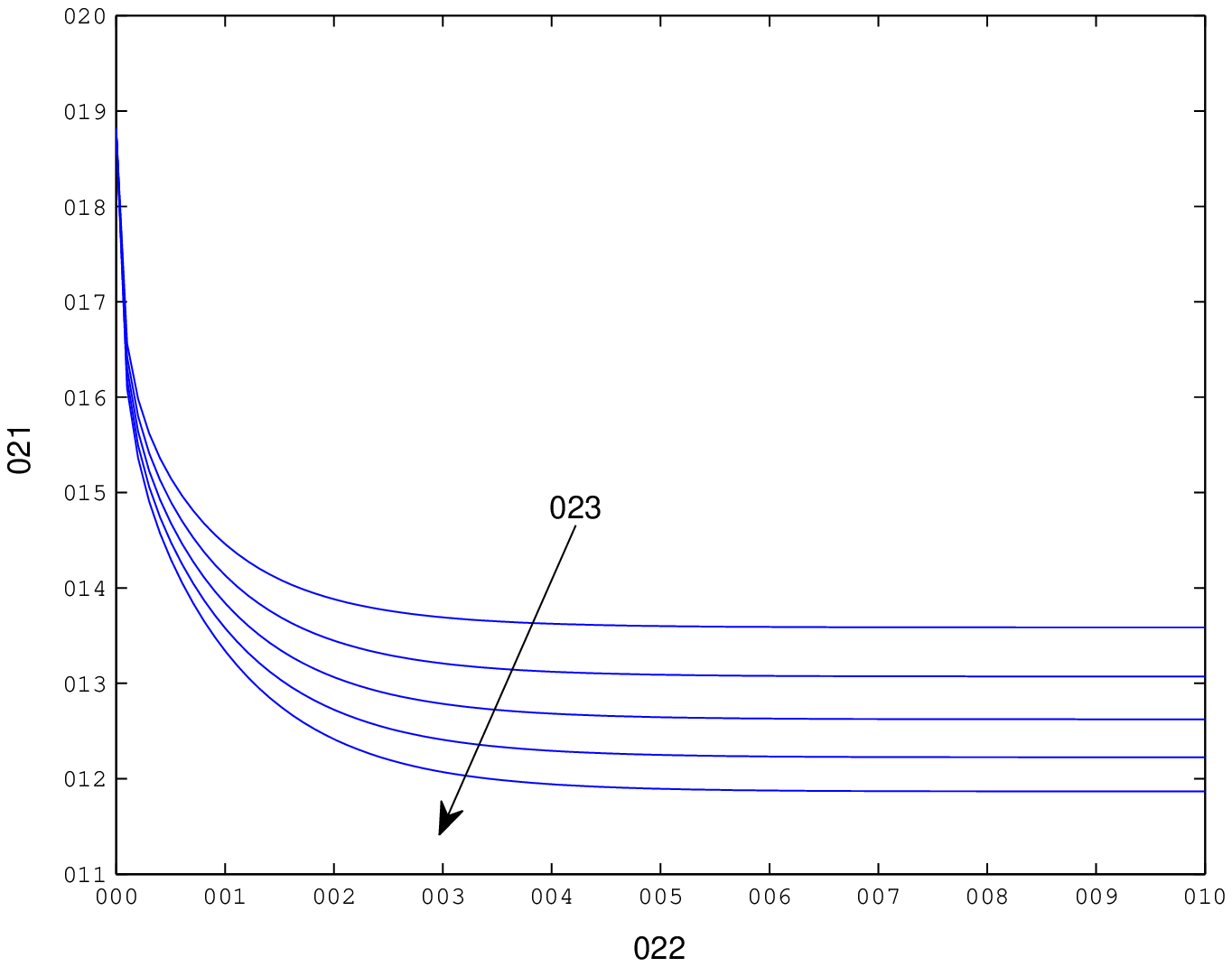}
  	}
	\subfloat[]{
	\label{fig:Moment-vs-t-n-changing}
 	\centering
	 \input{matlab-codes/figs/Moment-vs-t-n-changing.tex}
	\includegraphics[scale = 0.5]{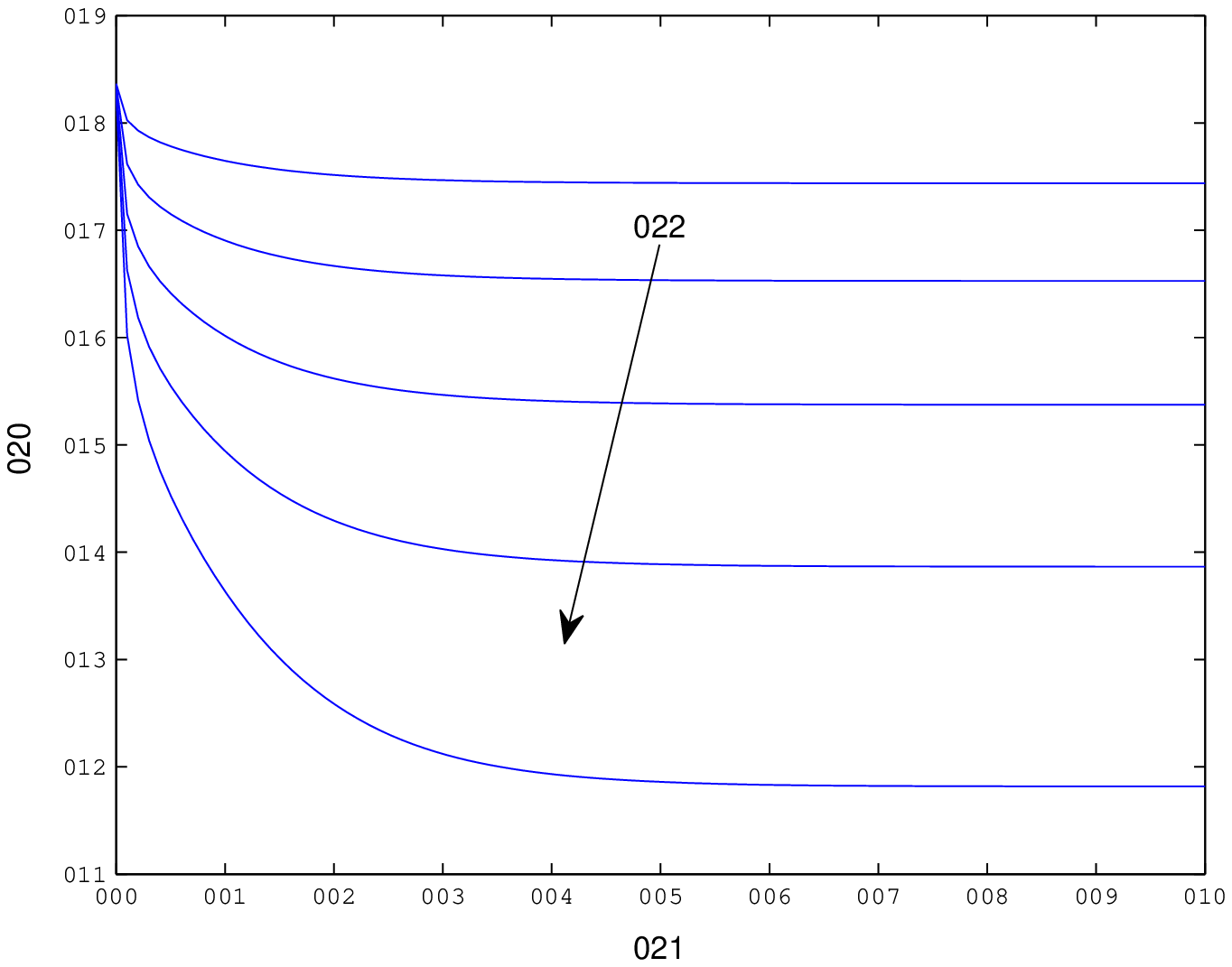}
	}
\caption{(a) Non-dimensional moment ($\bar{M}$) as a function of non-dimensional time ($\bar{t}$) for various values of $b_1$ varying from 0 to 0.8 in increments of 0.2 for the degradation case. Here, $R_i/R_o = 0.5$, $\bar{q} = 1$, $\bar{\psi} = 1$, $\bar{D} = 0.01$, $b_0 = n_0 = 1$, $\bar{\mu}_1 = 0.1$,  $n_1 = 0.1$. (b) Non-dimensional moment ($\bar{M}$) as a function of non-dimensional time ($\bar{t}$) for various values of $n_1$ starting at 0 to 0.8 in increments of 0.2 for the degradation case. Here, $R_i/R_o = 0.5$, $\bar{q} = 1$, $\bar{\psi} = 1$, $\bar{D} = 0.01$, $b_0 = n_0 = 1$, $b_1 = 0.1$, $\bar{\mu}_1 = 0.1$.}
\label{fig:2}
\end{figure}

\begin{figure}
	\subfloat[]{
	\label{fig:Moment-vs-t-mu1-changing-healing}
 	\centering
	 \input{matlab-codes/figs/Moment-vs-t-mu1-changing-healing.tex}
	\includegraphics[scale = 0.5]{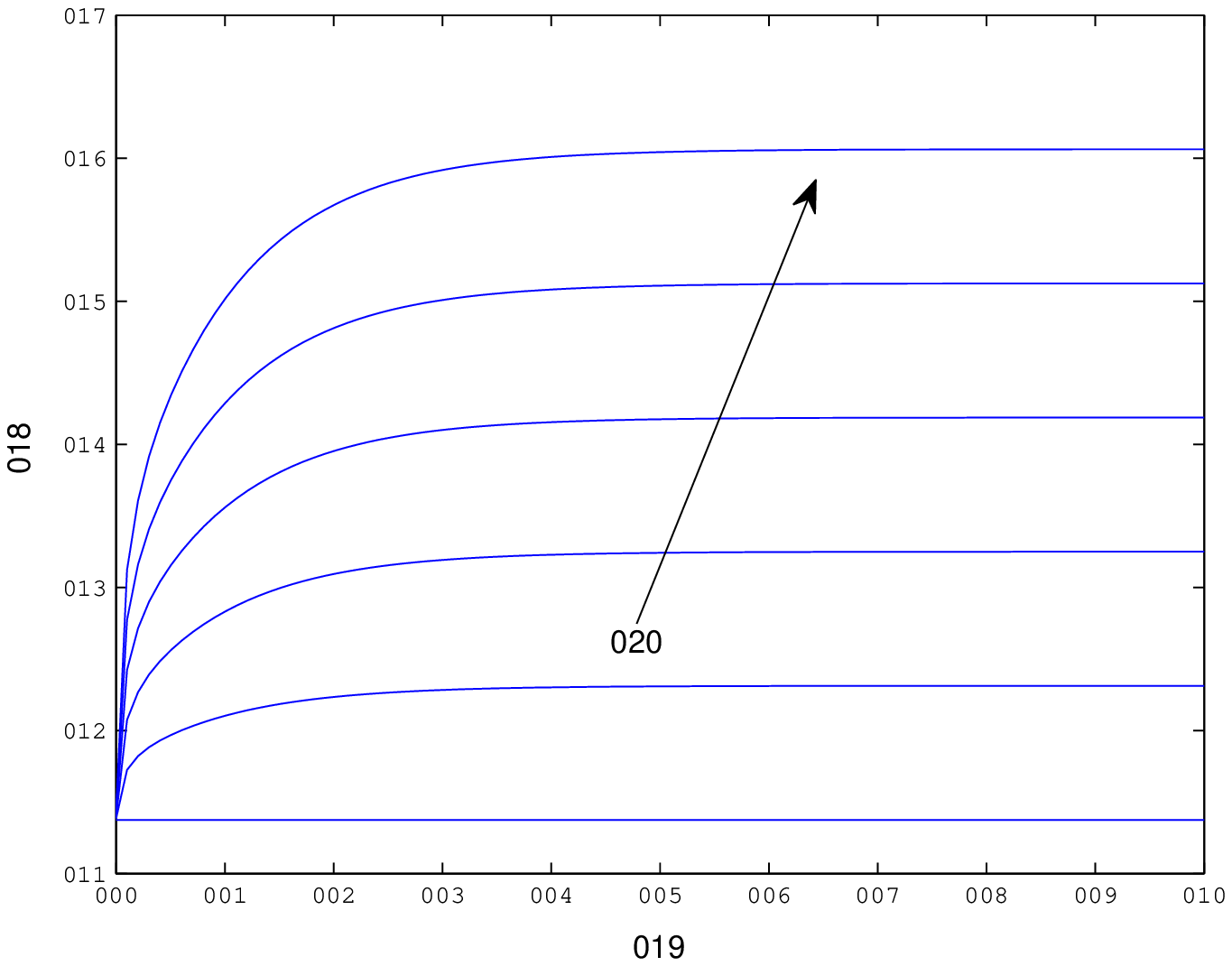}
  	}
	\subfloat[]{
	\label{fig:Moment-vs-t-mu1-changing-n1-nonzero-healing}
 	\centering
	 \input{matlab-codes/figs/Moment-vs-t-mu1-changing-n1-nonzero-healing.tex}
	\includegraphics[scale = 0.5]{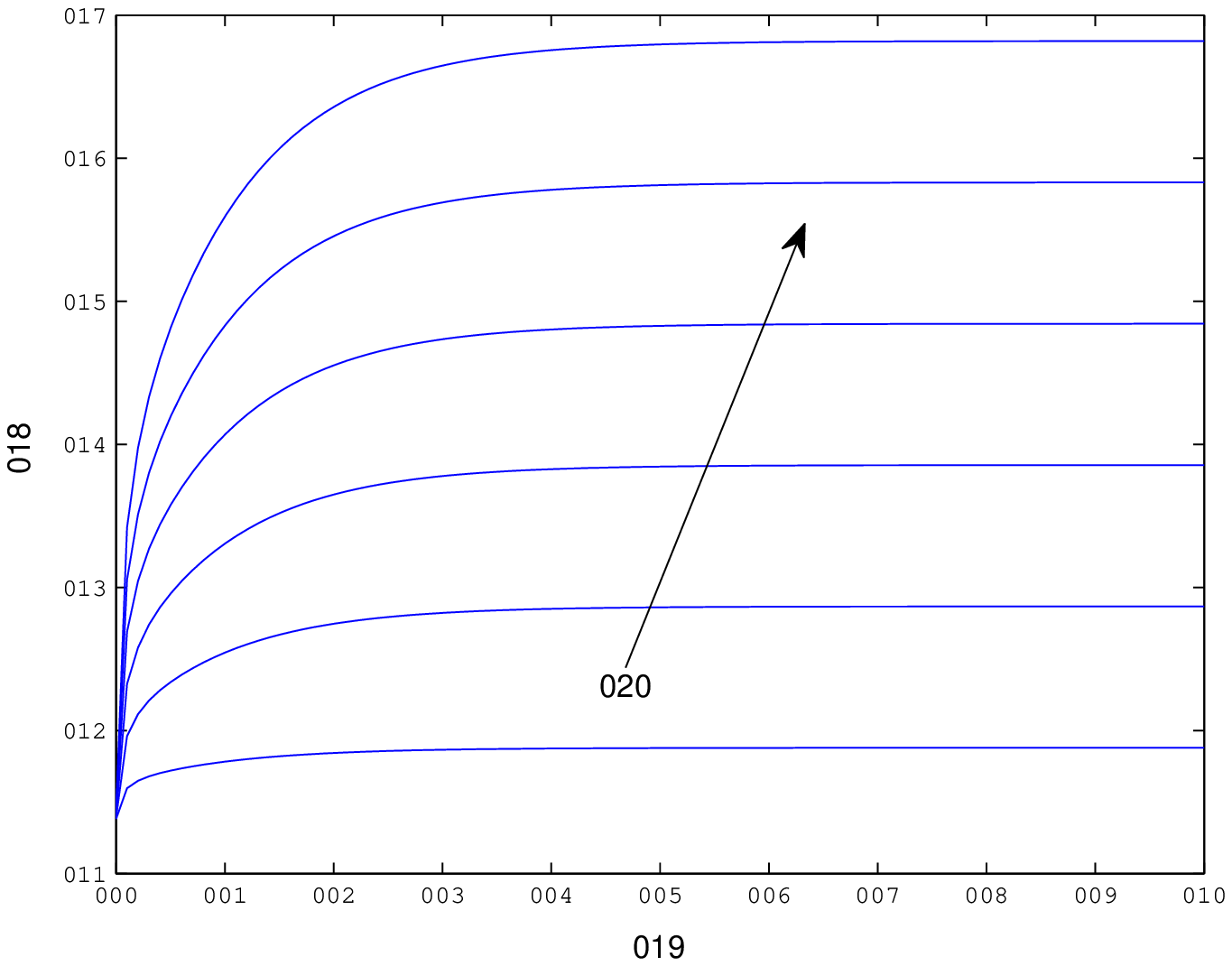}
  	}\\
	\subfloat[]{
	\label{fig:Moment-vs-t-b-changing-healing}
 	\centering
	 \input{matlab-codes/figs/Moment-vs-t-b-changing-healing.tex}
	\includegraphics[scale = 0.5]{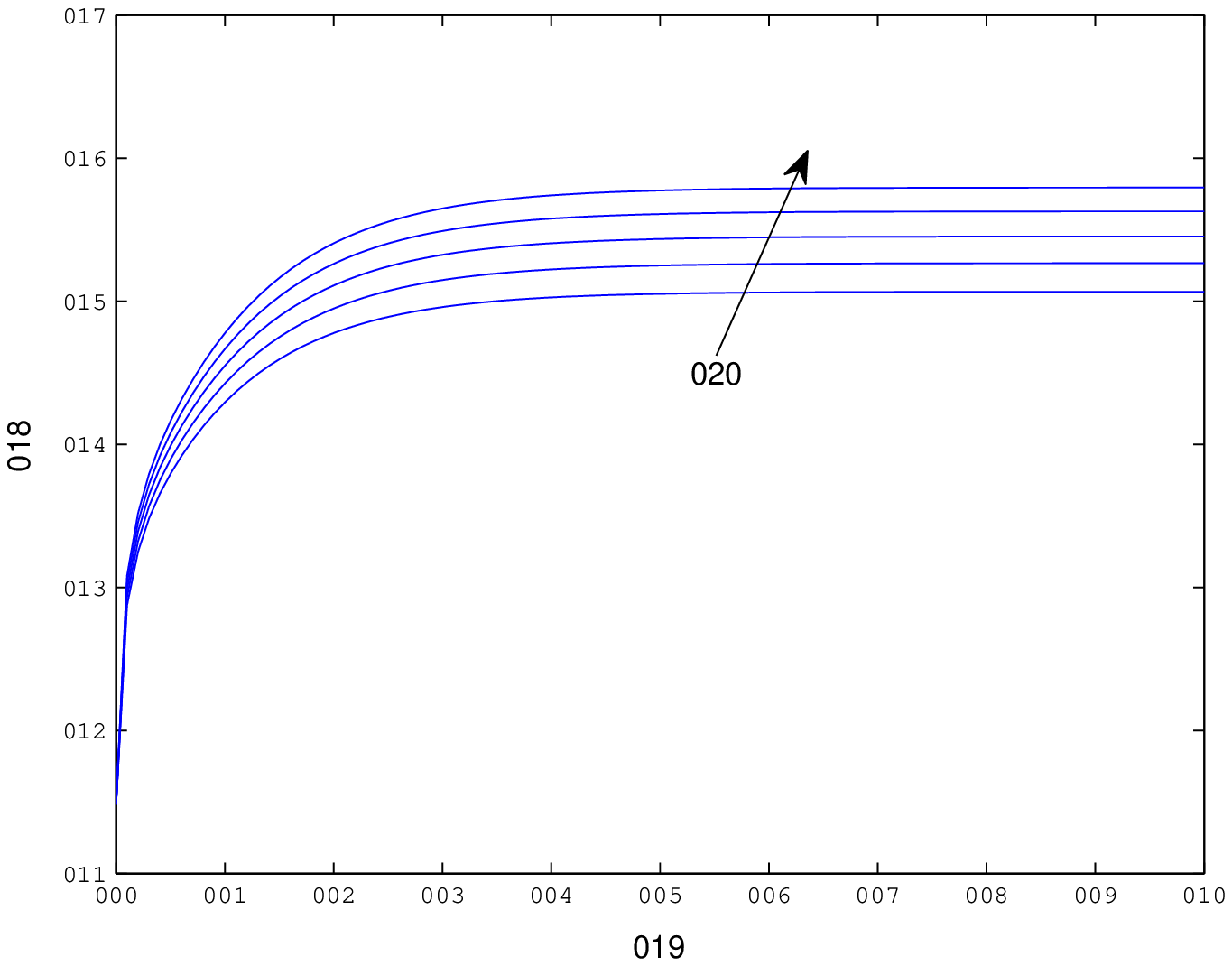}
  	}
	\subfloat[]{
	\label{fig:Moment-vs-t-n-changing-healing}
 	\centering
	 \input{matlab-codes/figs/Moment-vs-t-n-changing-healing.tex}
	\includegraphics[scale = 0.5]{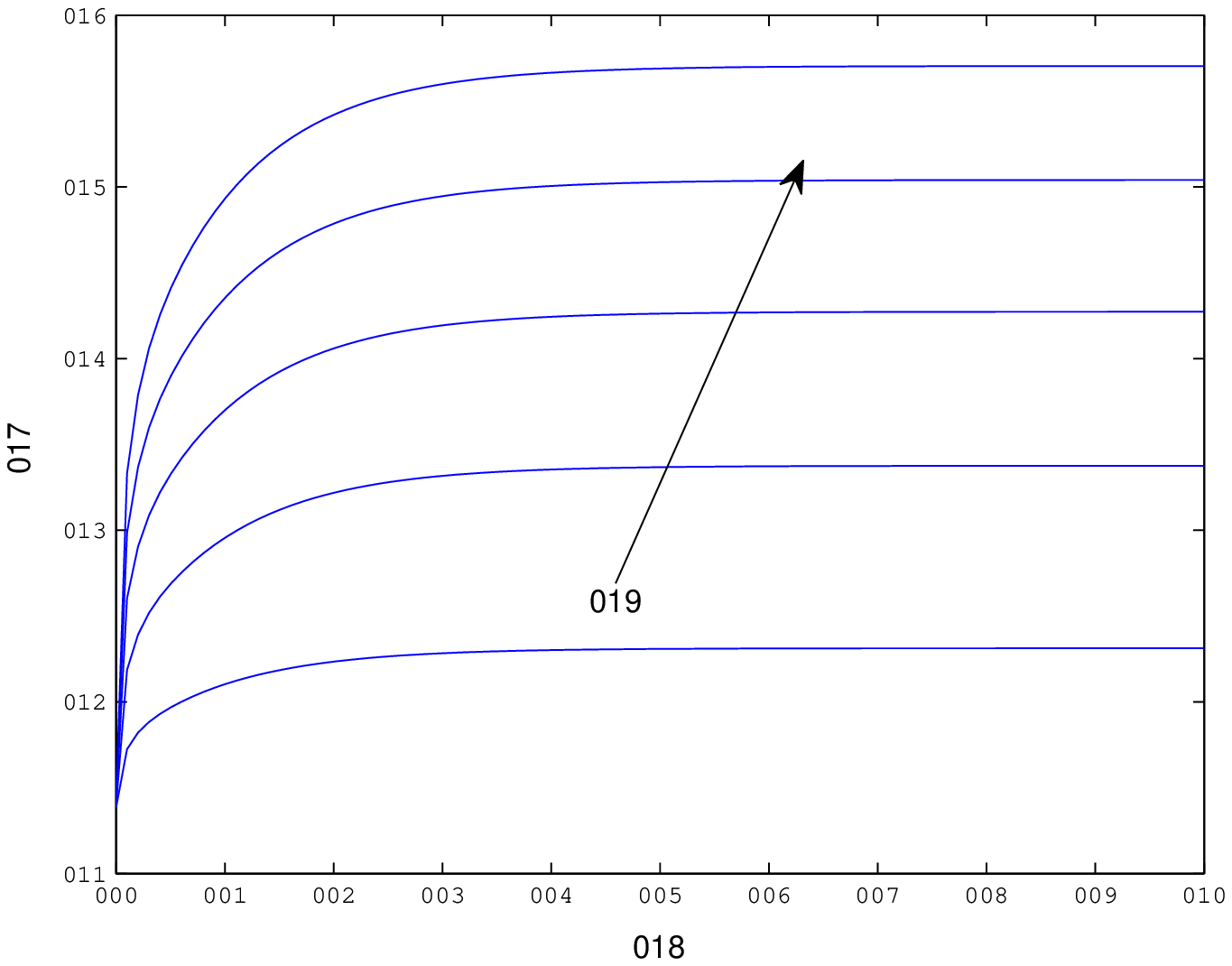}
	}
\caption{Non-dimensional moment ($\bar{M}$) as a function of non-dimensional time ($\bar{t}$) for various values of the material parameters when the body is healing. (a) $\bar{\mu}_1$ varying from 0 to 0.5 in increments of 0.1 with $b_0 = n_0 = 1$, $b_1 = 0$, $n_1 = 0$ (neo-Hookean model). (b) $\bar{\mu}_1$ varying from 0 to 0.5 in increments of 0.1 with $b_0 = n_0 = 1$,  $b_1 = 0.1$,  $n_1 = 0.1$. (c) $b_1$ varying from 0 to 0.8 in increments of 0.2 with $b_0 = n_0 = 1$, $\bar{\mu}_1 = 0.1$,  $n_1 = 0.1$. (d) $n_1$ varying from 0 to 0.8 in increments of 0.2 with $b_0 = n_0 = 1$, $b_1 = 0.1$, $\bar{\mu}_1 = 0.1$. Other values chosen in (a), (b), (c), (d) were $R_i/R_o = 0.5$, $\bar{q} = 1$, $\bar{\psi} = 1$, $\bar{D} = 0.01$.}
\label{fig:3}
\end{figure}

The solution depicted in figure \eqref{fig:Moment-vs-t-mu1-changing} was obtained by setting $n_0$  to 1 along with $n_1$ to 0 when degradation is assumed. Under these conditions the model reduces to a neo-Hookean model as $n=1$; furthermore, note that $\bar{\mu}_1 = 0$ corresponds to case when the neo-Hookean body is neither degrading nor healing. Notice that as $\bar{\mu}_1$ increases, the stress required to maintain the angular displacement decreases, i.e., the higher the degradation of the material, the greater is the stress relaxation in that less moment is necessary to maintain the angular displacement. Of course, since the moment is related to the shear stress through (\ref{mom}), we can call this stress relaxation due to degradation. In figures \eqref{fig:1}, \eqref{fig:2} the non-dimensional moment is portrayed as a function of non-dimensional time by varying $\mu_1$, $b_1$  and $n_1$ along with other quantities as indicated. In all these cases, it is seen that as the degradation increases, less moment is needed to maintain the deformation, as is to be expected. Next, we assume that the material is healing as the fluid diffuses (see figure \eqref{fig:3}), and the non-dimensional moment is observed as a function of time for a fixed value of non-dimensional angular displacement. As the body heals with time, the moment required to maintain the deformation increases.  Furthermore, as seen from figures \eqref{fig:1}, \eqref{fig:2}, \eqref{fig:3} the non-dimensional moment reaches a steady value to maintain the angular displacement since the concentration of the diffusant reaches a steady value after which there is no further degradation or healing.

\begin{figure}
\subfloat[]{
	\label{fig:psi-vs-t-mu1-changing}
 	\centering
	 \input{matlab-codes/figs/psi-vs-t-mu1-changing.tex}
	\includegraphics[scale = 0.5]{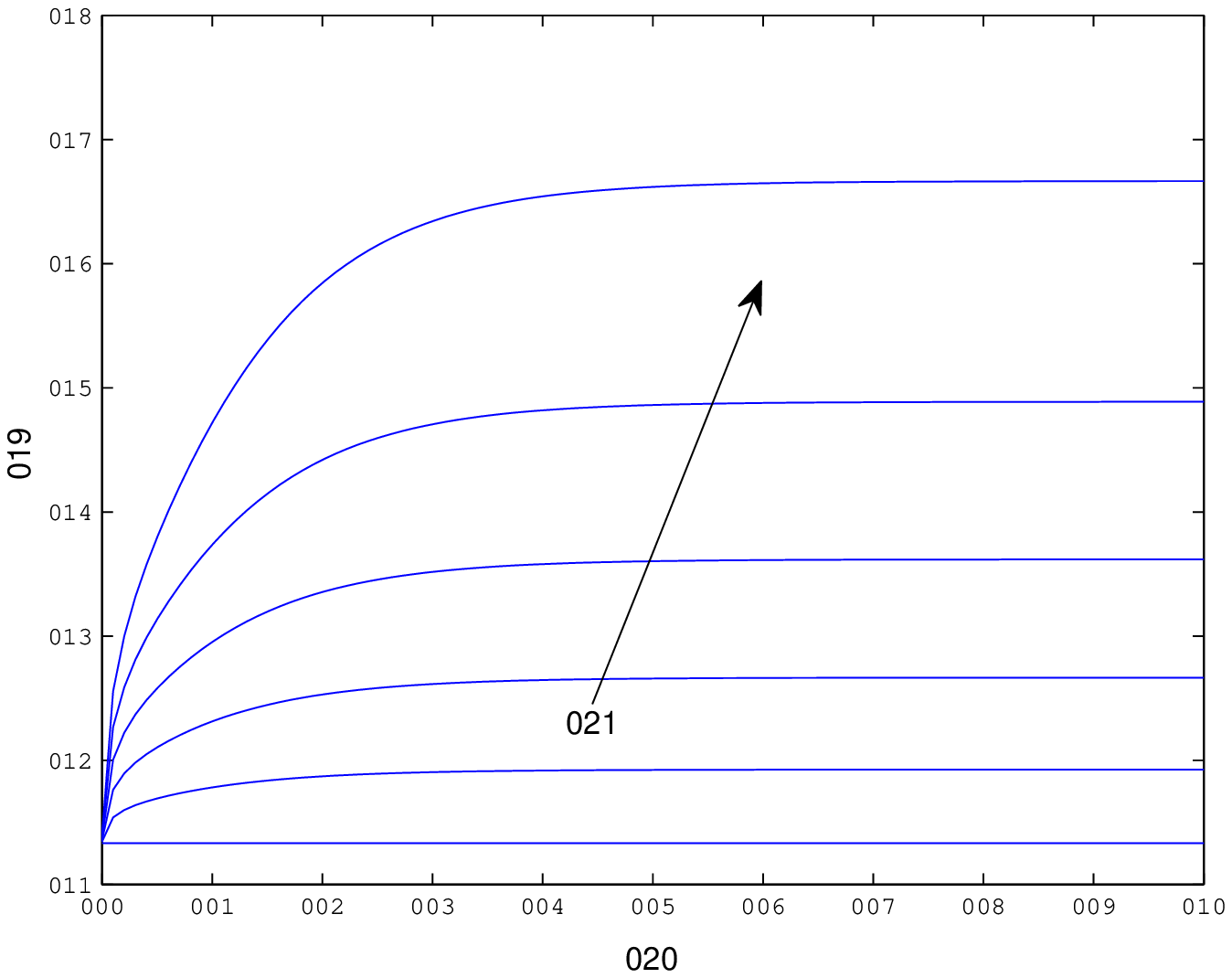}
  	}
	\subfloat[]{
	\label{fig:psi-vs-t-mu1-changing-n-nonzero}
 	\centering
	 \input{matlab-codes/figs/psi-vs-t-mu1-changing-n-nonzero.tex}
	\includegraphics[scale = 0.5]{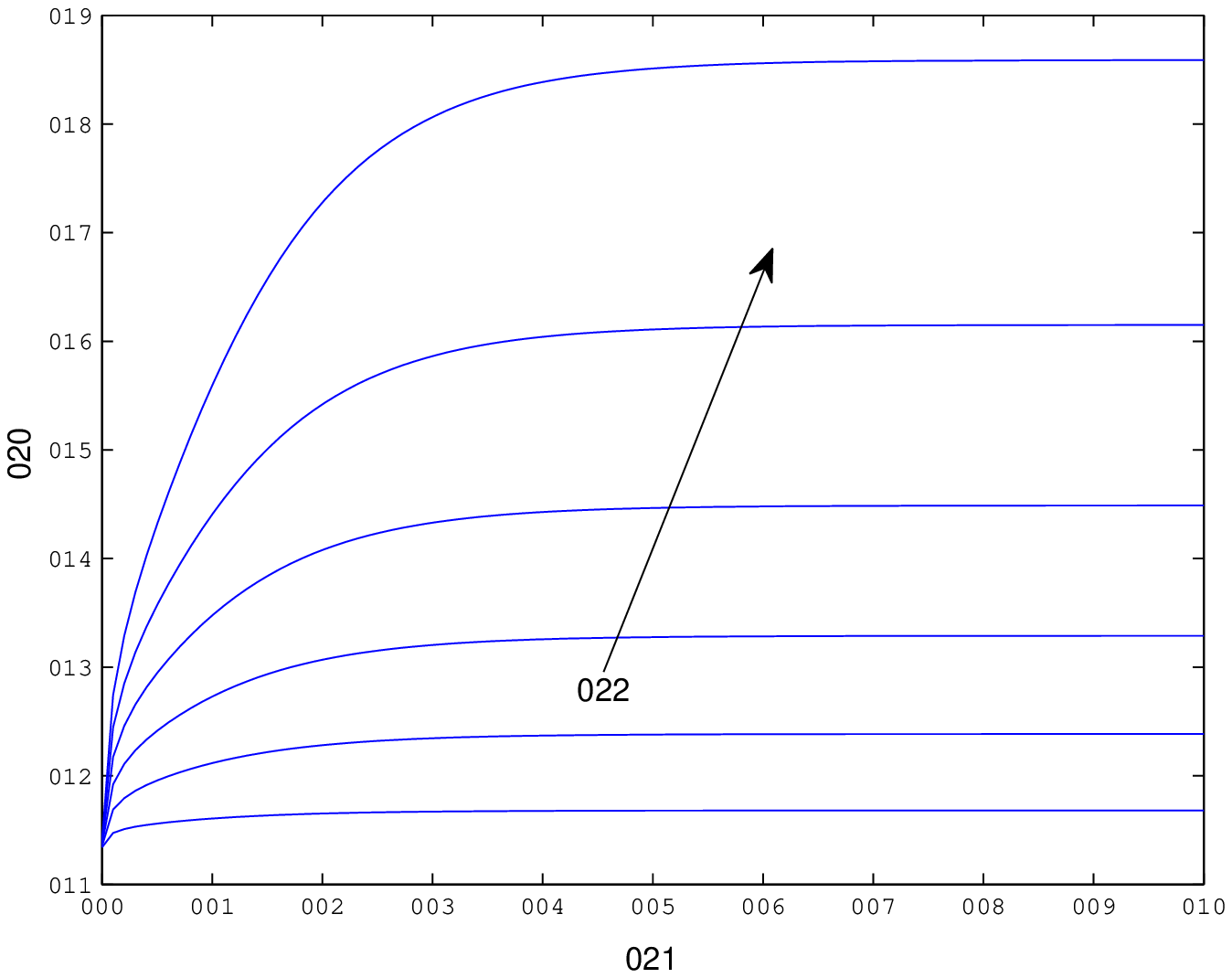}
  	}\\
\centering
\subfloat[]{
	\label{fig:psi-vs-t-b1-changing}
 	\centering
	 \input{matlab-codes/figs/psi-vs-t-b1-changing.tex}
	\includegraphics[scale = 0.5]{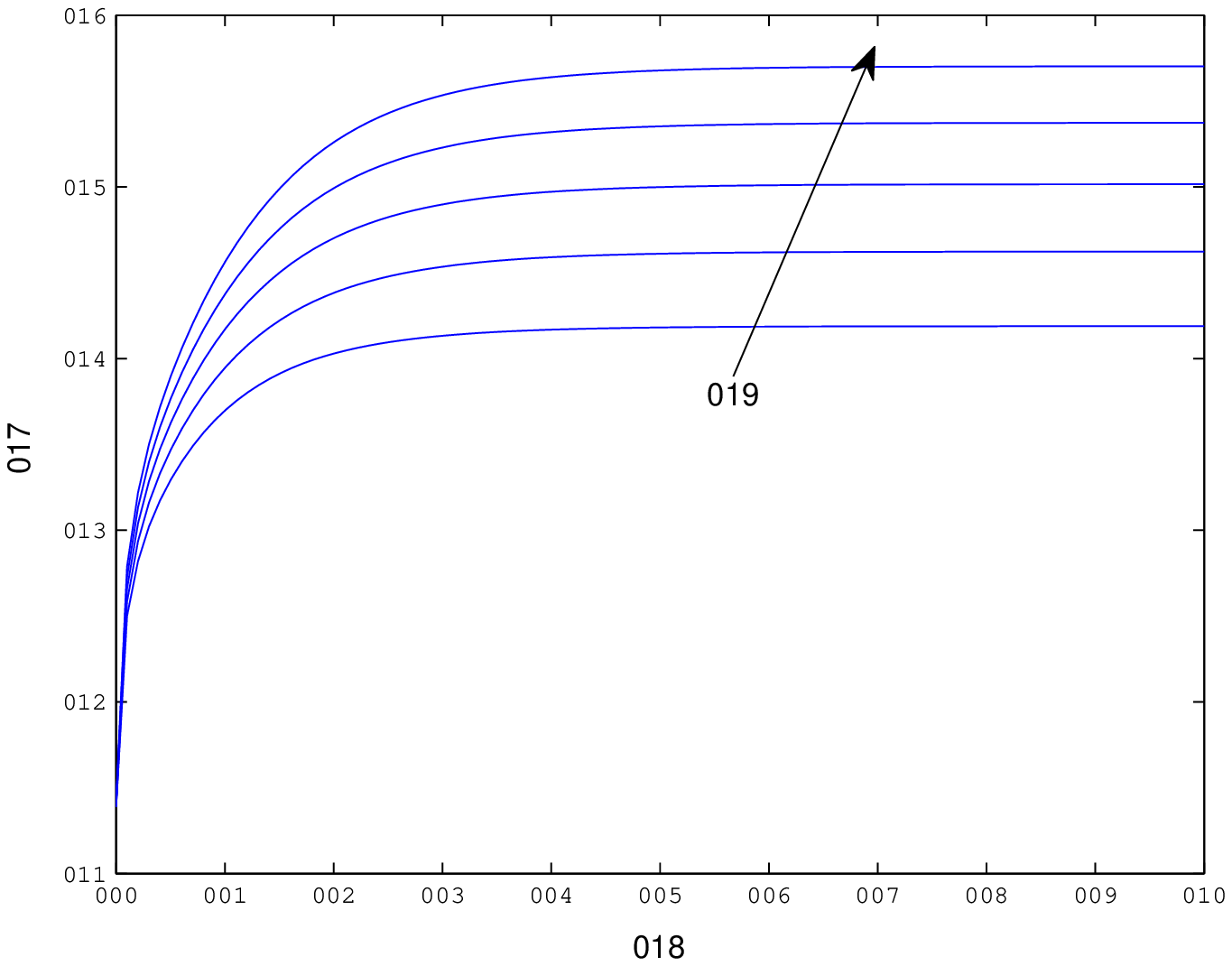}
  	}
	\subfloat[]{
	\label{fig:psi-vs-t-n1-changing}
 	\centering
	 \input{matlab-codes/figs/psi-vs-t-n1-changing.tex}
	\includegraphics[scale = 0.5]{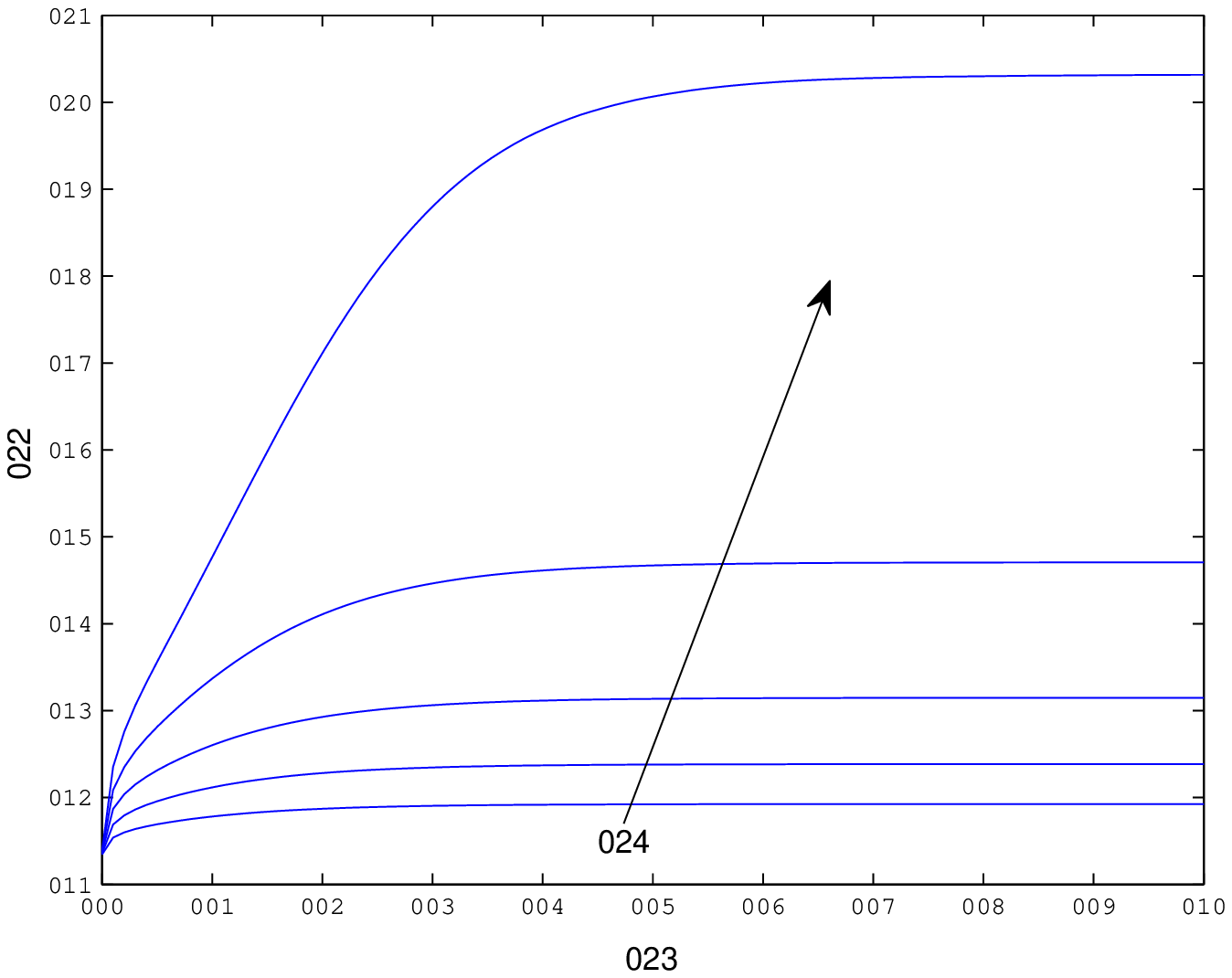}
	}
\caption{Non-dimensional angular displacement ($\bar{\psi}$) as a function of non-dimensional time ($\bar{t}$) for various values of the material parameters when the body is degrading. (a) $\bar{\mu}_1$ varying from 0 to 0.5 in increments of 0.1 with $b_0 = n_0 = 1$, $b_1 = 0$, $n_1 = 0$ (neo-Hookean model). (b) $\bar{\mu}_1$ varying from 0 to 0.5 in increments of 0.1 with $b_0 = n_0 = 1$,  $b_1 = 0.1$,  $n_1 = 0.1$. (c) $b_1$ varying from 0 to 0.8 in increments of 0.2 with $b_0 = n_0 = 1$, $\bar{\mu}_1 = 0.1$,  $n_1 = 0.1$. (d) $n_1$ varying from 0 to 0.4 in increments of 0.1 with $b_0 = n_0 = 1$, $b_1 = 0.1$, $\bar{\mu}_1 = 0.1$. Other values chosen in (a), (b), (c), (d) were $R_i/R_o = 0.5$, $\bar{q} = 1$, $\bar{M} = 0.5$, $\bar{D} = 0.01$.}
\label{fig:4}
\end{figure}

\begin{figure}
\subfloat[]{
	\label{fig:psi-vs-t-mu1-changing-healing}
 	\centering
	 \input{matlab-codes/figs/psi-vs-t-mu1-changing-healing.tex}
	\includegraphics[scale = 0.5]{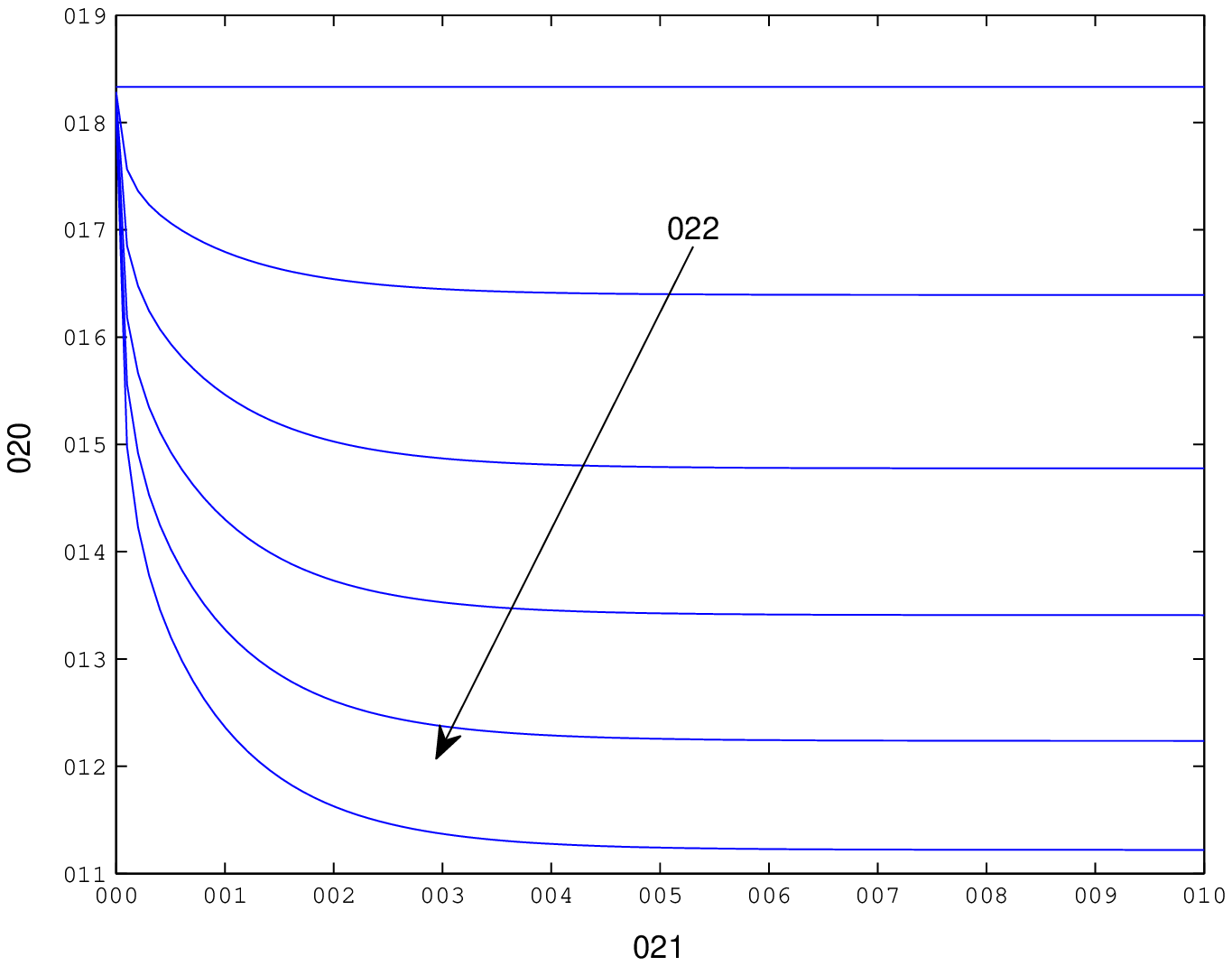}
  	}
	\subfloat[]{
	\label{fig:psi-vs-t-mu1-changing-n-nonzero-healing}
 	\centering
	 \input{matlab-codes/figs/psi-vs-t-mu1-changing-n-nonzero-healing.tex}
	\includegraphics[scale = 0.5]{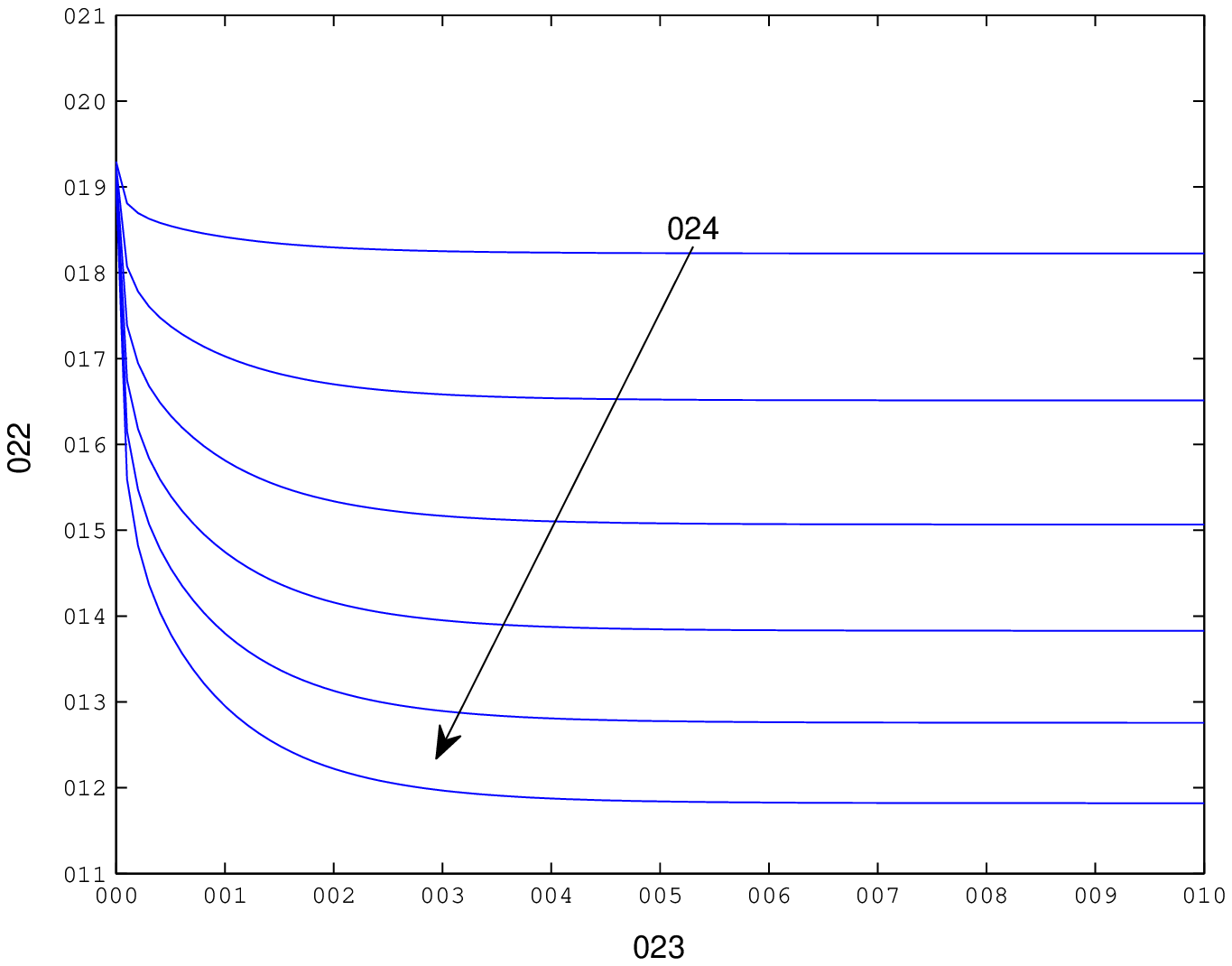}
  	}\\
\subfloat[]{
	\label{fig:psi-vs-t-b1-changing-healing}
 	\centering
	 \input{matlab-codes/figs/psi-vs-t-b1-changing-healing.tex}
	\includegraphics[scale = 0.5]{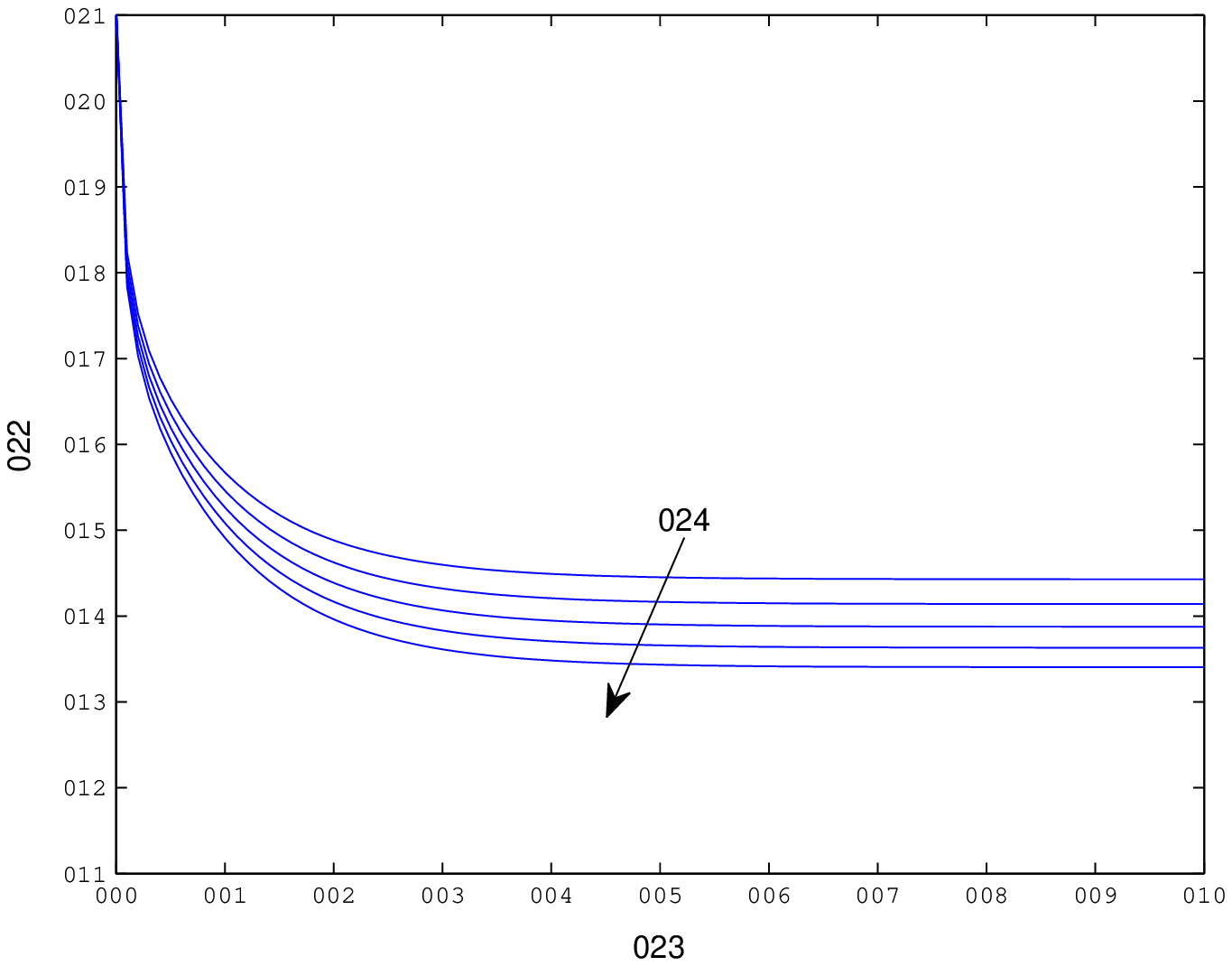}
  	}
	\subfloat[]{
	\label{fig:psi-vs-t-n1-changing-healing}
 	\centering
	 \input{matlab-codes/figs/psi-vs-t-n1-changing-healing.tex}
	\includegraphics[scale = 0.5]{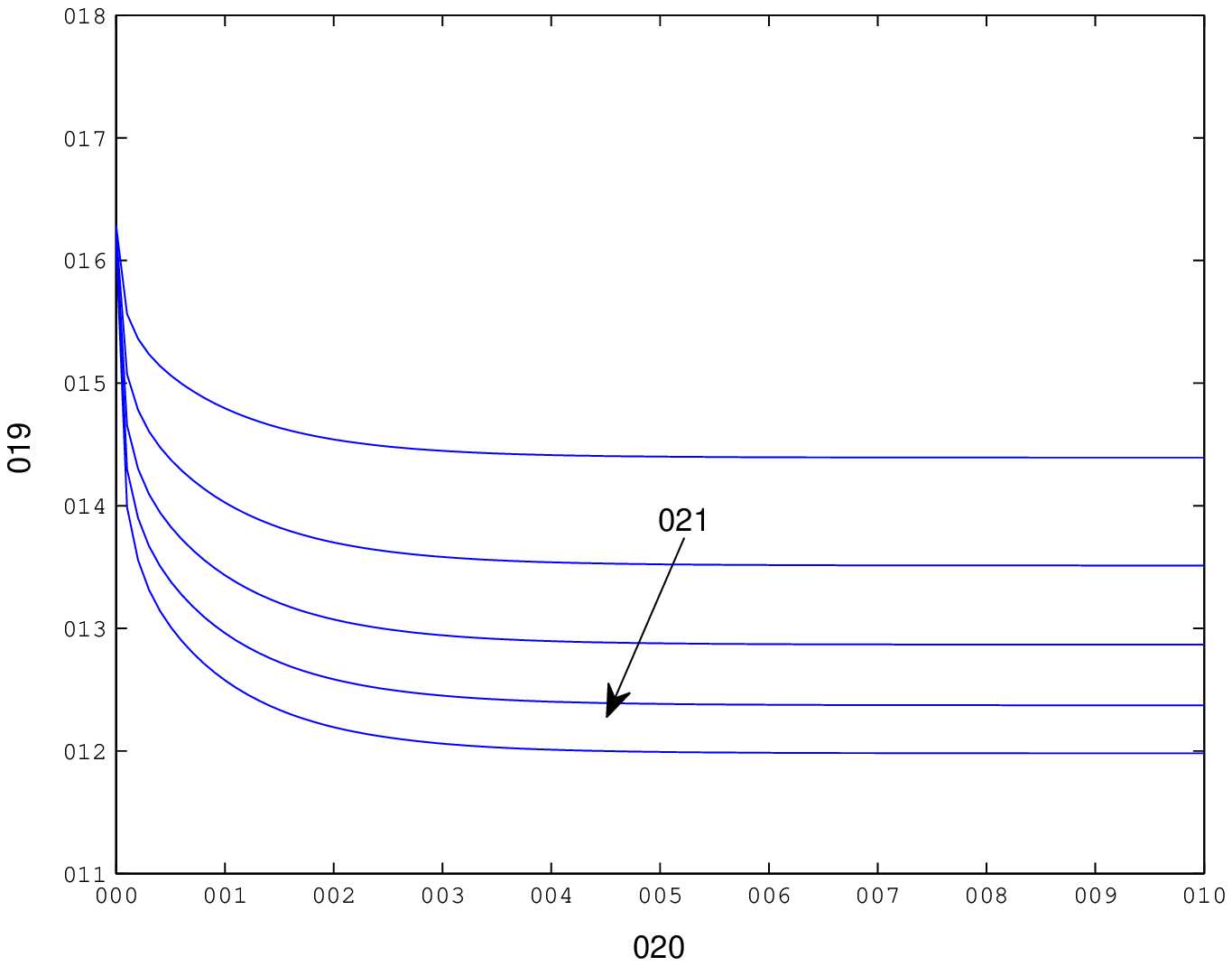}
	}
\caption{ Non-dimensional angular displacement ($\bar{\psi}$) as a function of non-dimensional time ($\bar{t}$) for various values of the material parameters when the body is healing. (a) $\bar{\mu}_1$ varying from 0 to 0.5 in increments of 0.1 with $b_0 = n_0 = 1$, $b_1 = 0$, $n_1 = 0$ (neo-Hookean model). (b) $\bar{\mu}_1$ varying from 0 to 0.5 in increments of 0.1 with $b_0 = n_0 = 1$,  $b_1 = 0.1$,  $n_1 = 0.1$. (c) $b_1$ varying from 0 to 0.8 in increments of 0.2 with $b_0 = n_0 = 1$, $\bar{\mu}_1 = 0.1$,  $n_1 = 0.1$. (d) $n_1$ varying from 0 to 0.4 in increments of 0.1 with $b_0 = n_0 = 1$, $b_1 = 0.1$, $\bar{\mu}_1 = 0.1$. Other values chosen in (a), (b), (c), (d) were $R_i/R_o = 0.5$, $\bar{q} = 1$, $\bar{M} = 0.5$, $\bar{D} = 0.01$.}
\label{fig:5}
\end{figure}

Figure \eqref{fig:4} shows that the non-dimensional angular displacement of the cylinder increases with time for an applied non-dimensional moment when the generalized neo-Hookean body is degrading i.e., the body "creeps" for sometime, and then the angular displacement reaches a steady value. This is very different from the creep in a viscoelastic solid, wherein the angular displacement continuously increasing with time when an external moment is applied. This is one way of determining if the creep undergone by a body is either due to the viscoelastic nature of the body or due to degradation. Now, when one considers healing, the non-dimensional angular displacement of the body decreases with time as shown in figure \eqref{fig:5} for an applied moment, and then remains steady. 

\begin{figure}[htp]
	\subfloat[]{
	\label{fig:moment-vs-t-comparison-degrading-radius75}
 	\centering
	 \input{matlab-codes/figs/Moment-vs-t-mu1-changing-degrading-radius-point75.tex}
	\includegraphics[scale = 0.5]{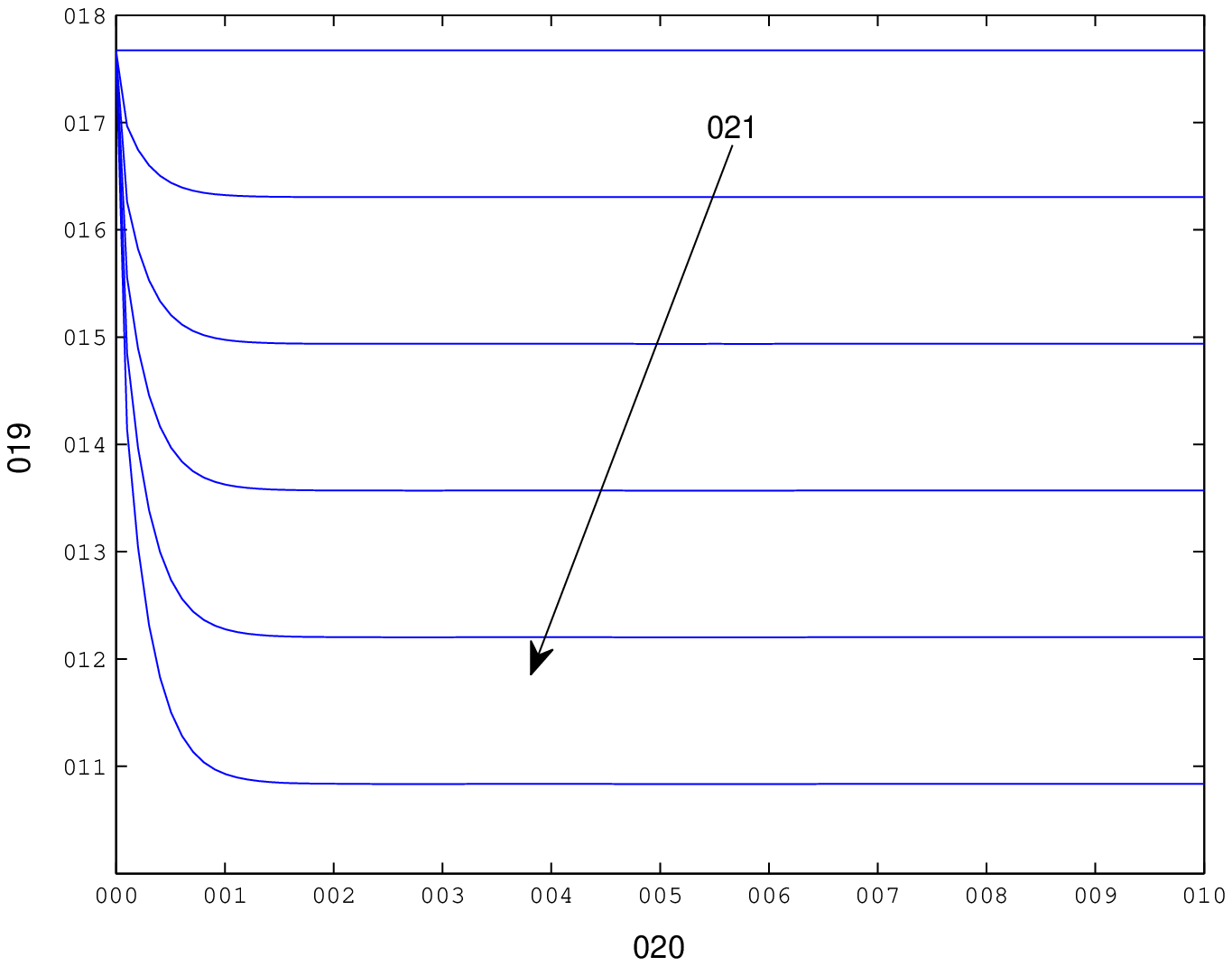}
  	}
	\subfloat[]{
	\label{fig:moment-vs-t-comparison-degrading-radius35}
 	\centering
	 \input{matlab-codes/figs/Moment-vs-t-mu1-changing-degrading-radius-point35.tex}
	\includegraphics[scale = 0.5]{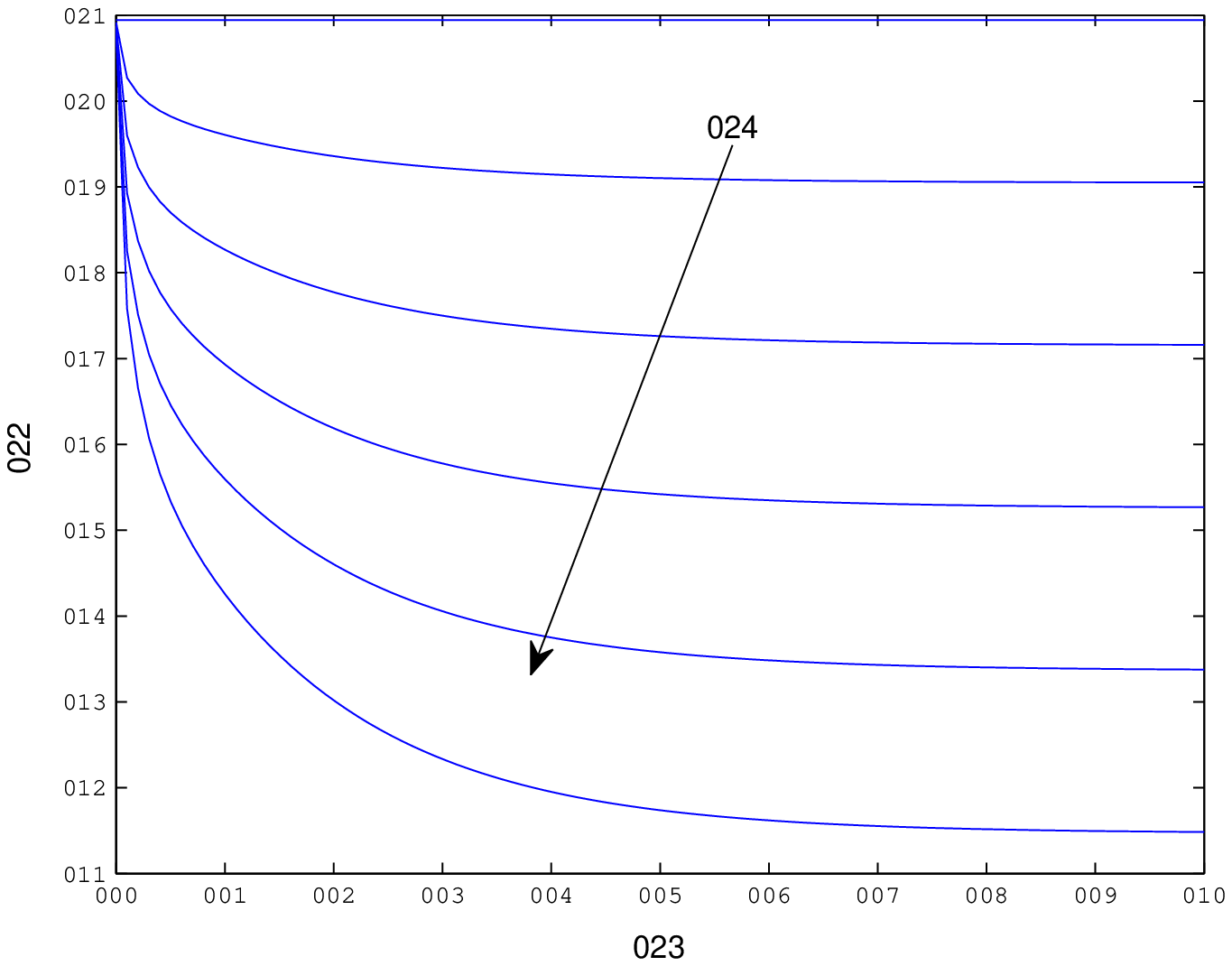}
  	}\\
  	\subfloat[]{
	\label{fig:psi-vs-t-comparison-degrading-radius75}
 	\centering
	 \input{matlab-codes/figs/psi-vs-t-mu1-changing-degrading-point75.tex}
	\includegraphics[scale = 0.5]{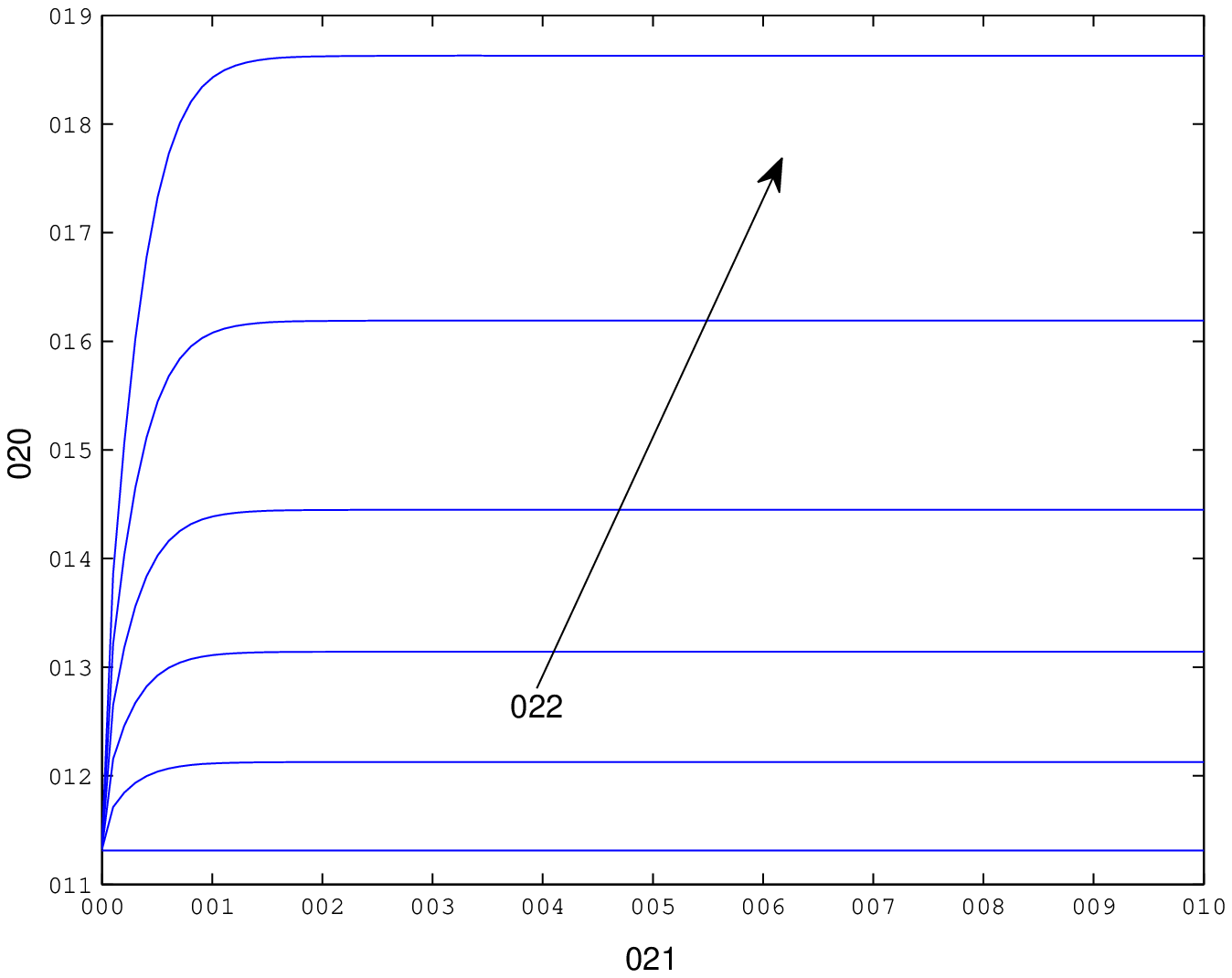}
  	}
	\subfloat[]{
	\label{fig:psi-vs-t-comparison-degrading-radius35}
 	\centering
	 \input{matlab-codes/figs/psi-vs-t-mu1-changing-degrading-point35.tex}
	\includegraphics[scale = 0.5]{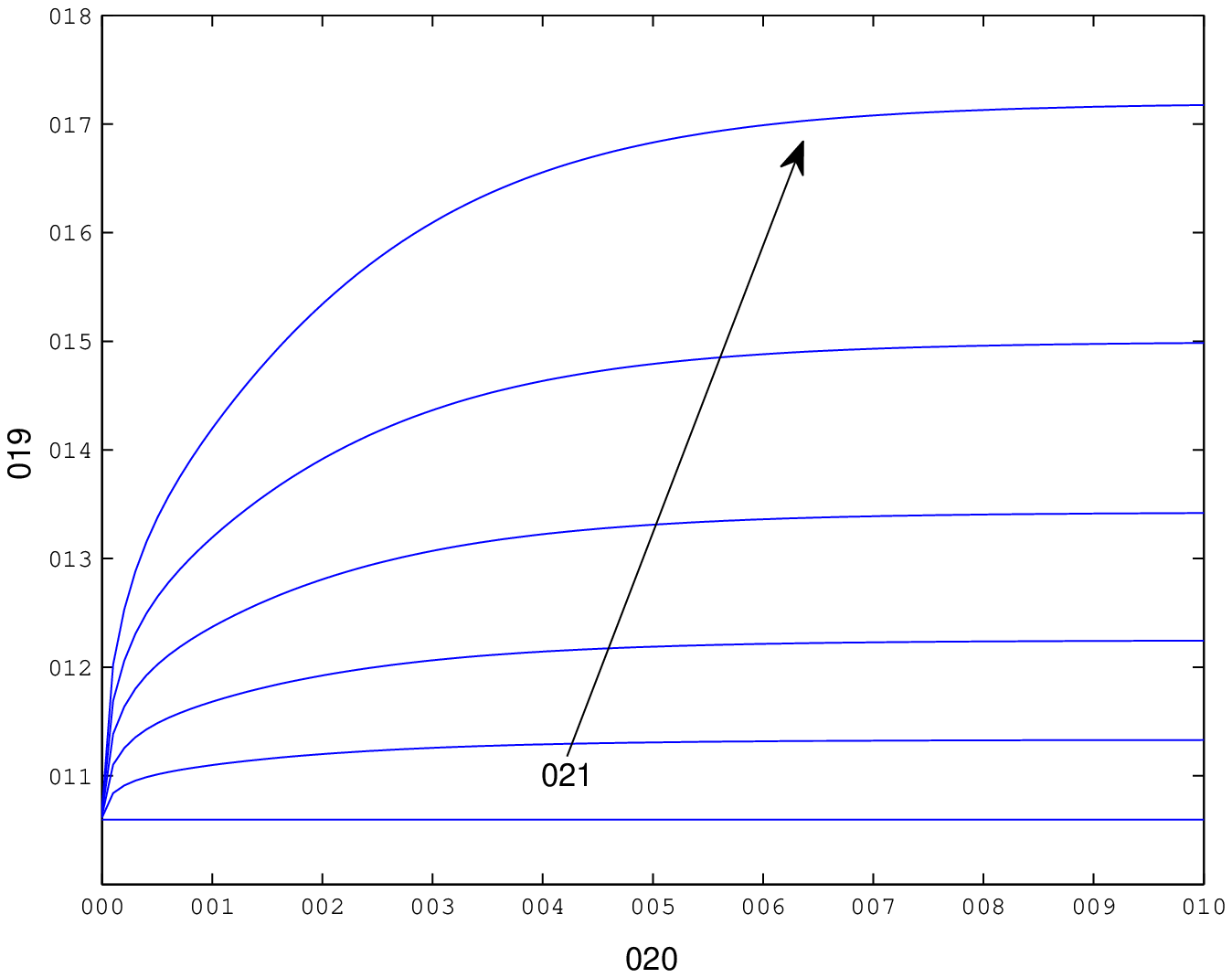}
  	}
	
\caption{ (a) Non-dimensional moment ($\bar{M}$) as a function of non-dimensional time ($\bar{t}$) for various values of $\bar{\mu}_1$ starting from 0 to 0.5 in increments of 0.1 for the degradation case with $R_i/R_o = 0.75$. (a)  Non-dimensional moment ($\bar{M}$) as a function of non-dimensional time ($\bar{t}$) for various values of $\bar{\mu}_1$ starting from 0 to 0.5 in increments of 0.1 for the degradation case with $R_i/R_o = 0.35$. (c) Non-dimensional angular displacement ($\bar{\psi}$) as a function of non-dimensional time ($\bar{t}$) for various values of $\bar{\mu}_1$ starting from 0 to 0.5 in increments of 0.1 for the degradation case with $R_i/R_o = 0.75$. (d) Non-dimensional angular displacement ($\bar{\psi}$) as a function of non-dimensional time ($\bar{t}$) for various values of $\bar{\mu}_1$ starting from 0 to 0.5 in increments of 0.1 for the degradation case with $R_i/R_o = 0.35$. In all cases, $\bar{q} = 1$, $b_0 = n_0 = 1$, $b_1 = 0.0$, $n_1 = 0.0$. Initial and boundary conditions considered are given by (\ref{case1_1}--\ref{case1_3}).}
\label{fig:creep-sr-radius}
\end{figure}

As mentioned in the introduction, Rajagopal and Wineman \cite{rajagopal.kr.wineman.as:note2004} showed that the stress relaxation due to aging depends on the material geometry, and this characteristic differentiates the stress relaxation due to aging/degradation from the stress relaxation due to viscoelasticity. To illustrate this phenomenon in our work, results were obtained at the ratio of the inner radius of the annulus to the outer radius ($\bar{r}_i$) being 0.35 and 0.75 (see figure \eqref{fig:creep-sr-radius}), and it is seen that not only stress relaxation but also creep due to degradation depends on the geometry. Comparing figures \eqref{fig:moment-vs-t-comparison-degrading-radius75} and \eqref{fig:moment-vs-t-comparison-degrading-radius35}, figures \eqref{fig:psi-vs-t-comparison-degrading-radius75} and \eqref{fig:psi-vs-t-comparison-degrading-radius35}, one can also see that the steady state values are attained at a faster rate when $\bar{r}_i = 0.75$. This is because the annulus has a smaller thickness for this case and so the concentration reaches steady state faster.

\begin{figure}[htp]
	\subfloat[]{
	\label{fig:moment-vs-t-comparison-diffusivities}
 	\centering
	 \input{matlab-codes/figs/moment-vs-t-comparison-diffusivities.tex}
	\includegraphics[scale = 0.5]{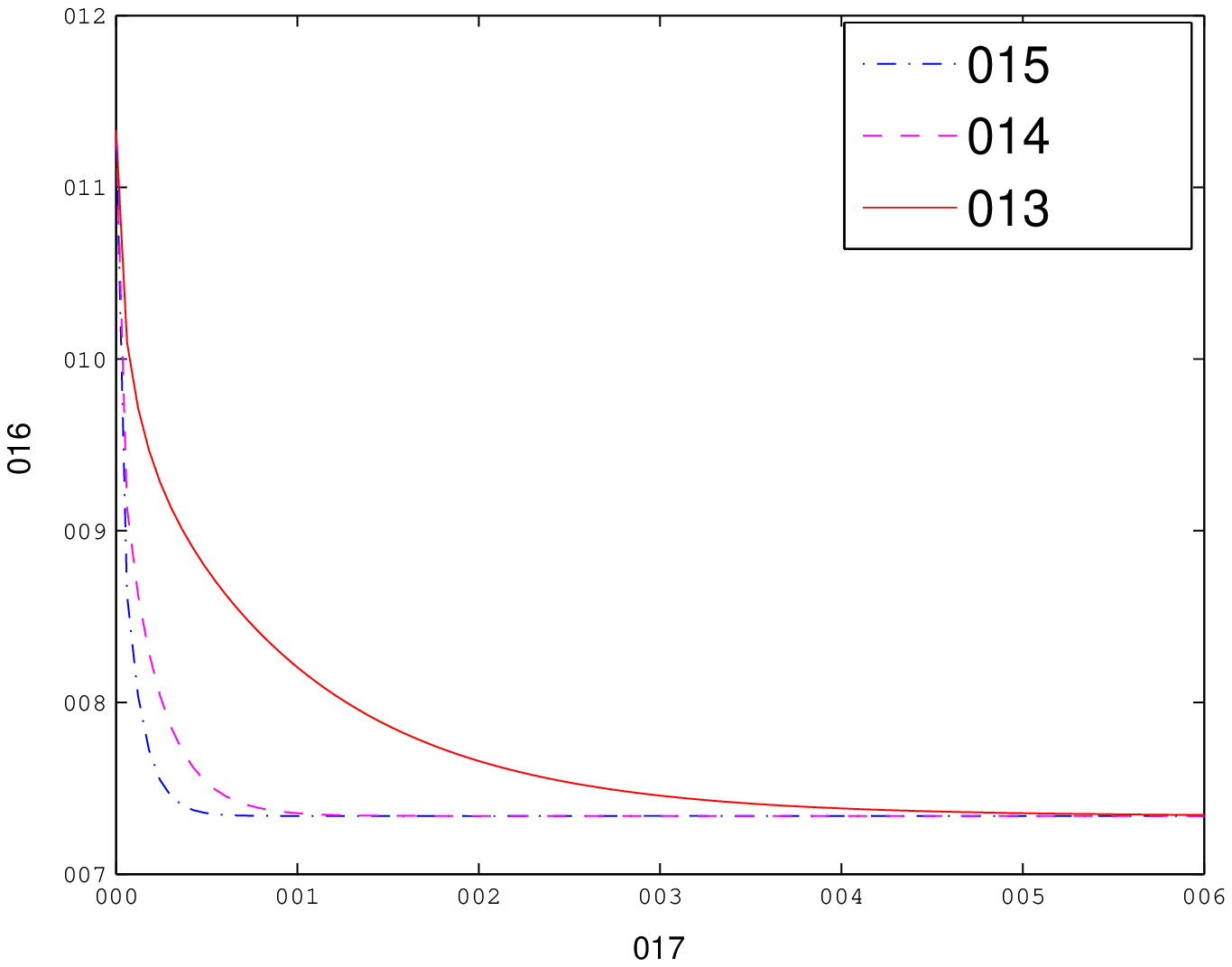}
  	}
	\subfloat[]{
	\label{fig:psi-vs-t-comparison-diffusivities}
 	\centering
	 \input{matlab-codes/figs/psi-vs-t-comparison-diffusivities.tex}
	\includegraphics[scale = 0.5]{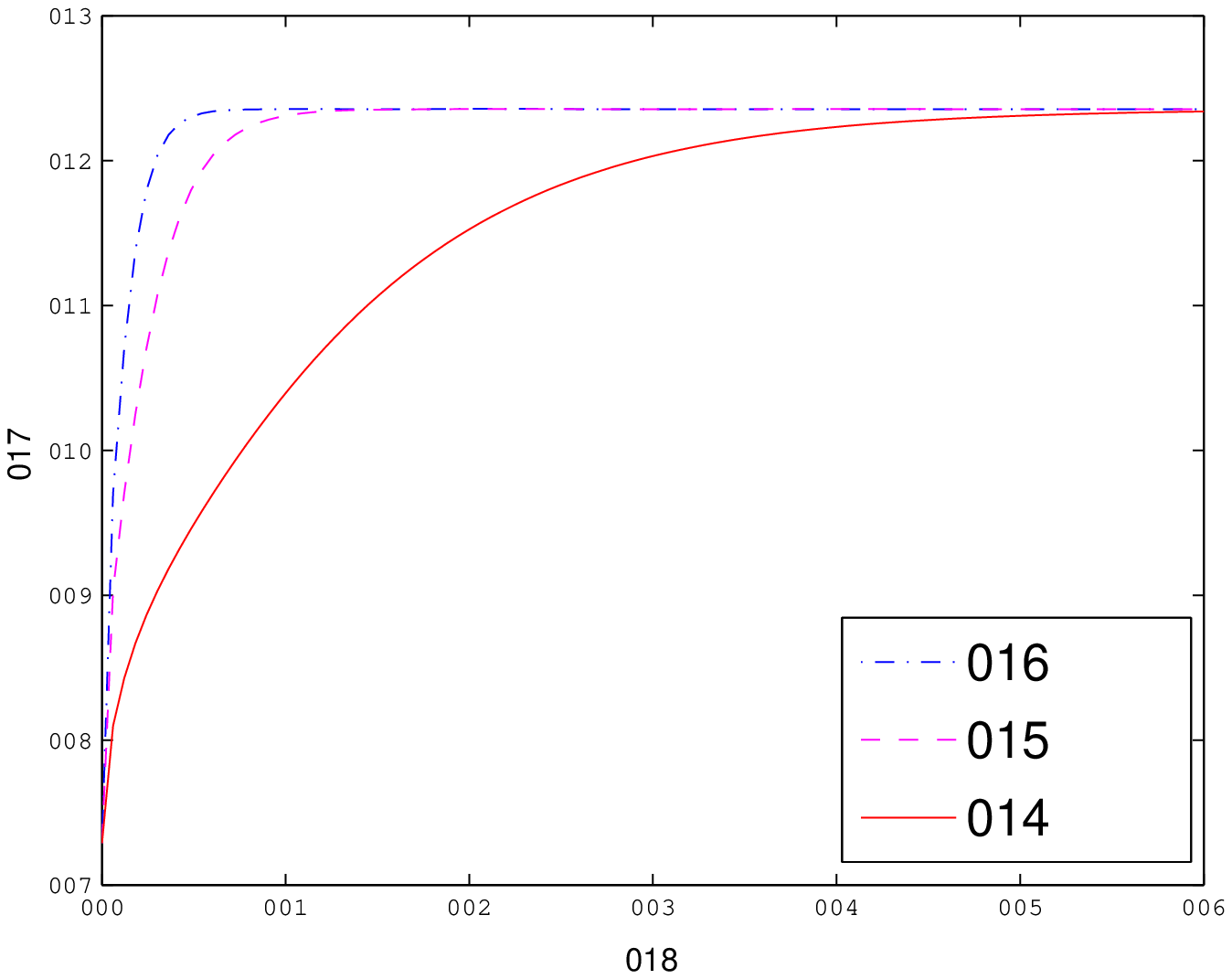}
	}
\caption{ (a) Comparison of non-dimensional moment ($\bar{M}$) as a function of non-dimensional time ($\bar{t}$)  for various constant diffusivity values for $\bar{\psi} = 1$. (b) Comparison of non-dimensional angular displacement ($\bar{\psi}$) as a function of non-dimensional time ($\bar{t}$)  for various constant diffusivity values for $\bar{M} = 1$ . Also, $R_i/R_o = 0.5$, $\bar{q} = 1$, $b_0 = n_0 = 1$, $b_1 = 0.1$, $n_1 = 0.1$, $\bar{\mu}_1 = 0.4$. Initial and boundary conditions considered are given by (\ref{case1_1}--\ref{case1_3}).}
\label{fig:6}
\end{figure}

To see how the stress relaxation and creep depend on the diffusivity, we changed the non-dimesional diffusivity ($\bar{D}$) from 0.01 to 0.1. As can be seen in figure \eqref{fig:6} the moment relaxes faster as the diffusivity increases, and the angular diplacement undergone by the body increases faster as the diffusivity increases. This is because higher diffusivity means that the concentration at any given location increases faster as it tends towards the steady state value, and as the material parameters decrease or increase (based on degradation or healing) in value with increasing concentration,  the material degrades or heals faster. Also, notice that the steady state non-dimensional moment and the non-dimensional angular displacement  values are the same for the three values of the non-dimensional diffusivity. This is due to the fact that the steady state solution for concentration of the diffusant is the same in all the cases, which is $c(r,t) = 1$, for $t$ sufficiently large so that steady state is reached. 

Next, we shall look at how the results vary when the diffusivity depends on the strain;  we choose the Almansi-Hamel strain as the measure of the non-linear strain. This is to capture the fact that the pore structure in the body depends on the strain and hence will lead to a change in the diffusivity. However, apriori, it is not clear whether the diffusivity has to increase or decrease. In this study, we shall assume that the diffusivity increases with the strain. 

Now, the Almansi-Hamel strain is given by
\begin{align}
 \mathbf{e} = \dfrac{1}{2}\left(\mathbf{I} - \mathbf{B}^{-1}\right) = \left[ \begin{array}{ccc}
                                               0 &0 &0\\
						0& 0 & \dfrac{1}{2}r\psi\\
						0 & \dfrac{1}{2}r\psi & -\dfrac{1}{2}r^2\psi^2
                                              \end{array}\right].
\end{align}
The Frobenius norm of $\mathbf{e}$ is given by
\begin{align}
 \|\mathbf{e}\| := \sqrt{tr\left(\mathbf{e}^{\text{T}}\mathbf{e} \right)} &= \sqrt{\dfrac{1}{4}(r\psi)^4 + \dfrac{1}{2}(r\psi)^2}\notag\\
&= \sqrt{\dfrac{1}{4}\bar{q}^2\left(\bar{r}\bar{\psi} \right)^4 +  \dfrac{1}{2} \bar{q}\left(\bar{r}\bar{\psi} \right)^2} .
\end{align}
We shall assume that the diffusivity $\bar{D}$ depends on the Almansi-Hamel strain in the following manner
\begin{align}
 \bar{D} = \bar{D}_0 + \left(\bar{D}_{\infty}- \bar{D}_0 \right) \left(1-e^{-\lambda\|\mathbf{e}\|}  \right). \label{al}
\end{align}

\begin{figure}[htp]
	\subfloat[]{
	\label{fig:conc-profile-constant-diff-flux-bc}
 	\centering
	 \input{matlab-codes/figs/conc-profile-constant-diff-flux-bc.tex}
	\includegraphics[scale = 0.5]{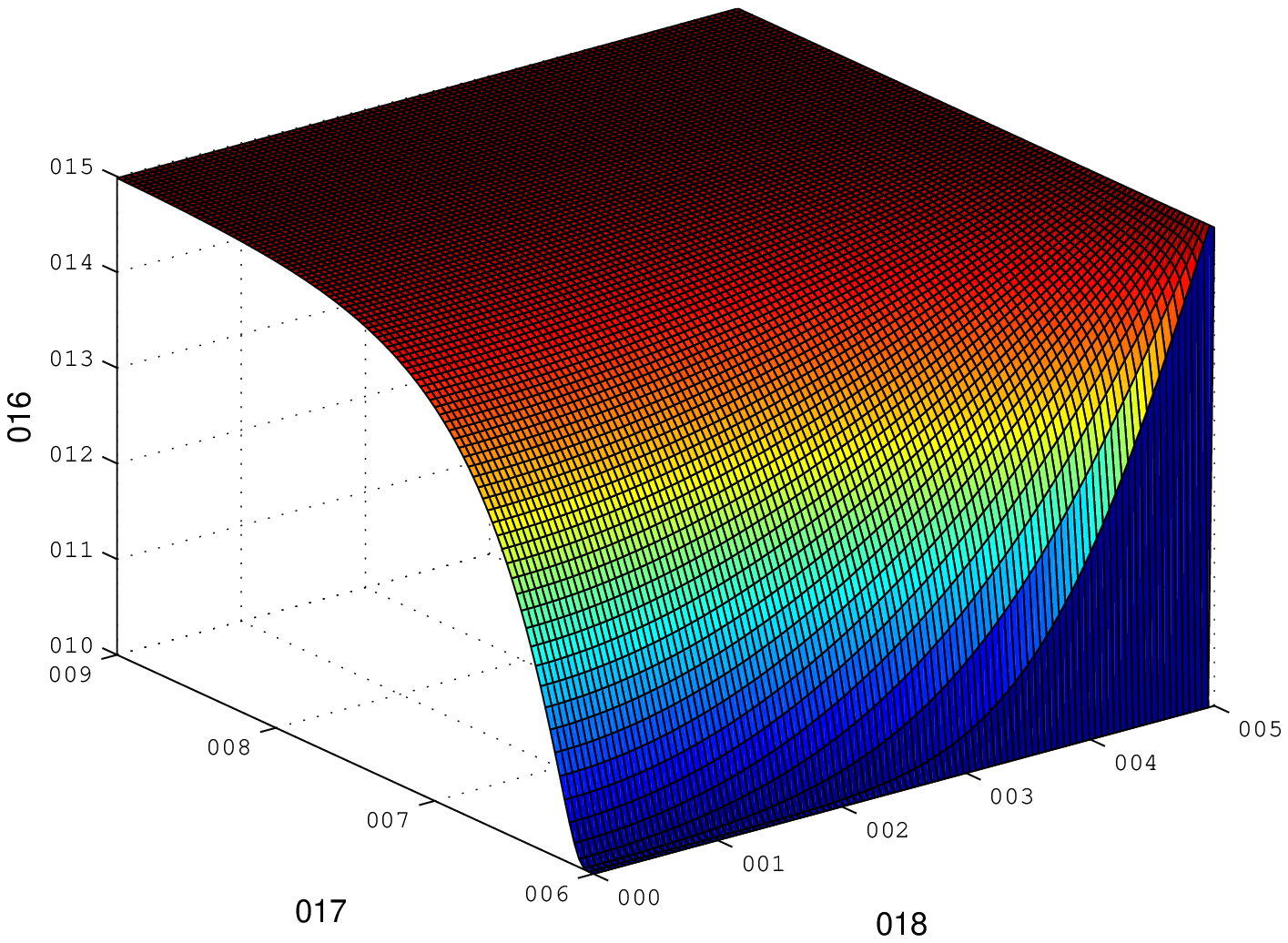}
  	}
	\subfloat[]{
	\label{fig:conc-profile-strain-diff-flux-bc}
 	\centering
	 \input{matlab-codes/figs/conc-profile-strain-diff-flux-bc.tex}
	\includegraphics[scale = 0.5]{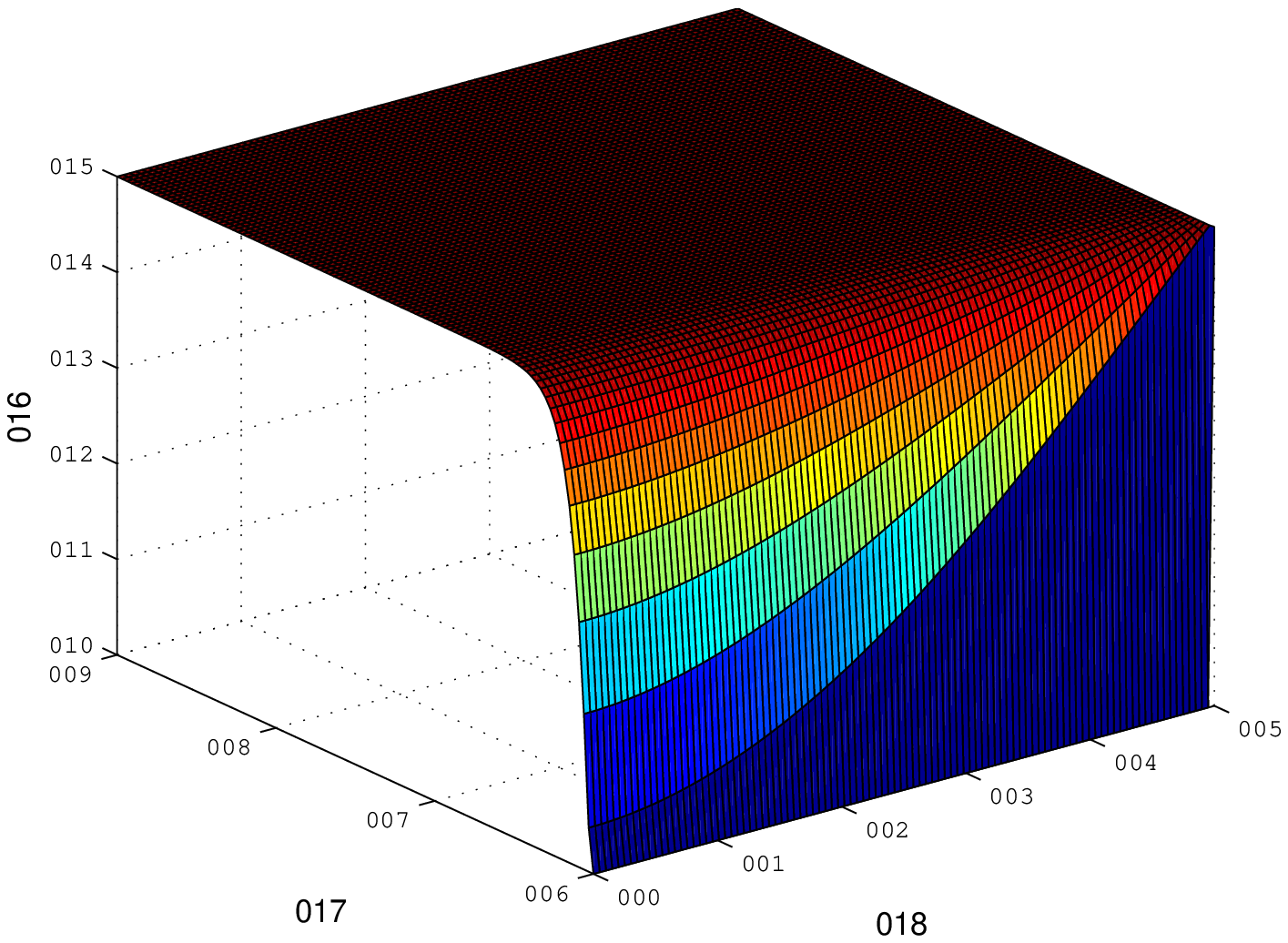}
	}\\
	\subfloat[]{
	\label{fig:moment-vs-t-comparison-flux-bc}
 	\centering
	 \input{matlab-codes/figs/moment-vs-t-comparison-flux-bc.tex}
	\includegraphics[scale = 0.5]{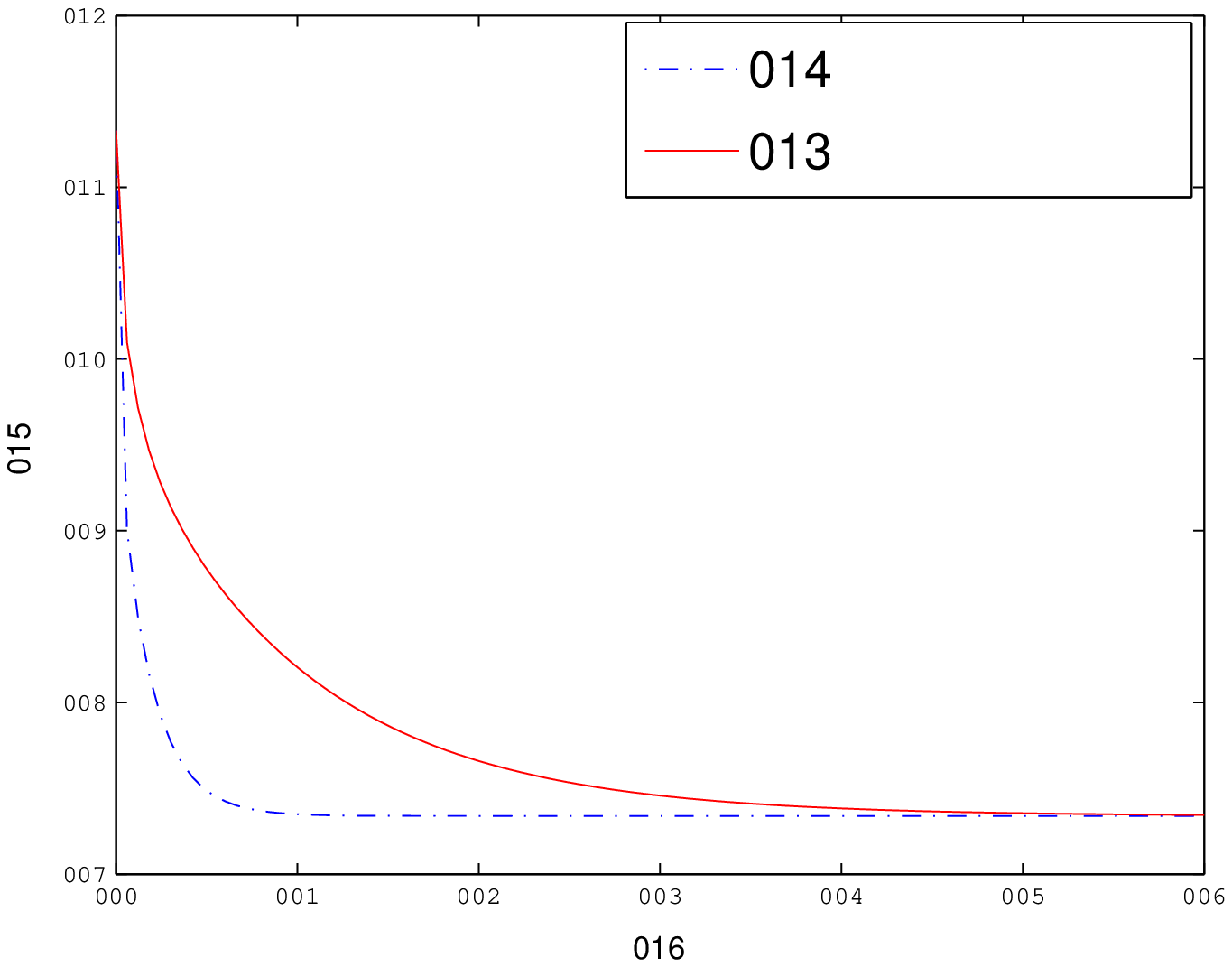}
	}
	\subfloat[]{
	\label{fig:psi-vs-t-comparison-flux-bc}
 	\centering
	 \input{matlab-codes/figs/psi-vs-t-comparison-flux-bc.tex}
	\includegraphics[scale = 0.5]{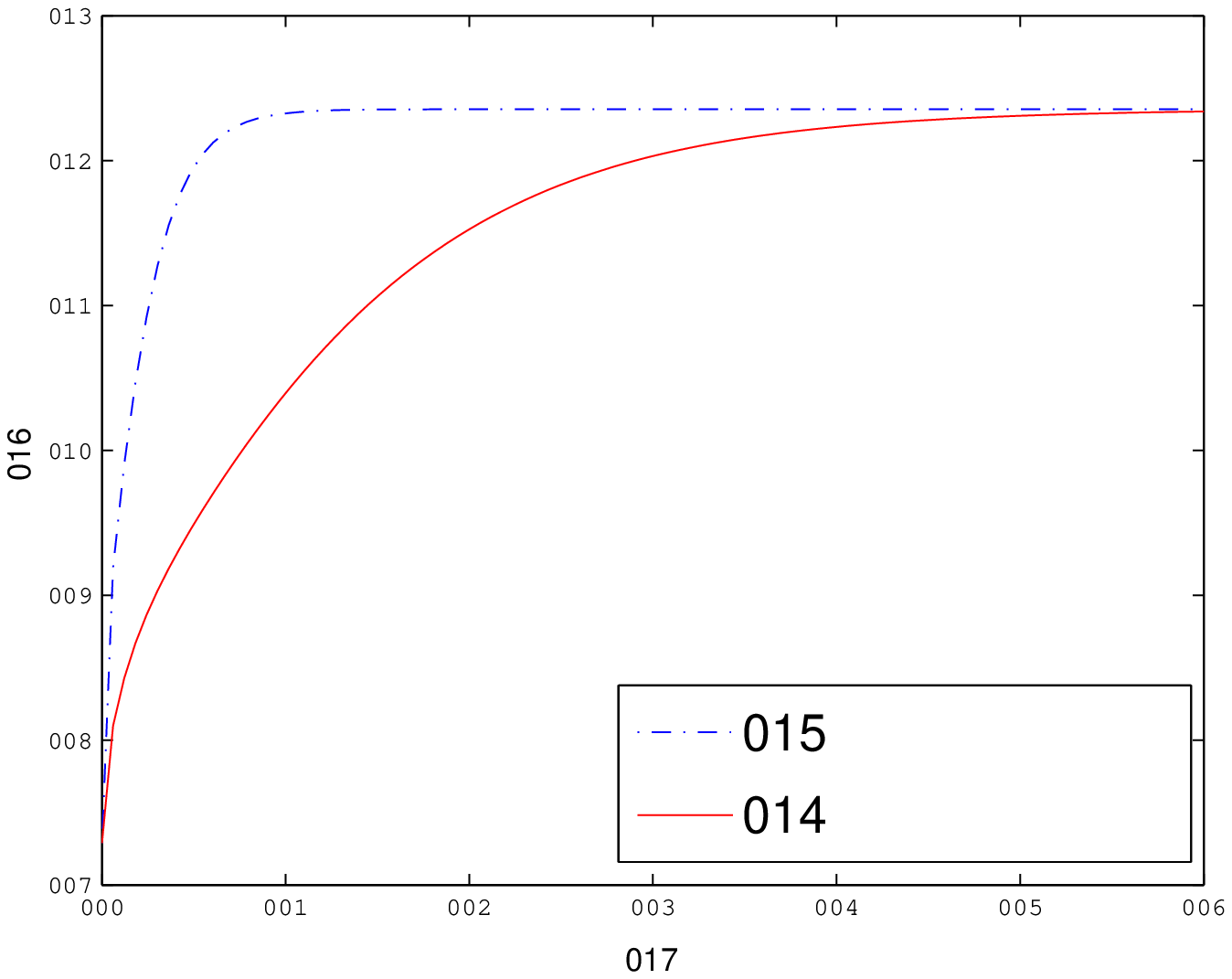}
	}
\caption{(a) Solution to the convection-diffusion equation (\ref{nondiff})  when the diffusivity is held constant.  (b) Solution to (\ref{nondiff}) when the diffusivity depends on the Almansi-Hamel strain. (c) Non-dimensional moment ($\bar{M}$) as a function of non-dimensional time ($\bar{t}$) with and without diffusivity depending on the Almansi-Hamel strain for $\bar{\psi} = 1$. (d) Non-dimensional angular displacement ($\bar{\psi}$) as a function of non-dimensional time ($\bar{t}$) with and without diffusivity depending on the Almansi-Hamel strain for $\bar{M} = 1$. Here, $R_i/R_o = 0.5$, $\bar{q} = 1$, $b_0 = n_0 = 1$, $b_1 = 0.1$, $n_1 = 0.1$, $\bar{\mu}_1 = 0.4$. For constant diffusivity case $\bar{D} = 0.01$, and for diffusivity depending on the Almansi-strain, relation (\ref{al}) is assumed with  $\bar{D}_0 = 0.01$, $\bar{D}_{\infty} = 0.1$, $\lambda  = 1$. Initial and boundary conditions considered are given by (\ref{case1_1}--\ref{case1_3}).}
\label{fig:7}
\end{figure}

Figures \eqref{fig:moment-vs-t-comparison-flux-bc}, \eqref{fig:psi-vs-t-comparison-flux-bc} compares the stress relaxation and creep for the case with constant diffusivity against the case when the diffusivity depends on the Almansi-Hamel strain with the initial and boundary conditions considered are given by (\ref{case1_1}--\ref{case1_3}). The stress relaxation and creep are faster when diffusivity is assumed to be a function of the Almansi-Hamel strain, as the diffusivity is higher in this case and hence the degradation is faster. 

\begin{figure}[htp]
	\subfloat[]{
	\label{fig:conc-profile-constant-diff-diri-bc}
 	\centering
	 \input{matlab-codes/figs/conc-profile-constant-diff-diri-bc.tex}
	\includegraphics[scale = 0.5]{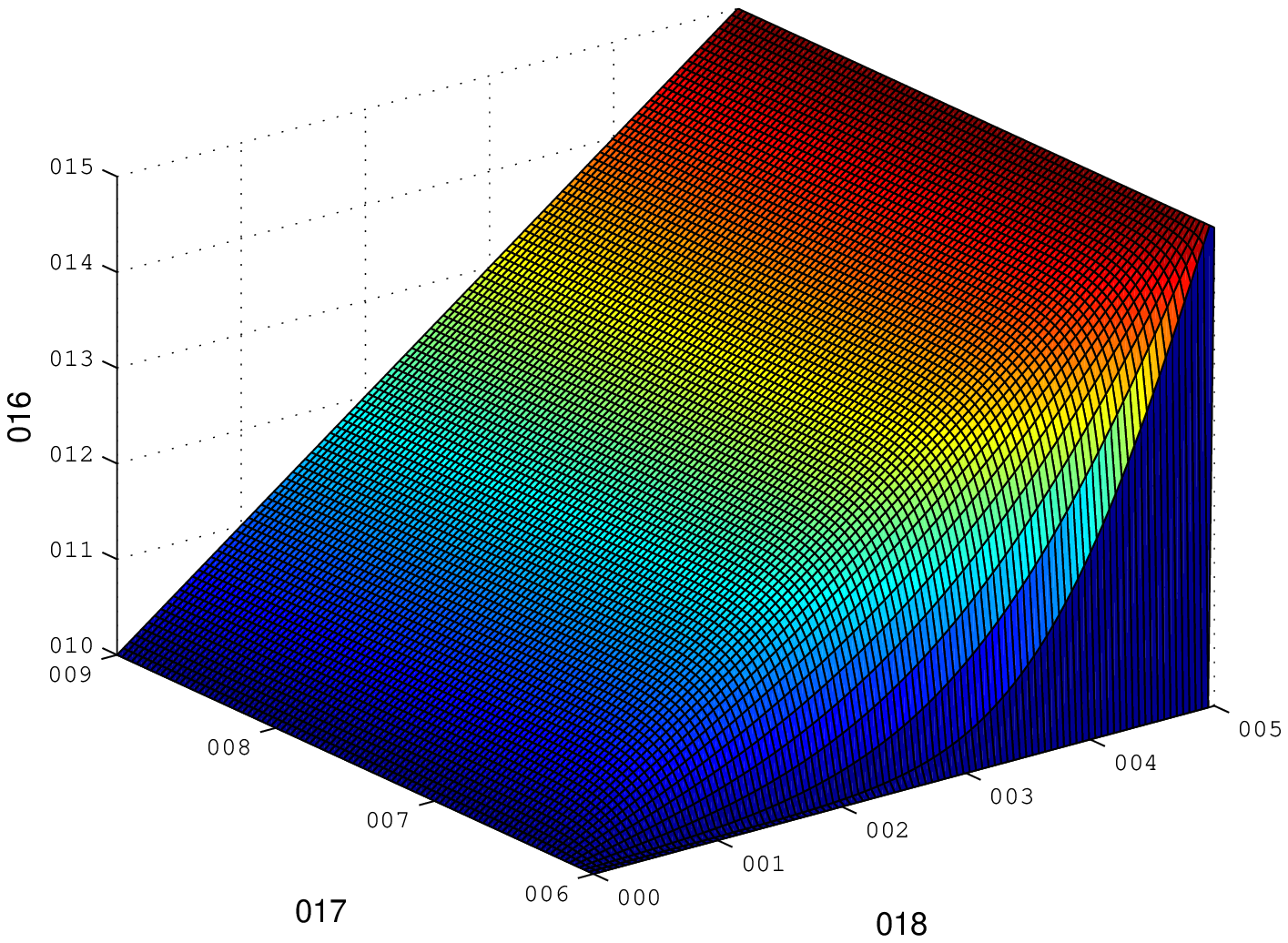}
  	}
	\subfloat[]{
	\label{fig:conc-profile-strain-diff-diri-bc}
 	\centering
	 \input{matlab-codes/figs/conc-profile-strain-diff-diri-bc.tex}
	\includegraphics[scale = 0.5]{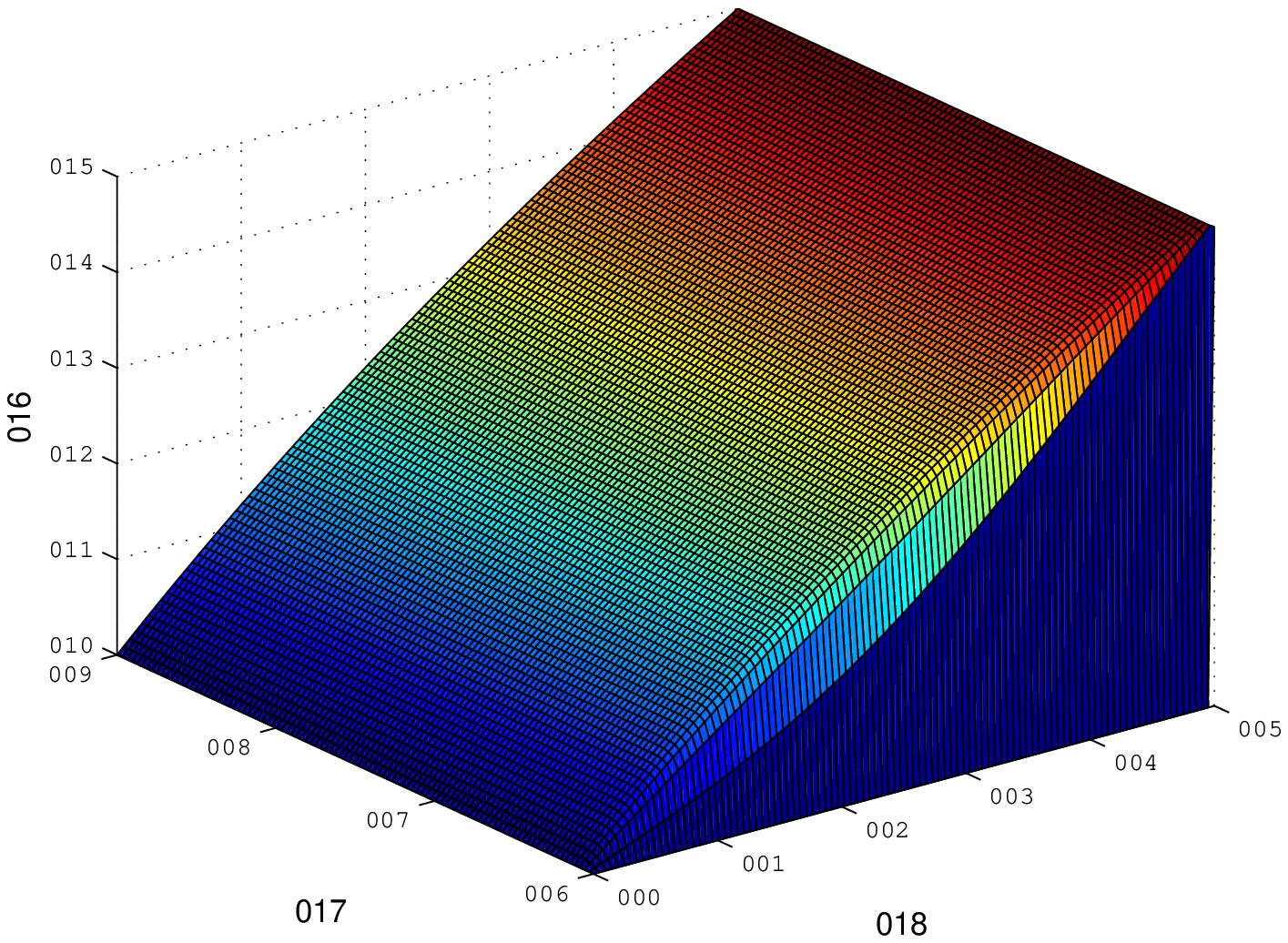}
	}\\
	\subfloat[]{
	\label{fig:moment-vs-t-comparison-diri-bc}
 	\centering
	 \input{matlab-codes/figs/moment-vs-t-comparison-diri-bc.tex}
	\includegraphics[scale = 0.5]{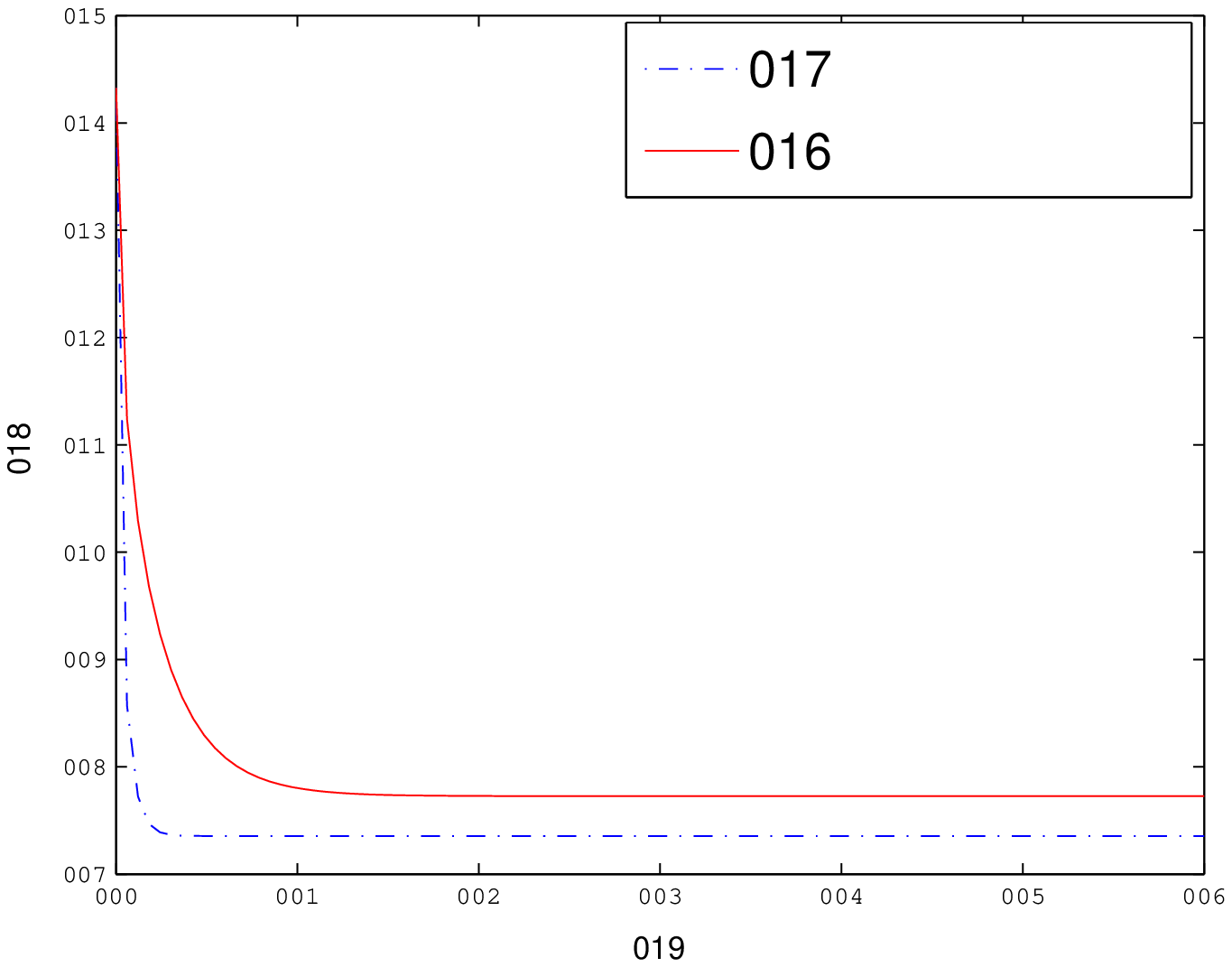}
	}
	\subfloat[]{
	\label{fig:psi-vs-t-comparison-diri-bc}
 	\centering
	 \input{matlab-codes/figs/psi-vs-t-comparison-diri-bc.tex}
	\includegraphics[scale = 0.5]{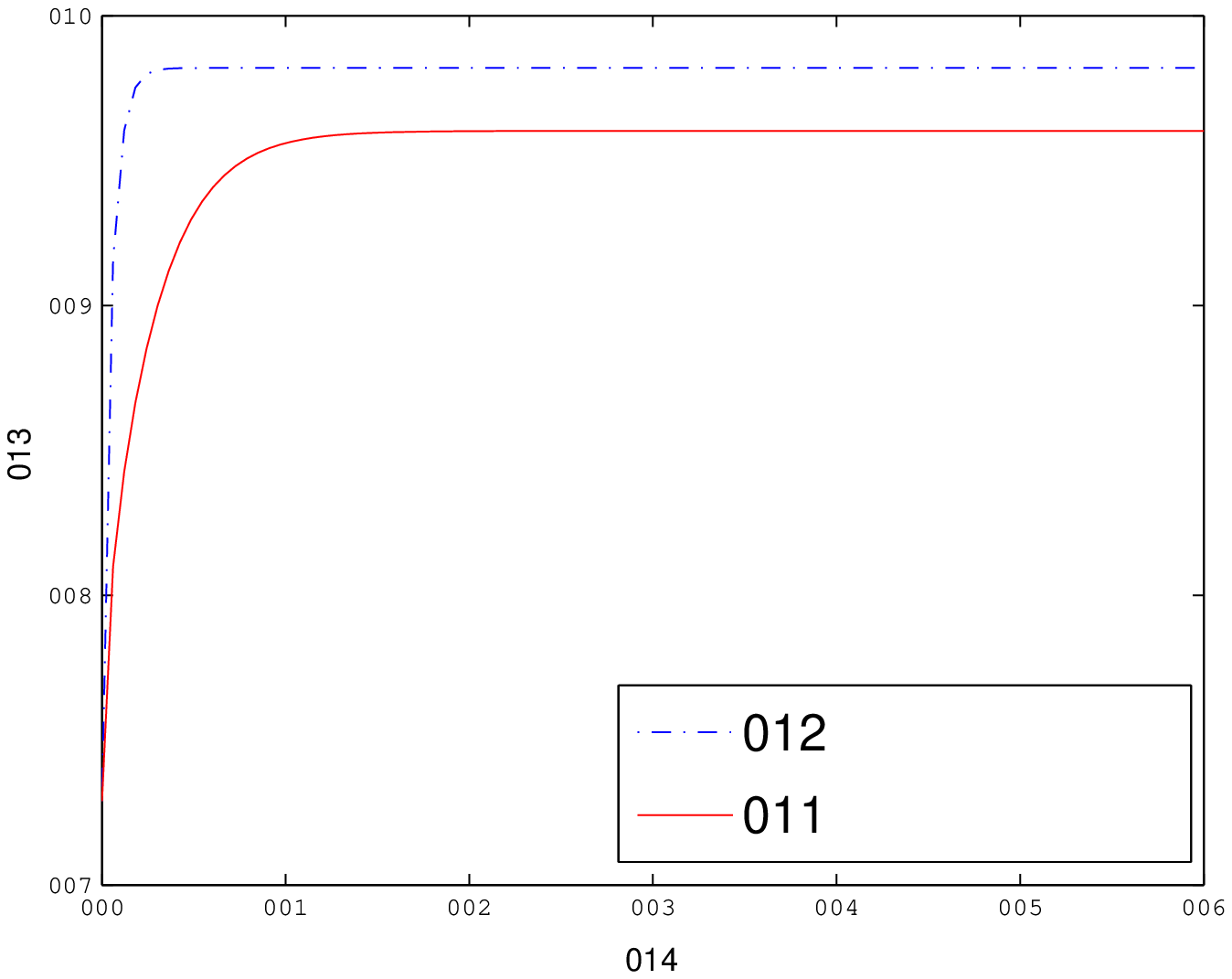}
	}
\caption{(a) Solution to the convection-diffusion equation (\ref{nondiff})  when the diffusivity is held constant.  (b) Solution to (\ref{nondiff}) when the diffusivity depends on the Almansi-Hamel strain. (c) Non-dimensional moment ($\bar{M}$) as a function of non-dimensional time ($\bar{t}$) with and without diffusivity depending on the Almansi-Hamel strain for $\bar{\psi} = 1$. (d) Non-dimensional displacement ($\bar{\psi}$) as a function of non-dimensional time ($\bar{t}$) with and without diffusivity depending on the Almansi-Hamel strain for $\bar{M} = 1$. Here, $R_i/R_o = 0.5$, $\bar{q} = 1$, $b_0 = n_0 = 1$, $b_1 = 0.1$, $n_1 = 0.1$, $\bar{\mu}_1 = 0.4$. For constant diffusivity case $\bar{D} = 0.01$, and for diffusivity depending on the Almansi-strain, relation (\ref{al}) is assumed with  $\bar{D}_0 = 0.01$, $\bar{D}_{\infty} = 0.1$, $\lambda  = 1$. Initial and boundary conditions considered are given by (\ref{case2_1}--\ref{case2_3}).}
\label{fig:8}
\end{figure}

\begin{figure}[htp]
 	\centering
	\input{matlab-codes/figs/conc-steady-state-diri-bc.tex}
	\includegraphics[scale = 0.5]{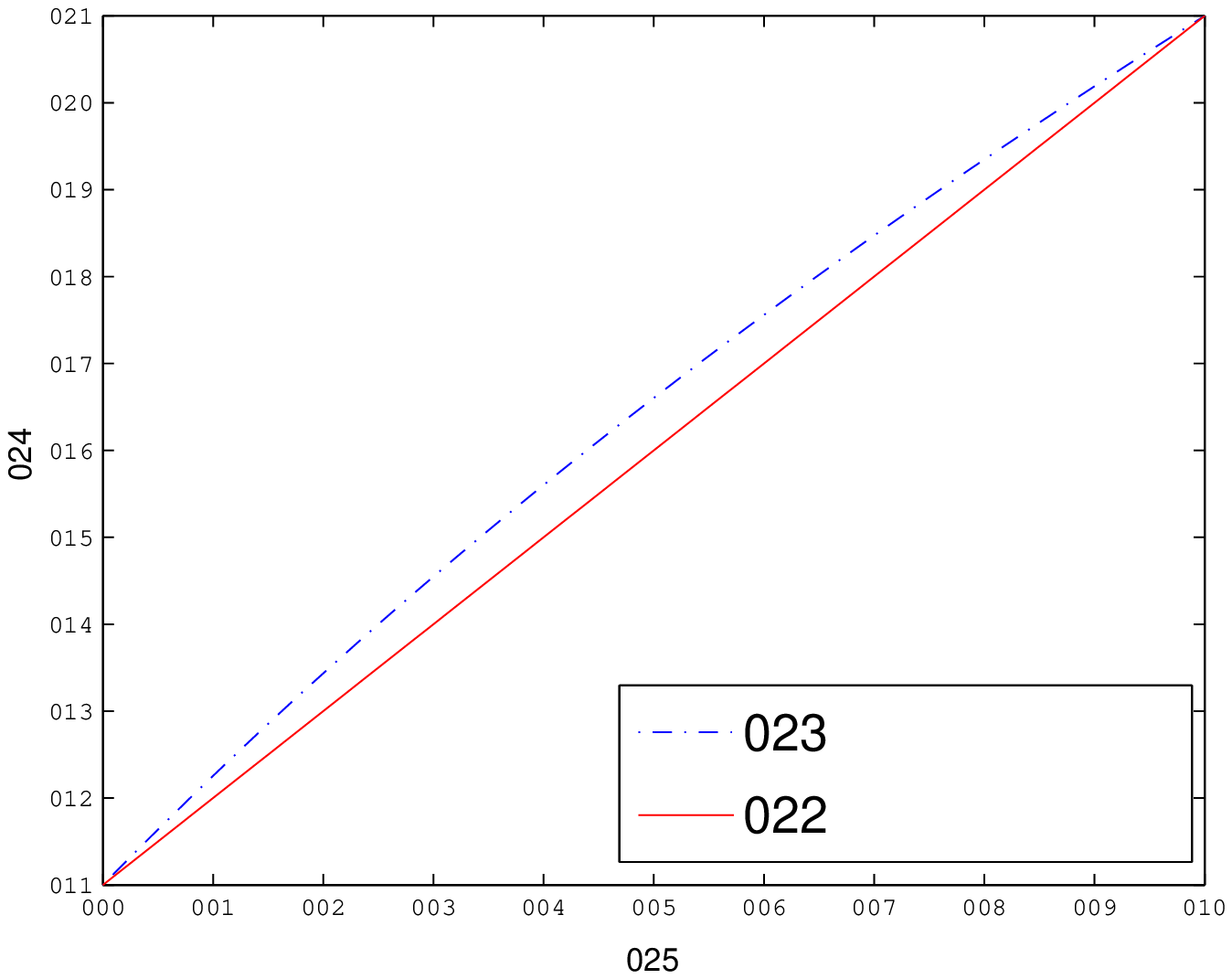}
\caption{ Comparison of the steady state solutions ($\bar{t} = 40$) to (\ref{nondiff})  with constant diffusivity and when the diffusivity depends on the Almansi-Hamel strain, with initial and boundary conditions given by (\ref{case2_1}--\ref{case2_3}). $R_i/R_o = 0.5$, $\bar{q} = 1$, $\bar{\psi} = 1$, $b_0 = n_0 = 1$, $b_1 = 0.1$, $n_1 = 0.1$, $\bar{\mu}_1 = 0.4$. For constant diffusivity case $\bar{D} = 0.01$, and for diffusivity depending on the Almansi-strain, relation (\ref{al}) is assumed with  $\bar{D}_0 = 0.01$, $\bar{D}_{\infty} = 0.1$, $\lambda  = 1$.}
\label{fig:9}
\end{figure}

Next, we shall consider the situation when we have the initial and boundary conditions  given by (\ref{case2_1}--\ref{case2_3}). Here too, the stress relaxation and creep are faster when the diffusivity depends on the Almansi-Hamel strain  is assumed (see figures \eqref{fig:moment-vs-t-comparison-diri-bc}, \eqref{fig:psi-vs-t-comparison-diri-bc}). However, the steady state values of the non-dimensional moment and non-dimensional angular displacement when the diffusivity is maintained constant is not the same as that when the diffusivity  depends on the Almansi-Hamel strain. One can explain this by looking at the steady state solutions for the convection-diffusion equation in figure (\ref{fig:9}). Unlike the situation corresponding to (\ref{case1_1}--\ref{case1_3}), the steady state solutions for the concentration are not the same and the steady state concentrations  when the diffusivity depends on the strain are higher than the steady state values when the diffusivity is constant. This induces the stresses to  relax to a lower value in \eqref{fig:moment-vs-t-comparison-diri-bc} and the angular diplacement to reach a higher value in \eqref{fig:psi-vs-t-comparison-diri-bc}. These values are also attained faster as the steady state solution is reached faster when the diffusivity depends on the strain.

\section{Conclusions}\label{sec4}

Using the generalized neo-Hookean solid as a vehicle to illustrate the effects of degradation of a body that is initially elastic we have shown that degradation due to the infusion of a fluid can cause stress relaxation and creep  until a steady state is reached. In contrast to a viscoelastic body which creeps continuously upon application of a load, the body considered here stops to creep after a certain steady state value is attained. This is one important characteristic that can be used to differentiate the creep due to the viscoelastic nature of a body, and the creep due to degradation. We have also considered the case of healing or strengthening of a generalized neo-Hookean body due to the infusion of a fluid.  One can easily extend this work to other forms of degradation in a generalized neo-Hookean material or for that matter in any elastic body whose properties change due to the infusion of a fluid, by providing an appropriate equation that governs the evolution of the degradation parameter. 

\section*{Acknowledgements}
The authors would like to thank AFOSR for their financial support through contract number  FA 9550-04-1-0137, and the financial support of AFOSR for a grant through the Universal Technology Corporation. Satish Karra would like to thank Maruti Mudunuru of Texas A\&M University for helpful suggestions.

\bibliographystyle{plain}
\bibliography{torsion_diffusion} 

\newpage


\end{document}

%% file: matlab-codes/figs/concentration-profile.tex
%
%
\providecommand\matlabtextA{\color[rgb]{0.000,0.000,0.000}\fontsize{10}{10}\selectfont\strut}%
\psfrag{018}[bc][bc]{\matlabtextA $c$}%
\psfrag{019}[tr][tr]{\matlabtextA $\bar{t}$}%
\psfrag{020}[tl][tl]{\matlabtextA $\bar{r}$}%
%
%
%
\def\matlabfragNegXTick{\mathord{\makebox[0pt][r]{$-$}}}
\providecommand\matlabtextB{\color[rgb]{0.000,0.000,0.000}\fontsize{8}{8}\selectfont\strut}%
\psfrag{000}[ct][ct]{\matlabtextB $0.5$}%
\psfrag{001}[ct][ct]{\matlabtextB $0.6$}%
\psfrag{002}[ct][ct]{\matlabtextB $0.7$}%
\psfrag{003}[ct][ct]{\matlabtextB $0.8$}%
\psfrag{004}[ct][ct]{\matlabtextB $0.9$}%
\psfrag{005}[ct][ct]{\matlabtextB $1$}%
%
%
%
\psfrag{006}[rc][rc]{\matlabtextB $0$}%
\psfrag{007}[rc][rc]{\matlabtextB $20$}%
\psfrag{008}[rc][rc]{\matlabtextB $40$}%
\psfrag{009}[rc][rc]{\matlabtextB $60$}%
\psfrag{010}[rc][rc]{\matlabtextB $80$}%
\psfrag{011}[rc][rc]{\matlabtextB $100$}%
%
%
%
\psfrag{012}[cr][cr]{\matlabtextB $0$}%
\psfrag{013}[cr][cr]{\matlabtextB $0.2$}%
\psfrag{014}[cr][cr]{\matlabtextB $0.4$}%
\psfrag{015}[cr][cr]{\matlabtextB $0.6$}%
\psfrag{016}[cr][cr]{\matlabtextB $0.8$}%
\psfrag{017}[cr][cr]{\matlabtextB $1$}%
%

%% file: matlab-codes/figs/Moment-vs-t-mu1-changing.tex
%
%
\providecommand\matlabtextA{\color[rgb]{0.000,0.000,0.000}\fontsize{10}{10}\selectfont\strut}%
\psfrag{018}[bc][bc]{\matlabtextA $\bar{M}$}%
\psfrag{019}[tc][tc]{\matlabtextA $\bar{t}$}%
\psfrag{020}[bc][bc]{\matlabtextA $\bar{\mu}_1$ varying from 0 to 0.5}%
%
%
%
\def\matlabfragNegXTick{\mathord{\makebox[0pt][r]{$-$}}}
\providecommand\matlabtextB{\color[rgb]{0.000,0.000,0.000}\fontsize{8}{8}\selectfont\strut}%
\psfrag{000}[ct][ct]{\matlabtextB $0$}%
\psfrag{001}[ct][ct]{\matlabtextB $10$}%
\psfrag{002}[ct][ct]{\matlabtextB $20$}%
\psfrag{003}[ct][ct]{\matlabtextB $30$}%
\psfrag{004}[ct][ct]{\matlabtextB $40$}%
\psfrag{005}[ct][ct]{\matlabtextB $50$}%
\psfrag{006}[ct][ct]{\matlabtextB $60$}%
\psfrag{007}[ct][ct]{\matlabtextB $70$}%
\psfrag{008}[ct][ct]{\matlabtextB $80$}%
\psfrag{009}[ct][ct]{\matlabtextB $90$}%
\psfrag{010}[ct][ct]{\matlabtextB $100$}%
%
%
%
\psfrag{011}[rc][rc]{\matlabtextB $0.2$}%
\psfrag{012}[rc][rc]{\matlabtextB $0.25$}%
\psfrag{013}[rc][rc]{\matlabtextB $0.3$}%
\psfrag{014}[rc][rc]{\matlabtextB $0.35$}%
\psfrag{015}[rc][rc]{\matlabtextB $0.4$}%
\psfrag{016}[rc][rc]{\matlabtextB $0.45$}%
\psfrag{017}[rc][rc]{\matlabtextB $0.5$}%
%

%% file: matlab-codes/figs/Moment-vs-t-mu1-changing-n1-nonzero.tex
%
%
\providecommand\matlabtextA{\color[rgb]{0.000,0.000,0.000}\fontsize{10}{10}\selectfont\strut}%
\psfrag{018}[bc][bc]{\matlabtextA $\bar{M}$}%
\psfrag{019}[tc][tc]{\matlabtextA $\bar{t}$}%
\psfrag{020}[bc][bc]{\matlabtextA $\bar{\mu}_1$ varying from 0 to 0.5}%
%
%
%
\def\matlabfragNegXTick{\mathord{\makebox[0pt][r]{$-$}}}
\providecommand\matlabtextB{\color[rgb]{0.000,0.000,0.000}\fontsize{8}{8}\selectfont\strut}%
\psfrag{000}[ct][ct]{\matlabtextB $0$}%
\psfrag{001}[ct][ct]{\matlabtextB $10$}%
\psfrag{002}[ct][ct]{\matlabtextB $20$}%
\psfrag{003}[ct][ct]{\matlabtextB $30$}%
\psfrag{004}[ct][ct]{\matlabtextB $40$}%
\psfrag{005}[ct][ct]{\matlabtextB $50$}%
\psfrag{006}[ct][ct]{\matlabtextB $60$}%
\psfrag{007}[ct][ct]{\matlabtextB $70$}%
\psfrag{008}[ct][ct]{\matlabtextB $80$}%
\psfrag{009}[ct][ct]{\matlabtextB $90$}%
\psfrag{010}[ct][ct]{\matlabtextB $100$}%
%
%
%
\psfrag{011}[rc][rc]{\matlabtextB $0.2$}%
\psfrag{012}[rc][rc]{\matlabtextB $0.25$}%
\psfrag{013}[rc][rc]{\matlabtextB $0.3$}%
\psfrag{014}[rc][rc]{\matlabtextB $0.35$}%
\psfrag{015}[rc][rc]{\matlabtextB $0.4$}%
\psfrag{016}[rc][rc]{\matlabtextB $0.45$}%
\psfrag{017}[rc][rc]{\matlabtextB $0.5$}%
%

%% file: matlab-codes/figs/Moment-vs-t-b-changing.tex
%
%
\providecommand\matlabtextA{\color[rgb]{0.000,0.000,0.000}\fontsize{10}{10}\selectfont\strut}%
\psfrag{021}[bc][bc]{\matlabtextA $\bar{M}$}%
\psfrag{022}[tc][tc]{\matlabtextA $\bar{t}$}%
\psfrag{023}[bc][bc]{\matlabtextA $b_1$ varying from 0 to 0.8}%
%
%
%
\def\matlabfragNegXTick{\mathord{\makebox[0pt][r]{$-$}}}
\providecommand\matlabtextB{\color[rgb]{0.000,0.000,0.000}\fontsize{8}{8}\selectfont\strut}%
\psfrag{000}[ct][ct]{\matlabtextB $0$}%
\psfrag{001}[ct][ct]{\matlabtextB $10$}%
\psfrag{002}[ct][ct]{\matlabtextB $20$}%
\psfrag{003}[ct][ct]{\matlabtextB $30$}%
\psfrag{004}[ct][ct]{\matlabtextB $40$}%
\psfrag{005}[ct][ct]{\matlabtextB $50$}%
\psfrag{006}[ct][ct]{\matlabtextB $60$}%
\psfrag{007}[ct][ct]{\matlabtextB $70$}%
\psfrag{008}[ct][ct]{\matlabtextB $80$}%
\psfrag{009}[ct][ct]{\matlabtextB $90$}%
\psfrag{010}[ct][ct]{\matlabtextB $100$}%
%
%
%
\psfrag{011}[rc][rc]{\matlabtextB $0.39$}%
\psfrag{012}[rc][rc]{\matlabtextB $0.4$}%
\psfrag{013}[rc][rc]{\matlabtextB $0.41$}%
\psfrag{014}[rc][rc]{\matlabtextB $0.42$}%
\psfrag{015}[rc][rc]{\matlabtextB $0.43$}%
\psfrag{016}[rc][rc]{\matlabtextB $0.44$}%
\psfrag{017}[rc][rc]{\matlabtextB $0.45$}%
\psfrag{018}[rc][rc]{\matlabtextB $0.46$}%
\psfrag{019}[rc][rc]{\matlabtextB $0.47$}%
\psfrag{020}[rc][rc]{\matlabtextB $0.48$}%
%

%% file: matlab-codes/figs/Moment-vs-t-n-changing.tex
%
%
\providecommand\matlabtextA{\color[rgb]{0.000,0.000,0.000}\fontsize{10}{10}\selectfont\strut}%
\psfrag{020}[bc][bc]{\matlabtextA $\bar{M}$}%
\psfrag{021}[tc][tc]{\matlabtextA $\bar{t}$}%
\psfrag{022}[bc][bc]{\matlabtextA $n_1$ varying from 0 to 0.8}%
%
%
%
\def\matlabfragNegXTick{\mathord{\makebox[0pt][r]{$-$}}}
\providecommand\matlabtextB{\color[rgb]{0.000,0.000,0.000}\fontsize{8}{8}\selectfont\strut}%
\psfrag{000}[ct][ct]{\matlabtextB $0$}%
\psfrag{001}[ct][ct]{\matlabtextB $10$}%
\psfrag{002}[ct][ct]{\matlabtextB $20$}%
\psfrag{003}[ct][ct]{\matlabtextB $30$}%
\psfrag{004}[ct][ct]{\matlabtextB $40$}%
\psfrag{005}[ct][ct]{\matlabtextB $50$}%
\psfrag{006}[ct][ct]{\matlabtextB $60$}%
\psfrag{007}[ct][ct]{\matlabtextB $70$}%
\psfrag{008}[ct][ct]{\matlabtextB $80$}%
\psfrag{009}[ct][ct]{\matlabtextB $90$}%
\psfrag{010}[ct][ct]{\matlabtextB $100$}%
%
%
%
\psfrag{011}[rc][rc]{\matlabtextB $0.1$}%
\psfrag{012}[rc][rc]{\matlabtextB $0.15$}%
\psfrag{013}[rc][rc]{\matlabtextB $0.2$}%
\psfrag{014}[rc][rc]{\matlabtextB $0.25$}%
\psfrag{015}[rc][rc]{\matlabtextB $0.3$}%
\psfrag{016}[rc][rc]{\matlabtextB $0.35$}%
\psfrag{017}[rc][rc]{\matlabtextB $0.4$}%
\psfrag{018}[rc][rc]{\matlabtextB $0.45$}%
\psfrag{019}[rc][rc]{\matlabtextB $0.5$}%
%

%% file: matlab-codes/figs/Moment-vs-t-mu1-changing-healing.tex
%
%
\providecommand\matlabtextA{\color[rgb]{0.000,0.000,0.000}\fontsize{10}{10}\selectfont\strut}%
\psfrag{018}[bc][bc]{\matlabtextA $\bar{M}$}%
\psfrag{019}[tc][tc]{\matlabtextA $\bar{t}$}%
\psfrag{020}[tc][tc]{\matlabtextA $\bar{\mu}_1$ varying from 0 to 0.5}%
%
%
%
\def\matlabfragNegXTick{\mathord{\makebox[0pt][r]{$-$}}}
\providecommand\matlabtextB{\color[rgb]{0.000,0.000,0.000}\fontsize{8}{8}\selectfont\strut}%
\psfrag{000}[ct][ct]{\matlabtextB $0$}%
\psfrag{001}[ct][ct]{\matlabtextB $10$}%
\psfrag{002}[ct][ct]{\matlabtextB $20$}%
\psfrag{003}[ct][ct]{\matlabtextB $30$}%
\psfrag{004}[ct][ct]{\matlabtextB $40$}%
\psfrag{005}[ct][ct]{\matlabtextB $50$}%
\psfrag{006}[ct][ct]{\matlabtextB $60$}%
\psfrag{007}[ct][ct]{\matlabtextB $70$}%
\psfrag{008}[ct][ct]{\matlabtextB $80$}%
\psfrag{009}[ct][ct]{\matlabtextB $90$}%
\psfrag{010}[ct][ct]{\matlabtextB $100$}%
%
%
%
\psfrag{011}[rc][rc]{\matlabtextB $0.45$}%
\psfrag{012}[rc][rc]{\matlabtextB $0.5$}%
\psfrag{013}[rc][rc]{\matlabtextB $0.55$}%
\psfrag{014}[rc][rc]{\matlabtextB $0.6$}%
\psfrag{015}[rc][rc]{\matlabtextB $0.65$}%
\psfrag{016}[rc][rc]{\matlabtextB $0.7$}%
\psfrag{017}[rc][rc]{\matlabtextB $0.75$}%
%

%% file: matlab-codes/figs/Moment-vs-t-mu1-changing-n1-nonzero-healing.tex
%
%
\providecommand\matlabtextA{\color[rgb]{0.000,0.000,0.000}\fontsize{10}{10}\selectfont\strut}%
\psfrag{018}[bc][bc]{\matlabtextA $\bar{M}$}%
\psfrag{019}[tc][tc]{\matlabtextA $\bar{t}$}%
\psfrag{020}[tc][tc]{\matlabtextA $\bar{\mu}_1$ varying from 0 to 0.5}%
%
%
%
\def\matlabfragNegXTick{\mathord{\makebox[0pt][r]{$-$}}}
\providecommand\matlabtextB{\color[rgb]{0.000,0.000,0.000}\fontsize{8}{8}\selectfont\strut}%
\psfrag{000}[ct][ct]{\matlabtextB $0$}%
\psfrag{001}[ct][ct]{\matlabtextB $10$}%
\psfrag{002}[ct][ct]{\matlabtextB $20$}%
\psfrag{003}[ct][ct]{\matlabtextB $30$}%
\psfrag{004}[ct][ct]{\matlabtextB $40$}%
\psfrag{005}[ct][ct]{\matlabtextB $50$}%
\psfrag{006}[ct][ct]{\matlabtextB $60$}%
\psfrag{007}[ct][ct]{\matlabtextB $70$}%
\psfrag{008}[ct][ct]{\matlabtextB $80$}%
\psfrag{009}[ct][ct]{\matlabtextB $90$}%
\psfrag{010}[ct][ct]{\matlabtextB $100$}%
%
%
%
\psfrag{011}[rc][rc]{\matlabtextB $0.45$}%
\psfrag{012}[rc][rc]{\matlabtextB $0.5$}%
\psfrag{013}[rc][rc]{\matlabtextB $0.55$}%
\psfrag{014}[rc][rc]{\matlabtextB $0.6$}%
\psfrag{015}[rc][rc]{\matlabtextB $0.65$}%
\psfrag{016}[rc][rc]{\matlabtextB $0.7$}%
\psfrag{017}[rc][rc]{\matlabtextB $0.75$}%
%

%% file: matlab-codes/figs/Moment-vs-t-b-changing-healing.tex
%
%
\providecommand\matlabtextA{\color[rgb]{0.000,0.000,0.000}\fontsize{10}{10}\selectfont\strut}%
\psfrag{018}[bc][bc]{\matlabtextA $\bar{M}$}%
\psfrag{019}[tc][tc]{\matlabtextA $\bar{t}$}%

\psfrag{020}[tc][tc]{\matlabtextA $b_1$ varying from 0 to 0.8}%
%
%
%
\def\matlabfragNegXTick{\mathord{\makebox[0pt][r]{$-$}}}
\providecommand\matlabtextB{\color[rgb]{0.000,0.000,0.000}\fontsize{7}{7}\selectfont\strut}%
\psfrag{000}[ct][ct]{\matlabtextB $0$}%
\psfrag{001}[ct][ct]{\matlabtextB $10$}%
\psfrag{002}[ct][ct]{\matlabtextB $20$}%
\psfrag{003}[ct][ct]{\matlabtextB $30$}%
\psfrag{004}[ct][ct]{\matlabtextB $40$}%
\psfrag{005}[ct][ct]{\matlabtextB $50$}%
\psfrag{006}[ct][ct]{\matlabtextB $60$}%
\psfrag{007}[ct][ct]{\matlabtextB $70$}%
\psfrag{008}[ct][ct]{\matlabtextB $80$}%
\psfrag{009}[ct][ct]{\matlabtextB $90$}%
\psfrag{010}[ct][ct]{\matlabtextB $100$}%
%
%
%
\psfrag{011}[rc][rc]{\matlabtextB $0.46$}%
\psfrag{012}[rc][rc]{\matlabtextB $0.48$}%
\psfrag{013}[rc][rc]{\matlabtextB $0.5$}%
\psfrag{014}[rc][rc]{\matlabtextB $0.52$}%
\psfrag{015}[rc][rc]{\matlabtextB $0.54$}%
\psfrag{016}[rc][rc]{\matlabtextB $0.56$}%
\psfrag{017}[rc][rc]{\matlabtextB $0.58$}%
%

%% file: matlab-codes/figs/Moment-vs-t-n-changing-healing.tex
%
%
\providecommand\matlabtextA{\color[rgb]{0.000,0.000,0.000}\fontsize{10}{10}\selectfont\strut}%
\psfrag{017}[bc][bc]{\matlabtextA $\bar{M}$}%
\psfrag{018}[tc][tc]{\matlabtextA $\bar{t}$}%
\psfrag{019}[tc][tc]{\matlabtextA $n_1$ varying from 0 to 0.8}%
%
%
%
\def\matlabfragNegXTick{\mathord{\makebox[0pt][r]{$-$}}}
\providecommand\matlabtextB{\color[rgb]{0.000,0.000,0.000}\fontsize{8}{8}\selectfont\strut}%
\psfrag{000}[ct][ct]{\matlabtextB $0$}%
\psfrag{001}[ct][ct]{\matlabtextB $10$}%
\psfrag{002}[ct][ct]{\matlabtextB $20$}%
\psfrag{003}[ct][ct]{\matlabtextB $30$}%
\psfrag{004}[ct][ct]{\matlabtextB $40$}%
\psfrag{005}[ct][ct]{\matlabtextB $50$}%
\psfrag{006}[ct][ct]{\matlabtextB $60$}%
\psfrag{007}[ct][ct]{\matlabtextB $70$}%
\psfrag{008}[ct][ct]{\matlabtextB $80$}%
\psfrag{009}[ct][ct]{\matlabtextB $90$}%
\psfrag{010}[ct][ct]{\matlabtextB $100$}%
%
%
%
\psfrag{011}[rc][rc]{\matlabtextB $0.45$}%
\psfrag{012}[rc][rc]{\matlabtextB $0.5$}%
\psfrag{013}[rc][rc]{\matlabtextB $0.55$}%
\psfrag{014}[rc][rc]{\matlabtextB $0.6$}%
\psfrag{015}[rc][rc]{\matlabtextB $0.65$}%
\psfrag{016}[rc][rc]{\matlabtextB $0.7$}%
%

%% file: matlab-codes/figs/psi-vs-t-mu1-changing.tex
%
%
\providecommand\matlabtextA{\color[rgb]{0.000,0.000,0.000}\fontsize{10}{10}\selectfont\strut}%
\psfrag{019}[bc][bc]{\matlabtextA $\bar{\psi}$}%
\psfrag{020}[tc][tc]{\matlabtextA $\bar{t}$}%
\psfrag{021}[tc][tc]{\matlabtextA $\bar{\mu}_1$ varying from 0 to 0.5}%
%
%
%
\def\matlabfragNegXTick{\mathord{\makebox[0pt][r]{$-$}}}
\providecommand\matlabtextB{\color[rgb]{0.000,0.000,0.000}\fontsize{8}{8}\selectfont\strut}%
\psfrag{000}[ct][ct]{\matlabtextB $0$}%
\psfrag{001}[ct][ct]{\matlabtextB $10$}%
\psfrag{002}[ct][ct]{\matlabtextB $20$}%
\psfrag{003}[ct][ct]{\matlabtextB $30$}%
\psfrag{004}[ct][ct]{\matlabtextB $40$}%
\psfrag{005}[ct][ct]{\matlabtextB $50$}%
\psfrag{006}[ct][ct]{\matlabtextB $60$}%
\psfrag{007}[ct][ct]{\matlabtextB $70$}%
\psfrag{008}[ct][ct]{\matlabtextB $80$}%
\psfrag{009}[ct][ct]{\matlabtextB $90$}%
\psfrag{010}[ct][ct]{\matlabtextB $100$}%
%
%
%
\psfrag{011}[rc][rc]{\matlabtextB $1$}%
\psfrag{012}[rc][rc]{\matlabtextB $1.2$}%
\psfrag{013}[rc][rc]{\matlabtextB $1.4$}%
\psfrag{014}[rc][rc]{\matlabtextB $1.6$}%
\psfrag{015}[rc][rc]{\matlabtextB $1.8$}%
\psfrag{016}[rc][rc]{\matlabtextB $2$}%
\psfrag{017}[rc][rc]{\matlabtextB $2.2$}%
\psfrag{018}[rc][rc]{\matlabtextB $2.4$}%
%

%% file: matlab-codes/figs/psi-vs-t-mu1-changing-n-nonzero.tex
%
%
\providecommand\matlabtextA{\color[rgb]{0.000,0.000,0.000}\fontsize{10}{10}\selectfont\strut}%
\psfrag{020}[bc][bc]{\matlabtextA $\bar{\psi}$}%
\psfrag{021}[tc][tc]{\matlabtextA $\bar{t}$}%
\psfrag{022}[tc][tc]{\matlabtextA $\bar{\mu}_1$ varying from 0 to 0.5}%
%
%
%
\def\matlabfragNegXTick{\mathord{\makebox[0pt][r]{$-$}}}
\providecommand\matlabtextB{\color[rgb]{0.000,0.000,0.000}\fontsize{8}{8}\selectfont\strut}%
\psfrag{000}[ct][ct]{\matlabtextB $0$}%
\psfrag{001}[ct][ct]{\matlabtextB $10$}%
\psfrag{002}[ct][ct]{\matlabtextB $20$}%
\psfrag{003}[ct][ct]{\matlabtextB $30$}%
\psfrag{004}[ct][ct]{\matlabtextB $40$}%
\psfrag{005}[ct][ct]{\matlabtextB $50$}%
\psfrag{006}[ct][ct]{\matlabtextB $60$}%
\psfrag{007}[ct][ct]{\matlabtextB $70$}%
\psfrag{008}[ct][ct]{\matlabtextB $80$}%
\psfrag{009}[ct][ct]{\matlabtextB $90$}%
\psfrag{010}[ct][ct]{\matlabtextB $100$}%
%
%
%
\psfrag{011}[rc][rc]{\matlabtextB $1$}%
\psfrag{012}[rc][rc]{\matlabtextB $1.2$}%
\psfrag{013}[rc][rc]{\matlabtextB $1.4$}%
\psfrag{014}[rc][rc]{\matlabtextB $1.6$}%
\psfrag{015}[rc][rc]{\matlabtextB $1.8$}%
\psfrag{016}[rc][rc]{\matlabtextB $2$}%
\psfrag{017}[rc][rc]{\matlabtextB $2.2$}%
\psfrag{018}[rc][rc]{\matlabtextB $2.4$}%
\psfrag{019}[rc][rc]{\matlabtextB $2.6$}%
%

%% file: matlab-codes/figs/psi-vs-t-b1-changing.tex
%
%
\providecommand\matlabtextA{\color[rgb]{0.000,0.000,0.000}\fontsize{10}{10}\selectfont\strut}%
\psfrag{017}[bc][bc]{\matlabtextA $\bar{\psi}$}%
\psfrag{018}[tc][tc]{\matlabtextA $\bar{t}$}%
\psfrag{019}[tc][tc]{\matlabtextA $b_1$ varying from 0 to 0.8}%
%
%
%
\def\matlabfragNegXTick{\mathord{\makebox[0pt][r]{$-$}}}
\providecommand\matlabtextB{\color[rgb]{0.000,0.000,0.000}\fontsize{8}{8}\selectfont\strut}%
\psfrag{000}[ct][ct]{\matlabtextB $0$}%
\psfrag{001}[ct][ct]{\matlabtextB $10$}%
\psfrag{002}[ct][ct]{\matlabtextB $20$}%
\psfrag{003}[ct][ct]{\matlabtextB $30$}%
\psfrag{004}[ct][ct]{\matlabtextB $40$}%
\psfrag{005}[ct][ct]{\matlabtextB $50$}%
\psfrag{006}[ct][ct]{\matlabtextB $60$}%
\psfrag{007}[ct][ct]{\matlabtextB $70$}%
\psfrag{008}[ct][ct]{\matlabtextB $80$}%
\psfrag{009}[ct][ct]{\matlabtextB $90$}%
\psfrag{010}[ct][ct]{\matlabtextB $100$}%
%
%
%
\psfrag{011}[rc][rc]{\matlabtextB $1.05$}%
\psfrag{012}[rc][rc]{\matlabtextB $1.1$}%
\psfrag{013}[rc][rc]{\matlabtextB $1.15$}%
\psfrag{014}[rc][rc]{\matlabtextB $1.2$}%
\psfrag{015}[rc][rc]{\matlabtextB $1.25$}%
\psfrag{016}[rc][rc]{\matlabtextB $1.3$}%
%

%% file: matlab-codes/figs/psi-vs-t-n1-changing.tex
%
%
\providecommand\matlabtextA{\color[rgb]{0.000,0.000,0.000}\fontsize{10}{10}\selectfont\strut}%
\psfrag{022}[bc][bc]{\matlabtextA $\bar{\psi}$}%
\psfrag{023}[tc][tc]{\matlabtextA $\bar{t}$}%
\psfrag{024}[tc][tc]{\matlabtextA $n_1$ varying from 0 to 0.4}%
%
%
%
\def\matlabfragNegXTick{\mathord{\makebox[0pt][r]{$-$}}}
\providecommand\matlabtextB{\color[rgb]{0.000,0.000,0.000}\fontsize{8}{8}\selectfont\strut}%
\psfrag{000}[ct][ct]{\matlabtextB $0$}%
\psfrag{001}[ct][ct]{\matlabtextB $10$}%
\psfrag{002}[ct][ct]{\matlabtextB $20$}%
\psfrag{003}[ct][ct]{\matlabtextB $30$}%
\psfrag{004}[ct][ct]{\matlabtextB $40$}%
\psfrag{005}[ct][ct]{\matlabtextB $50$}%
\psfrag{006}[ct][ct]{\matlabtextB $60$}%
\psfrag{007}[ct][ct]{\matlabtextB $70$}%
\psfrag{008}[ct][ct]{\matlabtextB $80$}%
\psfrag{009}[ct][ct]{\matlabtextB $90$}%
\psfrag{010}[ct][ct]{\matlabtextB $100$}%
%
%
%
\psfrag{011}[rc][rc]{\matlabtextB $1$}%
\psfrag{012}[rc][rc]{\matlabtextB $1.2$}%
\psfrag{013}[rc][rc]{\matlabtextB $1.4$}%
\psfrag{014}[rc][rc]{\matlabtextB $1.6$}%
\psfrag{015}[rc][rc]{\matlabtextB $1.8$}%
\psfrag{016}[rc][rc]{\matlabtextB $2$}%
\psfrag{017}[rc][rc]{\matlabtextB $2.2$}%
\psfrag{018}[rc][rc]{\matlabtextB $2.4$}%
\psfrag{019}[rc][rc]{\matlabtextB $2.6$}%
\psfrag{020}[rc][rc]{\matlabtextB $2.8$}%
\psfrag{021}[rc][rc]{\matlabtextB $3$}%
%

%% file: matlab-codes/figs/psi-vs-t-mu1-changing-healing.tex
%
%
\providecommand\matlabtextA{\color[rgb]{0.000,0.000,0.000}\fontsize{10}{10}\selectfont\strut}%
\psfrag{020}[bc][bc]{\matlabtextA $\bar{\psi}$}%
\psfrag{021}[tc][tc]{\matlabtextA $\bar{t}$}%
\psfrag{022}[bc][bc]{\matlabtextA $\bar{\mu}_1$ varying from 0 to 0.5}%
%
%
%
\def\matlabfragNegXTick{\mathord{\makebox[0pt][r]{$-$}}}
\providecommand\matlabtextB{\color[rgb]{0.000,0.000,0.000}\fontsize{8}{8}\selectfont\strut}%
\psfrag{000}[ct][ct]{\matlabtextB $0$}%
\psfrag{001}[ct][ct]{\matlabtextB $10$}%
\psfrag{002}[ct][ct]{\matlabtextB $20$}%
\psfrag{003}[ct][ct]{\matlabtextB $30$}%
\psfrag{004}[ct][ct]{\matlabtextB $40$}%
\psfrag{005}[ct][ct]{\matlabtextB $50$}%
\psfrag{006}[ct][ct]{\matlabtextB $60$}%
\psfrag{007}[ct][ct]{\matlabtextB $70$}%
\psfrag{008}[ct][ct]{\matlabtextB $80$}%
\psfrag{009}[ct][ct]{\matlabtextB $90$}%
\psfrag{010}[ct][ct]{\matlabtextB $100$}%
%
%
%
\psfrag{011}[rc][rc]{\matlabtextB $0.7$}%
\psfrag{012}[rc][rc]{\matlabtextB $0.75$}%
\psfrag{013}[rc][rc]{\matlabtextB $0.8$}%
\psfrag{014}[rc][rc]{\matlabtextB $0.85$}%
\psfrag{015}[rc][rc]{\matlabtextB $0.9$}%
\psfrag{016}[rc][rc]{\matlabtextB $0.95$}%
\psfrag{017}[rc][rc]{\matlabtextB $1$}%
\psfrag{018}[rc][rc]{\matlabtextB $1.05$}%
\psfrag{019}[rc][rc]{\matlabtextB $1.1$}%
%

%% file: matlab-codes/figs/psi-vs-t-mu1-changing-n-nonzero-healing.tex
%
%
\providecommand\matlabtextA{\color[rgb]{0.000,0.000,0.000}\fontsize{10}{10}\selectfont\strut}%
\psfrag{022}[bc][bc]{\matlabtextA $\bar{\psi}$}%
\psfrag{023}[tc][tc]{\matlabtextA $\bar{t}$}%
\psfrag{024}[bc][bc]{\matlabtextA $\bar{\mu}_1$ varying from 0 to 0.5}%
%
%
%
\def\matlabfragNegXTick{\mathord{\makebox[0pt][r]{$-$}}}
\providecommand\matlabtextB{\color[rgb]{0.000,0.000,0.000}\fontsize{8}{8}\selectfont\strut}%
\psfrag{000}[ct][ct]{\matlabtextB $0$}%
\psfrag{001}[ct][ct]{\matlabtextB $10$}%
\psfrag{002}[ct][ct]{\matlabtextB $20$}%
\psfrag{003}[ct][ct]{\matlabtextB $30$}%
\psfrag{004}[ct][ct]{\matlabtextB $40$}%
\psfrag{005}[ct][ct]{\matlabtextB $50$}%
\psfrag{006}[ct][ct]{\matlabtextB $60$}%
\psfrag{007}[ct][ct]{\matlabtextB $70$}%
\psfrag{008}[ct][ct]{\matlabtextB $80$}%
\psfrag{009}[ct][ct]{\matlabtextB $90$}%
\psfrag{010}[ct][ct]{\matlabtextB $100$}%
%
%
%
\psfrag{011}[rc][rc]{\matlabtextB $0.65$}%
\psfrag{012}[rc][rc]{\matlabtextB $0.7$}%
\psfrag{013}[rc][rc]{\matlabtextB $0.75$}%
\psfrag{014}[rc][rc]{\matlabtextB $0.8$}%
\psfrag{015}[rc][rc]{\matlabtextB $0.85$}%
\psfrag{016}[rc][rc]{\matlabtextB $0.9$}%
\psfrag{017}[rc][rc]{\matlabtextB $0.95$}%
\psfrag{018}[rc][rc]{\matlabtextB $1$}%
\psfrag{019}[rc][rc]{\matlabtextB $1.05$}%
\psfrag{020}[rc][rc]{\matlabtextB $1.1$}%
\psfrag{021}[rc][rc]{\matlabtextB $1.15$}%
%

%% file: matlab-codes/figs/psi-vs-t-b1-changing-healing.tex
%
%
\providecommand\matlabtextA{\color[rgb]{0.000,0.000,0.000}\fontsize{10}{10}\selectfont\strut}%
\psfrag{022}[bc][bc]{\matlabtextA $\bar{\psi}$}%
\psfrag{023}[tc][tc]{\matlabtextA $\bar{t}$}%
\psfrag{024}[bc][bc]{\matlabtextA $b_1$ varying from 0 to 0.8}%
%
%
%
\def\matlabfragNegXTick{\mathord{\makebox[0pt][r]{$-$}}}
\providecommand\matlabtextB{\color[rgb]{0.000,0.000,0.000}\fontsize{8}{8}\selectfont\strut}%
\psfrag{000}[ct][ct]{\matlabtextB $0$}%
\psfrag{001}[ct][ct]{\matlabtextB $10$}%
\psfrag{002}[ct][ct]{\matlabtextB $20$}%
\psfrag{003}[ct][ct]{\matlabtextB $30$}%
\psfrag{004}[ct][ct]{\matlabtextB $40$}%
\psfrag{005}[ct][ct]{\matlabtextB $50$}%
\psfrag{006}[ct][ct]{\matlabtextB $60$}%
\psfrag{007}[ct][ct]{\matlabtextB $70$}%
\psfrag{008}[ct][ct]{\matlabtextB $80$}%
\psfrag{009}[ct][ct]{\matlabtextB $90$}%
\psfrag{010}[ct][ct]{\matlabtextB $100$}%
%
%
%
\psfrag{011}[rc][rc]{\matlabtextB $0.86$}%
\psfrag{012}[rc][rc]{\matlabtextB $0.88$}%
\psfrag{013}[rc][rc]{\matlabtextB $0.9$}%
\psfrag{014}[rc][rc]{\matlabtextB $0.92$}%
\psfrag{015}[rc][rc]{\matlabtextB $0.94$}%
\psfrag{016}[rc][rc]{\matlabtextB $0.96$}%
\psfrag{017}[rc][rc]{\matlabtextB $0.98$}%
\psfrag{018}[rc][rc]{\matlabtextB $1$}%
\psfrag{019}[rc][rc]{\matlabtextB $1.02$}%
\psfrag{020}[rc][rc]{\matlabtextB $1.04$}%
\psfrag{021}[rc][rc]{\matlabtextB $1.06$}%
%

%% file: matlab-codes/figs/psi-vs-t-n1-changing-healing.tex
%
%
\providecommand\matlabtextA{\color[rgb]{0.000,0.000,0.000}\fontsize{10}{10}\selectfont\strut}%
\psfrag{019}[bc][bc]{\matlabtextA $\bar{\psi}$}%
\psfrag{020}[tc][tc]{\matlabtextA $\bar{t}$}%
\psfrag{021}[bc][bc]{\matlabtextA $n_1$ varying from 0 to 0.4}%
%
%
%
\def\matlabfragNegXTick{\mathord{\makebox[0pt][r]{$-$}}}
\providecommand\matlabtextB{\color[rgb]{0.000,0.000,0.000}\fontsize{8}{8}\selectfont\strut}%
\psfrag{000}[ct][ct]{\matlabtextB $0$}%
\psfrag{001}[ct][ct]{\matlabtextB $10$}%
\psfrag{002}[ct][ct]{\matlabtextB $20$}%
\psfrag{003}[ct][ct]{\matlabtextB $30$}%
\psfrag{004}[ct][ct]{\matlabtextB $40$}%
\psfrag{005}[ct][ct]{\matlabtextB $50$}%
\psfrag{006}[ct][ct]{\matlabtextB $60$}%
\psfrag{007}[ct][ct]{\matlabtextB $70$}%
\psfrag{008}[ct][ct]{\matlabtextB $80$}%
\psfrag{009}[ct][ct]{\matlabtextB $90$}%
\psfrag{010}[ct][ct]{\matlabtextB $100$}%
%
%
%
\psfrag{011}[rc][rc]{\matlabtextB $0.8$}%
\psfrag{012}[rc][rc]{\matlabtextB $0.85$}%
\psfrag{013}[rc][rc]{\matlabtextB $0.9$}%
\psfrag{014}[rc][rc]{\matlabtextB $0.95$}%
\psfrag{015}[rc][rc]{\matlabtextB $1$}%
\psfrag{016}[rc][rc]{\matlabtextB $1.05$}%
\psfrag{017}[rc][rc]{\matlabtextB $1.1$}%
\psfrag{018}[rc][rc]{\matlabtextB $1.15$}%
%

%% file: matlab-codes/figs/Moment-vs-t-mu1-changing-degrading-radius-point75.tex
%
%
\providecommand\matlabtextA{\color[rgb]{0.000,0.000,0.000}\fontsize{10}{10}\selectfont\strut}%
\psfrag{019}[bc][bc]{\matlabtextA $\bar{M}$}%
\psfrag{020}[tc][tc]{\matlabtextA $\bar{t}$}%
\psfrag{021}[bc][bc]{\matlabtextA $\bar{\mu}_1$ varying from 0 to 0.5}%
%
%
%
\def\matlabfragNegXTick{\mathord{\makebox[0pt][r]{$-$}}}
\providecommand\matlabtextB{\color[rgb]{0.000,0.000,0.000}\fontsize{8}{8}\selectfont\strut}%
\psfrag{000}[ct][ct]{\matlabtextB $0$}%
\psfrag{001}[ct][ct]{\matlabtextB $10$}%
\psfrag{002}[ct][ct]{\matlabtextB $20$}%
\psfrag{003}[ct][ct]{\matlabtextB $30$}%
\psfrag{004}[ct][ct]{\matlabtextB $40$}%
\psfrag{005}[ct][ct]{\matlabtextB $50$}%
\psfrag{006}[ct][ct]{\matlabtextB $60$}%
\psfrag{007}[ct][ct]{\matlabtextB $70$}%
\psfrag{008}[ct][ct]{\matlabtextB $80$}%
\psfrag{009}[ct][ct]{\matlabtextB $90$}%
\psfrag{010}[ct][ct]{\matlabtextB $100$}%
%
%
%
\psfrag{011}[rc][rc]{\matlabtextB $0.35$}%
\psfrag{012}[rc][rc]{\matlabtextB $0.4$}%
\psfrag{013}[rc][rc]{\matlabtextB $0.45$}%
\psfrag{014}[rc][rc]{\matlabtextB $0.5$}%
\psfrag{015}[rc][rc]{\matlabtextB $0.55$}%
\psfrag{016}[rc][rc]{\matlabtextB $0.6$}%
\psfrag{017}[rc][rc]{\matlabtextB $0.65$}%
\psfrag{018}[rc][rc]{\matlabtextB $0.7$}%
%

%% file: matlab-codes/figs/Moment-vs-t-mu1-changing-degrading-radius-point35.tex
%
%
\providecommand\matlabtextA{\color[rgb]{0.000,0.000,0.000}\fontsize{10}{10}\selectfont\strut}%
\psfrag{022}[bc][bc]{\matlabtextA $\bar{M}$}%
\psfrag{023}[tc][tc]{\matlabtextA $\bar{t}$}%
\psfrag{024}[bc][bc]{\matlabtextA $\bar{\mu}_1$ varying from 0 to 0.5}%
%
%
%
\def\matlabfragNegXTick{\mathord{\makebox[0pt][r]{$-$}}}
\providecommand\matlabtextB{\color[rgb]{0.000,0.000,0.000}\fontsize{8}{8}\selectfont\strut}%
\psfrag{000}[ct][ct]{\matlabtextB $0$}%
\psfrag{001}[ct][ct]{\matlabtextB $10$}%
\psfrag{002}[ct][ct]{\matlabtextB $20$}%
\psfrag{003}[ct][ct]{\matlabtextB $30$}%
\psfrag{004}[ct][ct]{\matlabtextB $40$}%
\psfrag{005}[ct][ct]{\matlabtextB $50$}%
\psfrag{006}[ct][ct]{\matlabtextB $60$}%
\psfrag{007}[ct][ct]{\matlabtextB $70$}%
\psfrag{008}[ct][ct]{\matlabtextB $80$}%
\psfrag{009}[ct][ct]{\matlabtextB $90$}%
\psfrag{010}[ct][ct]{\matlabtextB $100$}%
%
%
%
\psfrag{011}[rc][rc]{\matlabtextB $0.18$}%
\psfrag{012}[rc][rc]{\matlabtextB $0.2$}%
\psfrag{013}[rc][rc]{\matlabtextB $0.22$}%
\psfrag{014}[rc][rc]{\matlabtextB $0.24$}%
\psfrag{015}[rc][rc]{\matlabtextB $0.26$}%
\psfrag{016}[rc][rc]{\matlabtextB $0.28$}%
\psfrag{017}[rc][rc]{\matlabtextB $0.3$}%
\psfrag{018}[rc][rc]{\matlabtextB $0.32$}%
\psfrag{019}[rc][rc]{\matlabtextB $0.34$}%
\psfrag{020}[rc][rc]{\matlabtextB $0.36$}%
\psfrag{021}[rc][rc]{\matlabtextB $0.38$}%
%

%% file: matlab-codes/figs/psi-vs-t-mu1-changing-degrading-point75.tex
%
%
\providecommand\matlabtextA{\color[rgb]{0.000,0.000,0.000}\fontsize{10}{10}\selectfont\strut}%
\psfrag{020}[bc][bc]{\matlabtextA $\bar{\psi}$}%
\psfrag{021}[tc][tc]{\matlabtextA $\bar{t}$}%
\psfrag{022}[tc][tc]{\matlabtextA $\bar{\mu}_1$ varying from 0 to 0.5}%
%
%
%
\def\matlabfragNegXTick{\mathord{\makebox[0pt][r]{$-$}}}
\providecommand\matlabtextB{\color[rgb]{0.000,0.000,0.000}\fontsize{8}{8}\selectfont\strut}%
\psfrag{000}[ct][ct]{\matlabtextB $0$}%
\psfrag{001}[ct][ct]{\matlabtextB $10$}%
\psfrag{002}[ct][ct]{\matlabtextB $20$}%
\psfrag{003}[ct][ct]{\matlabtextB $30$}%
\psfrag{004}[ct][ct]{\matlabtextB $40$}%
\psfrag{005}[ct][ct]{\matlabtextB $50$}%
\psfrag{006}[ct][ct]{\matlabtextB $60$}%
\psfrag{007}[ct][ct]{\matlabtextB $70$}%
\psfrag{008}[ct][ct]{\matlabtextB $80$}%
\psfrag{009}[ct][ct]{\matlabtextB $90$}%
\psfrag{010}[ct][ct]{\matlabtextB $100$}%
%
%
%
\psfrag{011}[rc][rc]{\matlabtextB $0.7$}%
\psfrag{012}[rc][rc]{\matlabtextB $0.8$}%
\psfrag{013}[rc][rc]{\matlabtextB $0.9$}%
\psfrag{014}[rc][rc]{\matlabtextB $1$}%
\psfrag{015}[rc][rc]{\matlabtextB $1.1$}%
\psfrag{016}[rc][rc]{\matlabtextB $1.2$}%
\psfrag{017}[rc][rc]{\matlabtextB $1.3$}%
\psfrag{018}[rc][rc]{\matlabtextB $1.4$}%
\psfrag{019}[rc][rc]{\matlabtextB $1.5$}%
%

%% file: matlab-codes/figs/psi-vs-t-mu1-changing-degrading-point35.tex
%
%
\providecommand\matlabtextA{\color[rgb]{0.000,0.000,0.000}\fontsize{10}{10}\selectfont\strut}%
\psfrag{019}[bc][bc]{\matlabtextA $\bar{\psi}$}%
\psfrag{020}[tc][tc]{\matlabtextA $\bar{t}$}%
\psfrag{021}[tc][tc]{\matlabtextA $\bar{\mu}_1$ varying from 0 to 0.5}%
%
%
%
\def\matlabfragNegXTick{\mathord{\makebox[0pt][r]{$-$}}}
\providecommand\matlabtextB{\color[rgb]{0.000,0.000,0.000}\fontsize{8}{8}\selectfont\strut}%
\psfrag{000}[ct][ct]{\matlabtextB $0$}%
\psfrag{001}[ct][ct]{\matlabtextB $10$}%
\psfrag{002}[ct][ct]{\matlabtextB $20$}%
\psfrag{003}[ct][ct]{\matlabtextB $30$}%
\psfrag{004}[ct][ct]{\matlabtextB $40$}%
\psfrag{005}[ct][ct]{\matlabtextB $50$}%
\psfrag{006}[ct][ct]{\matlabtextB $60$}%
\psfrag{007}[ct][ct]{\matlabtextB $70$}%
\psfrag{008}[ct][ct]{\matlabtextB $80$}%
\psfrag{009}[ct][ct]{\matlabtextB $90$}%
\psfrag{010}[ct][ct]{\matlabtextB $100$}%
%
%
%
\psfrag{011}[rc][rc]{\matlabtextB $1.4$}%
\psfrag{012}[rc][rc]{\matlabtextB $1.6$}%
\psfrag{013}[rc][rc]{\matlabtextB $1.8$}%
\psfrag{014}[rc][rc]{\matlabtextB $2$}%
\psfrag{015}[rc][rc]{\matlabtextB $2.2$}%
\psfrag{016}[rc][rc]{\matlabtextB $2.4$}%
\psfrag{017}[rc][rc]{\matlabtextB $2.6$}%
\psfrag{018}[rc][rc]{\matlabtextB $2.8$}%
%

%% file: matlab-codes/figs/moment-vs-t-comparison-diffusivities.tex
%
%
\providecommand\matlabtextA{\color[rgb]{0.000,0.000,0.000}\fontsize{10}{10}\selectfont\strut}%
\psfrag{016}[bc][bc]{\matlabtextA $\bar{M}$}%
\psfrag{017}[tc][tc]{\matlabtextA $\bar{t}$}%
\providecommand\matlabtextB{\color[rgb]{0.000,0.000,0.000}\fontsize{8}{8}\selectfont\strut}%
\psfrag{013}[cl][cl]{\matlabtextB $\bar{D}=0.01$}%
\psfrag{014}[cl][cl]{\matlabtextB $\bar{D}=0.05$}%
\psfrag{015}[cl][cl]{\matlabtextB $\bar{D}=0.1$}%
%
%
%
\def\matlabfragNegXTick{\mathord{\makebox[0pt][r]{$-$}}}
\providecommand\matlabtextC{\color[rgb]{0.000,0.000,0.000}\fontsize{8}{8}\selectfont\strut}%
\psfrag{000}[ct][ct]{\matlabtextC $0$}%
\psfrag{001}[ct][ct]{\matlabtextC $10$}%
\psfrag{002}[ct][ct]{\matlabtextC $20$}%
\psfrag{003}[ct][ct]{\matlabtextC $30$}%
\psfrag{004}[ct][ct]{\matlabtextC $40$}%
\psfrag{005}[ct][ct]{\matlabtextC $50$}%
\psfrag{006}[ct][ct]{\matlabtextC $60$}%
%
%
%
\psfrag{007}[rc][rc]{\matlabtextC $0.25$}%
\psfrag{008}[rc][rc]{\matlabtextC $0.3$}%
\psfrag{009}[rc][rc]{\matlabtextC $0.35$}%
\psfrag{010}[rc][rc]{\matlabtextC $0.4$}%
\psfrag{011}[rc][rc]{\matlabtextC $0.45$}%
\psfrag{012}[rc][rc]{\matlabtextC $0.5$}%
%

%% file: matlab-codes/figs/psi-vs-t-comparison-diffusivities.tex
%
%
\providecommand\matlabtextA{\color[rgb]{0.000,0.000,0.000}\fontsize{10}{10}\selectfont\strut}%
\psfrag{017}[bc][bc]{\matlabtextA $\bar{\psi}$}%
\psfrag{018}[tc][tc]{\matlabtextA $\bar{t}$}%
\providecommand\matlabtextB{\color[rgb]{0.000,0.000,0.000}\fontsize{8}{8}\selectfont\strut}%
\psfrag{014}[cl][cl]{\matlabtextB $\bar{D} = 0.01$}%
\psfrag{015}[cl][cl]{\matlabtextB $\bar{D} = 0.05$}%
\psfrag{016}[cl][cl]{\matlabtextB $\bar{D} = 0.1$}%
%
%
%
\def\matlabfragNegXTick{\mathord{\makebox[0pt][r]{$-$}}}
\providecommand\matlabtextC{\color[rgb]{0.000,0.000,0.000}\fontsize{8}{8}\selectfont\strut}%
\psfrag{000}[ct][ct]{\matlabtextC $0$}%
\psfrag{001}[ct][ct]{\matlabtextC $10$}%
\psfrag{002}[ct][ct]{\matlabtextC $20$}%
\psfrag{003}[ct][ct]{\matlabtextC $30$}%
\psfrag{004}[ct][ct]{\matlabtextC $40$}%
\psfrag{005}[ct][ct]{\matlabtextC $50$}%
\psfrag{006}[ct][ct]{\matlabtextC $60$}%
%
%
%
\psfrag{007}[rc][rc]{\matlabtextC $2$}%
\psfrag{008}[rc][rc]{\matlabtextC $2.5$}%
\psfrag{009}[rc][rc]{\matlabtextC $3$}%
\psfrag{010}[rc][rc]{\matlabtextC $3.5$}%
\psfrag{011}[rc][rc]{\matlabtextC $4$}%
\psfrag{012}[rc][rc]{\matlabtextC $4.5$}%
\psfrag{013}[rc][rc]{\matlabtextC $5$}%
%

%% file: matlab-codes/figs/conc-profile-constant-diff-flux-bc.tex
%
%
\providecommand\matlabtextA{\color[rgb]{0.000,0.000,0.000}\fontsize{10}{10}\selectfont\strut}%
\psfrag{016}[bc][bc]{\matlabtextA $c$}%
\psfrag{017}[tr][tr]{\matlabtextA $\bar{t}$}%
\psfrag{018}[tl][tl]{\matlabtextA $\bar{r}$}%
%
%
%
\def\matlabfragNegXTick{\mathord{\makebox[0pt][r]{$-$}}}
\providecommand\matlabtextB{\color[rgb]{0.000,0.000,0.000}\fontsize{8}{8}\selectfont\strut}%
\psfrag{000}[ct][ct]{\matlabtextB $0.5$}%
\psfrag{001}[ct][ct]{\matlabtextB $0.6$}%
\psfrag{002}[ct][ct]{\matlabtextB $0.7$}%
\psfrag{003}[ct][ct]{\matlabtextB $0.8$}%
\psfrag{004}[ct][ct]{\matlabtextB $0.9$}%
\psfrag{005}[ct][ct]{\matlabtextB $1$}%
%
%
%
\psfrag{006}[rc][rc]{\matlabtextB $0$}%
\psfrag{007}[rc][rc]{\matlabtextB $20$}%
\psfrag{008}[rc][rc]{\matlabtextB $40$}%
\psfrag{009}[rc][rc]{\matlabtextB $60$}%
%
%
%
\psfrag{010}[cr][cr]{\matlabtextB $0$}%
\psfrag{011}[cr][cr]{\matlabtextB $0.2$}%
\psfrag{012}[cr][cr]{\matlabtextB $0.4$}%
\psfrag{013}[cr][cr]{\matlabtextB $0.6$}%
\psfrag{014}[cr][cr]{\matlabtextB $0.8$}%
\psfrag{015}[cr][cr]{\matlabtextB $1$}%
%

%% file: matlab-codes/figs/conc-profile-strain-diff-flux-bc.tex
%
%
\providecommand\matlabtextA{\color[rgb]{0.000,0.000,0.000}\fontsize{10}{10}\selectfont\strut}%
\psfrag{016}[bc][bc]{\matlabtextA $c$}%
\psfrag{017}[tr][tr]{\matlabtextA $\bar{t}$}%
\psfrag{018}[tl][tl]{\matlabtextA $\bar{r}$}%
%
%
%
\def\matlabfragNegXTick{\mathord{\makebox[0pt][r]{$-$}}}
\providecommand\matlabtextB{\color[rgb]{0.000,0.000,0.000}\fontsize{8}{8}\selectfont\strut}%
\psfrag{000}[ct][ct]{\matlabtextB $0.5$}%
\psfrag{001}[ct][ct]{\matlabtextB $0.6$}%
\psfrag{002}[ct][ct]{\matlabtextB $0.7$}%
\psfrag{003}[ct][ct]{\matlabtextB $0.8$}%
\psfrag{004}[ct][ct]{\matlabtextB $0.9$}%
\psfrag{005}[ct][ct]{\matlabtextB $1$}%
%
%
%
\psfrag{006}[rc][rc]{\matlabtextB $0$}%
\psfrag{007}[rc][rc]{\matlabtextB $20$}%
\psfrag{008}[rc][rc]{\matlabtextB $40$}%
\psfrag{009}[rc][rc]{\matlabtextB $60$}%
%
%
%
\psfrag{010}[cr][cr]{\matlabtextB $0$}%
\psfrag{011}[cr][cr]{\matlabtextB $0.2$}%
\psfrag{012}[cr][cr]{\matlabtextB $0.4$}%
\psfrag{013}[cr][cr]{\matlabtextB $0.6$}%
\psfrag{014}[cr][cr]{\matlabtextB $0.8$}%
\psfrag{015}[cr][cr]{\matlabtextB $1$}%
%

%% file: matlab-codes/figs/moment-vs-t-comparison-flux-bc.tex
%
%
\providecommand\matlabtextA{\color[rgb]{0.000,0.000,0.000}\fontsize{10}{10}\selectfont\strut}%
\psfrag{015}[bc][bc]{\matlabtextA $\bar{M}$}%
\psfrag{016}[tc][tc]{\matlabtextA $\bar{t}$}%
\providecommand\matlabtextB{\color[rgb]{0.000,0.000,0.000}\fontsize{7}{7}\selectfont\strut}%
\psfrag{013}[cl][cl]{\matlabtextB constant diffusivity}%
\psfrag{014}[cl][cl]{\matlabtextB strain dependence}%
%
%
%
\def\matlabfragNegXTick{\mathord{\makebox[0pt][r]{$-$}}}
\providecommand\matlabtextC{\color[rgb]{0.000,0.000,0.000}\fontsize{8}{8}\selectfont\strut}%
\psfrag{000}[ct][ct]{\matlabtextC $0$}%
\psfrag{001}[ct][ct]{\matlabtextC $10$}%
\psfrag{002}[ct][ct]{\matlabtextC $20$}%
\psfrag{003}[ct][ct]{\matlabtextC $30$}%
\psfrag{004}[ct][ct]{\matlabtextC $40$}%
\psfrag{005}[ct][ct]{\matlabtextC $50$}%
\psfrag{006}[ct][ct]{\matlabtextC $60$}%
%
%
%
\psfrag{007}[rc][rc]{\matlabtextC $0.25$}%
\psfrag{008}[rc][rc]{\matlabtextC $0.3$}%
\psfrag{009}[rc][rc]{\matlabtextC $0.35$}%
\psfrag{010}[rc][rc]{\matlabtextC $0.4$}%
\psfrag{011}[rc][rc]{\matlabtextC $0.45$}%
\psfrag{012}[rc][rc]{\matlabtextC $0.5$}%
%

%% file: matlab-codes/figs/psi-vs-t-comparison-flux-bc.tex
%
%
\providecommand\matlabtextA{\color[rgb]{0.000,0.000,0.000}\fontsize{10}{10}\selectfont\strut}%
\psfrag{016}[bc][bc]{\matlabtextA $\bar{\psi}$}%
\psfrag{017}[tc][tc]{\matlabtextA $\bar{t}$}%
\providecommand\matlabtextB{\color[rgb]{0.000,0.000,0.000}\fontsize{7}{7}\selectfont\strut}%
\psfrag{014}[cl][cl]{\matlabtextB constant diffusivity}%
\psfrag{015}[cl][cl]{\matlabtextB strain dependence}%
%
%
%
\def\matlabfragNegXTick{\mathord{\makebox[0pt][r]{$-$}}}
\providecommand\matlabtextC{\color[rgb]{0.000,0.000,0.000}\fontsize{8}{8}\selectfont\strut}%
\psfrag{000}[ct][ct]{\matlabtextC $0$}%
\psfrag{001}[ct][ct]{\matlabtextC $10$}%
\psfrag{002}[ct][ct]{\matlabtextC $20$}%
\psfrag{003}[ct][ct]{\matlabtextC $30$}%
\psfrag{004}[ct][ct]{\matlabtextC $40$}%
\psfrag{005}[ct][ct]{\matlabtextC $50$}%
\psfrag{006}[ct][ct]{\matlabtextC $60$}%
%
%
%
\psfrag{007}[rc][rc]{\matlabtextC $2$}%
\psfrag{008}[rc][rc]{\matlabtextC $2.5$}%
\psfrag{009}[rc][rc]{\matlabtextC $3$}%
\psfrag{010}[rc][rc]{\matlabtextC $3.5$}%
\psfrag{011}[rc][rc]{\matlabtextC $4$}%
\psfrag{012}[rc][rc]{\matlabtextC $4.5$}%
\psfrag{013}[rc][rc]{\matlabtextC $5$}%
%

%% file: matlab-codes/figs/conc-profile-constant-diff-diri-bc.tex
%
%
\providecommand\matlabtextA{\color[rgb]{0.000,0.000,0.000}\fontsize{10}{10}\selectfont\strut}%
\psfrag{016}[bc][bc]{\matlabtextA $c$}%
\psfrag{017}[tr][tr]{\matlabtextA $\bar{t}$}%
\psfrag{018}[tl][tl]{\matlabtextA $\bar{r}$}%
%
%
%
\def\matlabfragNegXTick{\mathord{\makebox[0pt][r]{$-$}}}
\providecommand\matlabtextB{\color[rgb]{0.000,0.000,0.000}\fontsize{8}{8}\selectfont\strut}%
\psfrag{000}[ct][ct]{\matlabtextB $0.5$}%
\psfrag{001}[ct][ct]{\matlabtextB $0.6$}%
\psfrag{002}[ct][ct]{\matlabtextB $0.7$}%
\psfrag{003}[ct][ct]{\matlabtextB $0.8$}%
\psfrag{004}[ct][ct]{\matlabtextB $0.9$}%
\psfrag{005}[ct][ct]{\matlabtextB $1$}%
%
%
%
\psfrag{006}[rc][rc]{\matlabtextB $0$}%
\psfrag{007}[rc][rc]{\matlabtextB $20$}%
\psfrag{008}[rc][rc]{\matlabtextB $40$}%
\psfrag{009}[rc][rc]{\matlabtextB $60$}%
%
%
%
\psfrag{010}[cr][cr]{\matlabtextB $0$}%
\psfrag{011}[cr][cr]{\matlabtextB $0.2$}%
\psfrag{012}[cr][cr]{\matlabtextB $0.4$}%
\psfrag{013}[cr][cr]{\matlabtextB $0.6$}%
\psfrag{014}[cr][cr]{\matlabtextB $0.8$}%
\psfrag{015}[cr][cr]{\matlabtextB $1$}%
%

%% file: matlab-codes/figs/conc-profile-strain-diff-diri-bc.tex
%
%
\providecommand\matlabtextA{\color[rgb]{0.000,0.000,0.000}\fontsize{10}{10}\selectfont\strut}%
\psfrag{016}[bc][bc]{\matlabtextA $c$}%
\psfrag{017}[tr][tr]{\matlabtextA $\bar{t}$}%
\psfrag{018}[tl][tl]{\matlabtextA $\bar{r}$}%
%
%
%
\def\matlabfragNegXTick{\mathord{\makebox[0pt][r]{$-$}}}
\providecommand\matlabtextB{\color[rgb]{0.000,0.000,0.000}\fontsize{8}{8}\selectfont\strut}%
\psfrag{000}[ct][ct]{\matlabtextB $0.5$}%
\psfrag{001}[ct][ct]{\matlabtextB $0.6$}%
\psfrag{002}[ct][ct]{\matlabtextB $0.7$}%
\psfrag{003}[ct][ct]{\matlabtextB $0.8$}%
\psfrag{004}[ct][ct]{\matlabtextB $0.9$}%
\psfrag{005}[ct][ct]{\matlabtextB $1$}%
%
%
%
\psfrag{006}[rc][rc]{\matlabtextB $0$}%
\psfrag{007}[rc][rc]{\matlabtextB $20$}%
\psfrag{008}[rc][rc]{\matlabtextB $40$}%
\psfrag{009}[rc][rc]{\matlabtextB $60$}%
%
%
%
\psfrag{010}[cr][cr]{\matlabtextB $0$}%
\psfrag{011}[cr][cr]{\matlabtextB $0.2$}%
\psfrag{012}[cr][cr]{\matlabtextB $0.4$}%
\psfrag{013}[cr][cr]{\matlabtextB $0.6$}%
\psfrag{014}[cr][cr]{\matlabtextB $0.8$}%
\psfrag{015}[cr][cr]{\matlabtextB $1$}%
%

%% file: matlab-codes/figs/moment-vs-t-comparison-diri-bc.tex
%
%
\providecommand\matlabtextA{\color[rgb]{0.000,0.000,0.000}\fontsize{10}{10}\selectfont\strut}%
\psfrag{018}[bc][bc]{\matlabtextA $\bar{M}$}%
\psfrag{019}[tc][tc]{\matlabtextA $\bar{t}$}%
\providecommand\matlabtextB{\color[rgb]{0.000,0.000,0.000}\fontsize{7}{7}\selectfont\strut}%
\psfrag{016}[cl][cl]{\matlabtextB constant diffusivity}%
\psfrag{017}[cl][cl]{\matlabtextB strain dependence}%
%
%
%
\def\matlabfragNegXTick{\mathord{\makebox[0pt][r]{$-$}}}
\providecommand\matlabtextC{\color[rgb]{0.000,0.000,0.000}\fontsize{8}{8}\selectfont\strut}%
\psfrag{000}[ct][ct]{\matlabtextC $0$}%
\psfrag{001}[ct][ct]{\matlabtextC $10$}%
\psfrag{002}[ct][ct]{\matlabtextC $20$}%
\psfrag{003}[ct][ct]{\matlabtextC $30$}%
\psfrag{004}[ct][ct]{\matlabtextC $40$}%
\psfrag{005}[ct][ct]{\matlabtextC $50$}%
\psfrag{006}[ct][ct]{\matlabtextC $60$}%
%
%
%
\psfrag{007}[rc][rc]{\matlabtextC $0.32$}%
\psfrag{008}[rc][rc]{\matlabtextC $0.34$}%
\psfrag{009}[rc][rc]{\matlabtextC $0.36$}%
\psfrag{010}[rc][rc]{\matlabtextC $0.38$}%
\psfrag{011}[rc][rc]{\matlabtextC $0.4$}%
\psfrag{012}[rc][rc]{\matlabtextC $0.42$}%
\psfrag{013}[rc][rc]{\matlabtextC $0.44$}%
\psfrag{014}[rc][rc]{\matlabtextC $0.46$}%
\psfrag{015}[rc][rc]{\matlabtextC $0.48$}%
%

%% file: matlab-codes/figs/psi-vs-t-comparison-diri-bc.tex
%
%
\providecommand\matlabtextA{\color[rgb]{0.000,0.000,0.000}\fontsize{10}{10}\selectfont\strut}%
\psfrag{013}[bc][bc]{\matlabtextA $\bar{\psi}$}%
\psfrag{014}[tc][tc]{\matlabtextA $\bar{t}$}%
\providecommand\matlabtextB{\color[rgb]{0.000,0.000,0.000}\fontsize{7}{7}\selectfont\strut}%
\psfrag{011}[cl][cl]{\matlabtextB constant diffusivity}%
\psfrag{012}[cl][cl]{\matlabtextB strain dependence}%
%
%
%
\def\matlabfragNegXTick{\mathord{\makebox[0pt][r]{$-$}}}
\providecommand\matlabtextC{\color[rgb]{0.000,0.000,0.000}\fontsize{8}{8}\selectfont\strut}%
\psfrag{000}[ct][ct]{\matlabtextC $0$}%
\psfrag{001}[ct][ct]{\matlabtextC $10$}%
\psfrag{002}[ct][ct]{\matlabtextC $20$}%
\psfrag{003}[ct][ct]{\matlabtextC $30$}%
\psfrag{004}[ct][ct]{\matlabtextC $40$}%
\psfrag{005}[ct][ct]{\matlabtextC $50$}%
\psfrag{006}[ct][ct]{\matlabtextC $60$}%
%
%
%
\psfrag{007}[rc][rc]{\matlabtextC $2$}%
\psfrag{008}[rc][rc]{\matlabtextC $2.5$}%
\psfrag{009}[rc][rc]{\matlabtextC $3$}%
\psfrag{010}[rc][rc]{\matlabtextC $3.5$}%
%

%% file: matlab-codes/figs/conc-steady-state-diri-bc.tex
%
%
\providecommand\matlabtextA{\color[rgb]{0.000,0.000,0.000}\fontsize{10}{10}\selectfont\strut}%
\psfrag{024}[bc][bc]{\matlabtextA $c$}%
\psfrag{025}[tc][tc]{\matlabtextA $\bar{r}$}%
\providecommand\matlabtextB{\color[rgb]{0.000,0.000,0.000}\fontsize{7}{7}\selectfont\strut}%
\psfrag{022}[cl][cl]{\matlabtextB constant diffusivity}%
\psfrag{023}[cl][cl]{\matlabtextB strain dependence}%
%
%
%
\def\matlabfragNegXTick{\mathord{\makebox[0pt][r]{$-$}}}
\providecommand\matlabtextC{\color[rgb]{0.000,0.000,0.000}\fontsize{8}{8}\selectfont\strut}%
\psfrag{000}[ct][ct]{\matlabtextC $0.5$}%
\psfrag{001}[ct][ct]{\matlabtextC $0.55$}%
\psfrag{002}[ct][ct]{\matlabtextC $0.6$}%
\psfrag{003}[ct][ct]{\matlabtextC $0.65$}%
\psfrag{004}[ct][ct]{\matlabtextC $0.7$}%
\psfrag{005}[ct][ct]{\matlabtextC $0.75$}%
\psfrag{006}[ct][ct]{\matlabtextC $0.8$}%
\psfrag{007}[ct][ct]{\matlabtextC $0.85$}%
\psfrag{008}[ct][ct]{\matlabtextC $0.9$}%
\psfrag{009}[ct][ct]{\matlabtextC $0.95$}%
\psfrag{010}[ct][ct]{\matlabtextC $1$}%
%
%
%
\psfrag{011}[rc][rc]{\matlabtextC $0$}%
\psfrag{012}[rc][rc]{\matlabtextC $0.1$}%
\psfrag{013}[rc][rc]{\matlabtextC $0.2$}%
\psfrag{014}[rc][rc]{\matlabtextC $0.3$}%
\psfrag{015}[rc][rc]{\matlabtextC $0.4$}%
\psfrag{016}[rc][rc]{\matlabtextC $0.5$}%
\psfrag{017}[rc][rc]{\matlabtextC $0.6$}%
\psfrag{018}[rc][rc]{\matlabtextC $0.7$}%
\psfrag{019}[rc][rc]{\matlabtextC $0.8$}%
\psfrag{020}[rc][rc]{\matlabtextC $0.9$}%
\psfrag{021}[rc][rc]{\matlabtextC $1$}%
%